\documentclass[pdftex,twocolumn,epjc3]{svjour3}          % twocolumn

\RequirePackage[T1]{fontenc}

\smartqed  % flush right qed marks, e.g. at end of proof

\RequirePackage{graphicx}
\RequirePackage{mathptmx}      % use Times fonts if available on your TeX system
\RequirePackage{flushend}
\RequirePackage[numbers,sort&compress]{natbib}
\RequirePackage[colorlinks,citecolor=blue,urlcolor=blue,linkcolor=blue]{hyperref}

\journalname{Eur. Phys. J. A}

\newcommand{\lambdabar}{{\mkern0.75mu\mathchar '26\mkern -9.75mu\lambda}}

\begin{document}

\title{The Fusion-by-Diffusion model as a tool to calculate cross sections for the production of superheavy nuclei}

\author{T.~Cap\thanksref{e1,addr1}
        \and
        M. ~Kowal\thanksref{e2,addr1}
        \and
        K.~Siwek-Wilczy\'nska\thanksref{e3,addr2}}

%\thankstext[$\star$]{t1}{Thanks to the title}
\thankstext{e1}{e-mail: Tomasz.Cap@ncbj.gov.pl}
\thankstext{e2}{e-mail: Michal.Kowal@ncbj.gov.pl}
\thankstext{e3}{e-mail: siwek@fuw.edu.pl}

\institute{National Centre for Nuclear Research, Pasteura 7, 02-093 Warsaw, Poland\label{addr1}
          \and
          Faculty of Physics, University of Warsaw, Pasteura 5, 02-093 Warsaw, Poland\label{addr2}
}

\date{Received: date / Accepted: date}

\maketitle

\begin{abstract}
This article summarizes recent progress in our understanding of the reaction mechanisms leading to the formation of superheavy nuclei in cold and hot fusion reactions. Calculations are done within the Fusion-by-Diffusion (FBD) model using the new nuclear data tables by Jachimowicz \emph{et al.} [At. Data Nucl. Data Tables 138, 101393 (2021)]. The synthesis reaction is treated in a standard way as a three-step process (i.e., capture, fusion, and survival). Each reaction step is analysed separately. Model calculations are compared with selected experimental data on capture, fissionlike and fusion cross sections, fusion probabilities, and evaporation residue excitation functions. The role of the angular momentum in the fusion step is discussed in detail. A set of fusion excitation functions with corresponding fusion probabilities is provided for cold and hot synthesis reactions.

\end{abstract}

\section{Introduction}
One of the biggest challenges in low-energy nuclear physics is the synthesis and study of new superheavy nuclei (SHN). Systematic experimental research performed over the past 30 years has finally led, with the discovery of element 118, oganesson, to the completion of the 7th row of the periodic table~\cite{Hofmann,GSI-5,Riken,Oganessian+2011+429+439,Oganessian_2015}. Unfortunately, experimental attempts to go beyond Og have not been successful so far~\cite{GSI-5,NiU,PhysRevC.79.024603,CrCm,PhysRevC.102.064602}, mainly due to the extremely low production cross sections. Many theoretical models have been developed aiming at describing the SHN synthesis process (see, e.g. the review article by Bao~\cite{Bao} and references therein). The overriding goal for such models is to state the most suitable projectile-target combintion and predict the optimal bombarding energy in the entrance channel at which the production cross-section in a given exit channel is greatest. The other equally important goal is to give a physical explanation of the fusion process in collisions between heavy-ions.

This article provides an overview of the results obtained within the Fusion-by-Diffusion (FBD) model, in which the merging of the colliding ions is described using a diffusion approach. The presented results were obtained using new nuclear data tables for SHN~\cite{Jach2021}, providing a consistent set of masses, deformations, fission barriers, shell corrections, etc. Both cold and hot fusion reactions will be discussed in detail.

By cold fusion reactions, we understand reactions leading to the production of actinide and superheavy nuclei with atomic numbers $102 \le Z \le 113$ in the 1n evaporation channel~\cite{Hofmann,Riken}. In these reactions:
\begin{itemize}
  \item the strongly bound target nuclei ($^{208}$Pb or $^{209}$Bi) are bombarded with projectiles ranging from Ca to Zn;
  \item the excitation energy of the resulting compound nucleus is usually in the range of 10 to 20 MeV;
  \item as the target-projectile symmetry increases, the compound nucleus production cross section decreases.
\end{itemize}

By hot fusion reactions, we understand reactions leading to the synthesis of SHN with atomic numbers $112 \le Z \le 118$ in which~\cite{Oganessian+2011+429+439}:
\begin{itemize}
  \item the deformed actinide target nuclei (from U to Cm) are bombarded with a doubly magic $^{48}$Ca projectile;
  \item the excitation energy of the resulting compound nucleus is usually in the range of 30 to 40 MeV, and the dominant evaporation channels are the 3n and 4n channels;
  \item the evaporation residue cross sections do not show any strong dependence on the target-projectile symmetry and are at the picobarn level.
\end{itemize}

\begin{figure}[h]
\includegraphics[width=1.0\columnwidth,angle=-90]{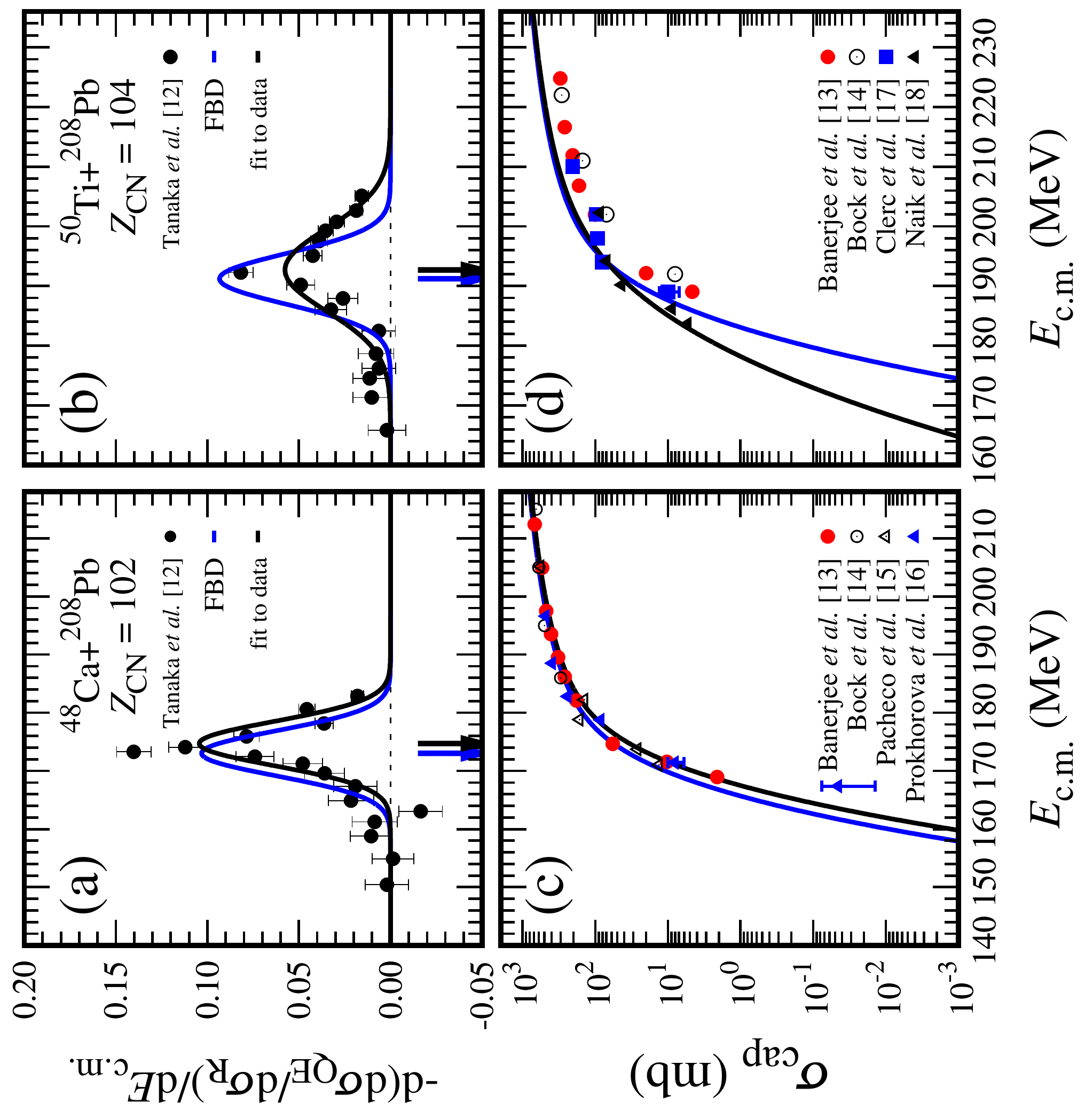}
\caption{\label{Fig_same_interf_factor} Panels (a), (b): entrance channel barrier distributions for $^{48}$Ca and $^{50}$Ti projectiles incident on $^{208}$Pb target. Black points show the experimental data of Tanaka \emph{et al.}~\cite{Tanaka_cold}, blue lines are the predictions of the FBD model, and black lines show Gaussian fits to the data. Panels (c), (d):points represent the experimentally measured capture or fissionlike cross sections taken from Refs.~\cite{HindePRL,BOCK,Pacheco,PROKH,CLERC,Naik}, solid lines show calculations corresponding to the barrier distributions in panels (a) and (b). See text for details.}\label{fig:CaPb_TiPb}
\end{figure}

\begin{figure}[h]
\includegraphics[width=1.0\columnwidth,angle=-90]{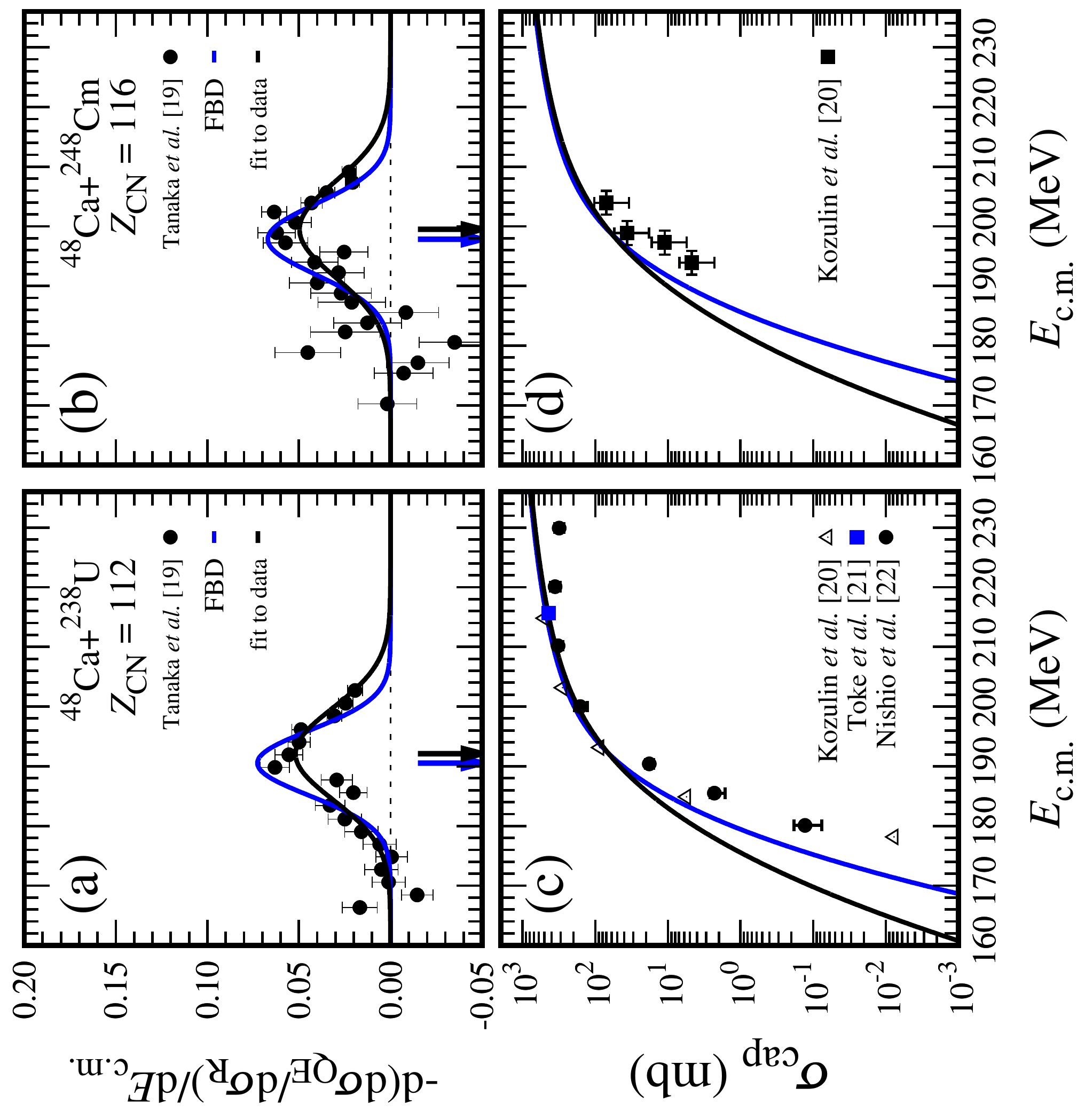}
\caption{\label{Fig_same_interf_factor} Panels (a), (b): entrance channel barrier distributions for the $^{48}$Ca+ $^{238}$U and $^{48}$Ca +$^{248}$Cm reactions. Black points show the experimental data of Tanaka \emph{et al.}~\cite{Tanaka_hot}, blue lines are the predictions of the FBD model, and black lines show Gaussian fits to the data. Panels (c), (d): points represent the experimentally measured capture or fissionlike cross sections taken from Refs.~\cite{KOZULIN2010227,TOKE1985327,Nishio_CaU,Kozlulin_CaUPuCm}, solid lines show calculations corresponding to the barrier distributions in panels (a) and (b). See text for details.}\label{fig:CaU_CaCm}
\end{figure}

\section{FBD model - physical background}

The Fusion-by-Diffusion (FBD) model in its first form was developed as a simple tool to calculate cross sections and optimum bombarding energies for the class of 1n cold fusion reactions \cite{FBD-Acta,FBD-05}. A significant development of this model was the incorporation of the angular momentum dependence, that is, the contributions from successive partial waves to the reaction cross section \cite{FBD-11}. Further development of the model included the incorporation of $x$n channels \cite{FBD-APP-xn}, which made the description of the class of hot fusion reactions possible \cite{PRC-hot,PRCOg}. Recently, the model was extended by including evaporation channels with light-charged particles emission \cite{PRC-ap}.

The fundamental assumption used to describe the formation of superheavy nuclei (SHN) in fusion reactions is Bohr's hypothesis, which implies that all stages of the process are independent. This hypothesis can be justified due to the different time scales of the consecutive steps. Therefore, the partial evaporation residue cross section, $\sigma_{\rm{ER}}(l)$ can be described as a product of the following factors: the partial capture cross section $\sigma_{\rm{cap}}(l) = \pi \lambdabar^2(2l + 1)T(l)$, the fusion probability $P_{\rm{fus}}(l)$, and the survival probability $P_{\rm{surv}}^{x\rm{n}}(l)$. Thus, the total evaporation residue cross section for the production of a given superheavy nucleus in its ground state is
\begin{equation}
\label{factorize}
\sigma_{\rm{ER}} = \pi \lambdabar^2 \sum_{l = 0}^{\infty}(2l+1)T(l)\times P_{\rm{fus}}(l)\times P_{\rm{surv}}^{x\rm{n}}(l),
\end{equation}
where $\lambdabar$ is the wavelength, and $\lambdabar^2=\hbar^2/2\mu E_{\rm{c.m.}}$. Here $\mu$ is the reduced mass of the colliding system, and $E_{\rm{c.m.}}$ is the center-of-mass energy at which the reaction takes place.

Details of the calculations using the FBD model are described in the following subsections. The capture and fusion cross sections are descibed in subsections \ref{sec:capture} and \ref{sec:fusion}, while subsection \ref{sec:surv} deals with the survival probability. Discussion of the results is provided in section \ref{sec:results}. Finally, section \ref{sec:summary} provides a summary and prospects.

\subsection{Capture cross section}\label{sec:capture}

In the FBD approach the capture transmission coefficients $T(l)$ in Eq.~\ref{factorize} are calculated in a simple sharp cut off approximation. The upper limit $l_{\rm{max}}$ of full transmission ($T(l)=1$) is determined from the empirical systematics of the capture cross sections for heavy nuclear systems~\cite{KSW04,FBD-05}.

Based on experimental results, it is assumed that the entrance channel barrier $B$ is not described by a single value but by a distribution that can be well approximated by a Gaussian function
\begin{eqnarray}
\label{B0}
D(B) = \frac{1}{\sqrt{2\pi}\omega}\exp\Big(-\frac{(B-B_0)^2}{2\omega^2}\Big),
\end{eqnarray}
described by two parameters, the mean barrier $B_{0}$ and the distribution width $\omega$~\cite{KSW04,FBD-05}.
By folding the Gaussian barrier distribution with the classical expression for the fusion cross section one can obtain the formula for the capture cross section
\begin{eqnarray}
\label{capture}
\sigma_{\rm{cap}} &=& \pi R^{2} \frac{\omega}{E_{\rm{c.m.}} \sqrt {2\pi}} \Big[X\sqrt{\pi} (1 + \rm{erf}(X))+\exp(-X^{2})\Big] \nonumber\\&=& \pi \lambdabar^2 (l_{\rm{max}}+1)^2,
\end{eqnarray}
where:  $X = \frac{E_{\rm{c.m.}}-B_{0}}{\omega \sqrt{2}}$, and $\textrm{erf}(X)$ is the Gaussian error function.

The free parameters of formula~\ref{capture}: $B_0$, $\omega$, and the normalization factor $R$ are calculated using empirical systematics obtained from analyzing experimentally measured fusion or capture excitation functions for about 50 heavy nuclear systems for which the fusion probability is equal or close to unity~\cite{KSW04}. The distribution width $\omega$ was parametrized, taking into account the $\beta_2$ deformations of both projectile and target nuclei. In this paper we use the parametrization of $B_0$, $\omega$, and $R$ of Ref.~\cite{FBD-11}.

\subsection{Entrance channel barrier distribution}

The entrance channel barrier distribution is a valuable source of information for assessing the impact of structural effects (such as vibrations) or the nucleon transfer processes on the reaction dynamics~\cite{NTSHANGASE200727,doi:10.1146/annurev.nucl.48.1.401,RevModPhys.86.317,RevModPhys7077}. The experimental barrier distributions $D(E_{\rm{c.m.}})$, which give the probability of encountering a barrier of height $B$ equal to $E_{\rm{c.m.}}$, are obtained from precisely measured fusion excitation functions \cite{ROWLEY199125} (for systems with the fusion probability equal or close to one) or quasielastic back-scattering cross section measurements \cite{TIMMERS1995190}.

The entrance channel barrier distribution depends on the deformations of the projectile and target nuclei involved in the reaction and their mutual arrangement. In the case of cold fusion reactions, both $^{208}$Pb and $^{209}$Bi target nuclei, and the vast majority of the projectiles have a spherical shape, which makes the barrier distributions for these reactions very similar to the Gaussian function. In hot fusion reactions, deformed target nuclei are bombarded with the spherical $^{48}$Ca projectile. However, the barrier distribution can still be well approximated by the Gaussian shape. As will be shown later, such an approach reproduces the experimentally measured capture or fissionlike cross-sections reasonably well.

In Fig.~\ref{fig:CaPb_TiPb}, panels (a) and (b), we compare barrier distributions derived from the quasielastic back-scattering data for two cold fusion reactions, $^{48}$Ca + $^{208}$Pb and $^{50}$Ti + $^{208}$Pb \cite{Tanaka_cold}, with formula~\ref{B0}. Experimental barrier distributions,\\
$-{\rm d}({\rm d}\sigma_{{\rm QE}}/{\rm d}\sigma_{\rm R})/{\rm d}E_{{\rm c.m.}}$, were obtained from the measured excitation functions for the quasi-elastic back scattering cross section ($\sigma_{{\rm QE}}$) relative to the Rutherford cross section ($\sigma_{\rm R}$). The blue line shows the barrier distribution obtained with the empirical systematics of $B_0$ and $\omega$ of Ref.~\cite{FBD-11}, while the black line is the fit of formula~\ref{B0} to the experimental data with $B_0$ and $\omega$ as free parameters.

The colored arrows in Fig.~\ref{fig:CaPb_TiPb} indicate the corresponding values of the mean barriers $B_0$ from the fit and the model. It is difficult unequivocally to decide which of the two approaches describes the experimental distributions better. For $^{48}$Ca induced reactions, both methods lead to a similar barrier distribution, while in the case of $^{50}$Ti, the two distributions have different widths. In both cases, calculated and fitted values of the mean barriers do not differ by more than 2 MeV.

The corresponding capture cross sections calculated using Eq.~\ref{capture} are shown in panels (c) and (d) in Fig.~\ref{fig:CaPb_TiPb}, along with the experimental data on capture or fissionlike cross sections from various measurements \cite{HindePRL,BOCK,Pacheco,PROKH,CLERC,Naik}. As before, the blue lines show calculations using parameters
from the systematics. The black lines are the capture cross sections calculated with the $B_0$ and $\omega$ obtained from the fits to the experimental barrier distributions. For the $^{48}$Ca+$^{208}$Pb reaction, the capture excitation functions are simply shifted relative to each other by the difference in $B_0$. For the $^{50}$Ti+$^{208}$Pb reaction, the observed discrepancies result mainly from the difference in the widths, $\omega$.
However, the discrepancies decrease with increasing energy and both functions are in good agreement at energies around and above $B_0$.

Panels (a) and (b) in Fig.~\ref{fig:CaU_CaCm} show the barrier distributions for two $^{48}$Ca induced reactions on actinide targets, $^{238}$U and $^{248}$Cm~\cite{Tanaka_hot}. The data were analyzed in the same way as in Fig.~\ref{fig:CaPb_TiPb}. In both cases, the calculated and fitted mean barrier values do not differ by more than 2 MeV. However, the widths are different, which leads to discrepancies in the predicted capture cross sections (see panels (c) and (d)). Comparison with the experimental data from Refs.~\cite{KOZULIN2010227,TOKE1985327,Nishio_CaU,Kozlulin_CaUPuCm} favors the parametrization used in the FBD model at energies below the mean barrier, $B_0$. For higher incident energies, both capture excitation functions overlap. Since in hot fusion reactions the incident energies are close to or above the mean barrier $B_0$, we consider the parametrization of Ref.~\cite{FBD-11} valid for these reactions.

The target and projectile deformations may generally lead to various configurations of colliding ions in the entrance channel~\cite{Tanaka_hot}. In our simple approach, these effects (although not directly) are partially taken into account in the width $\omega$ and the mean value of the barrier distribution $B_0$. The edges of the barrier distribution correspond to the tip-to-tip orientation on the lower energy side and the equatorial configuration of the two interacting ions on the higher energy side. Since, in hot fusion reactions, all possible orientations can appear, we interpret $B_{0}$ (disregarding vibrations of the nuclei and the couplings of these vibrations to rotations and other dynamical second-order effects) as the barrier height which corresponds to the optimal geometrical configuration in the entrance channel.

For superheavy systems, calculated capture cross sections (Eq.~\ref{capture}) usually exceed the experimental values. There are two reasons for this. The first is related to the difficulty in determining capture or fissionlike cross sections in the experiments. The measured cross sections for the same system made by different groups may differ significantly from each other due to the use of different experimental setups and analysis methods (see lower panels in Figs.~\ref{fig:CaPb_TiPb} and~\ref{fig:CaU_CaCm}).

The second reason is physical in nature. At higher incident energies, especially above $B_0$, the contribution from the higher partial waves to the total cross section increases. The more peripheral collisions lead to projectile-target geometric configurations close to the ``asymmetric'' fission saddle, bringing the system to a fast asymmetric split before reaching an equilibrium state. This effect increases with the increase of the charge asymmetry of the target-projectile system, making the discrepancies between calculations and experimental data greater, especially in the case of cold fusion reactions \cite{HindePRL}. Some authors suggest introducing a scaling of the capture excitation functions to account for this cross-section reduction (see, for example, Fig. 2 in Ref. \cite{HindePRL}). In the FBD model, all the phenomena that reduces the probability of reaching a compound nucleus
 configuration after overcoming the entrance channel barrier are included in the subsequent step of the calculations - the fusion probability.

\begin{figure}[t]
\center{\includegraphics[width=0.6\columnwidth,angle=-90]{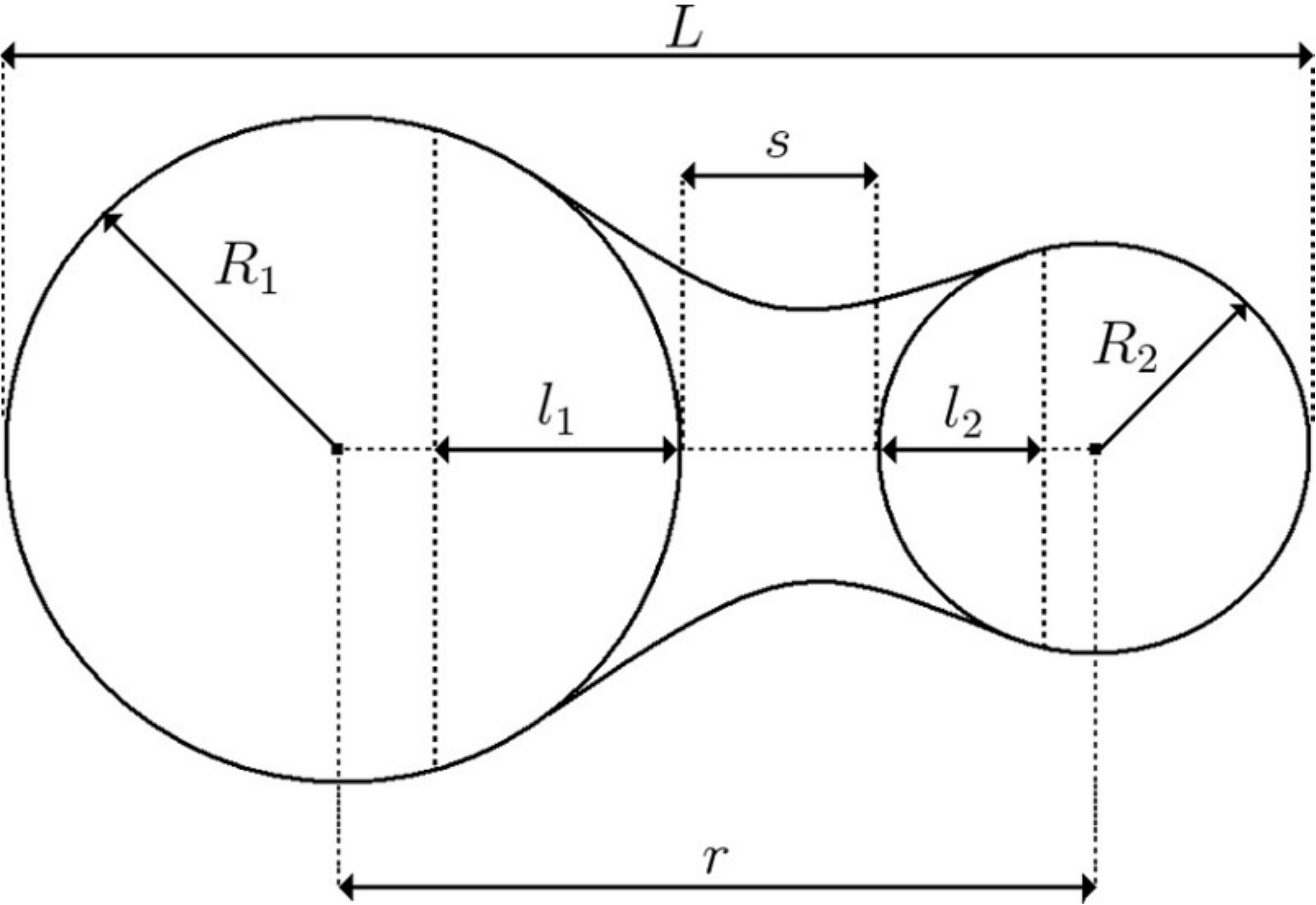}}
\caption{Shape parametrization adapted from Ref.~\cite{Blocki} of the two interacting fragments with radii $R_1$ and $R_2$, the centers of which are at the distance $r$. The total length of the system and the surface separation distance are denoted by $L$ and $s$, respectively. The thickness of missing lenses, $l_1$ and $l_2$, is a measure of the degree of opening of the neck or window through which the fragments communicate.}\label{fig:geo}
\end{figure}

\begin{figure}[t]
\center{\includegraphics[width=0.8\columnwidth,angle=-90]{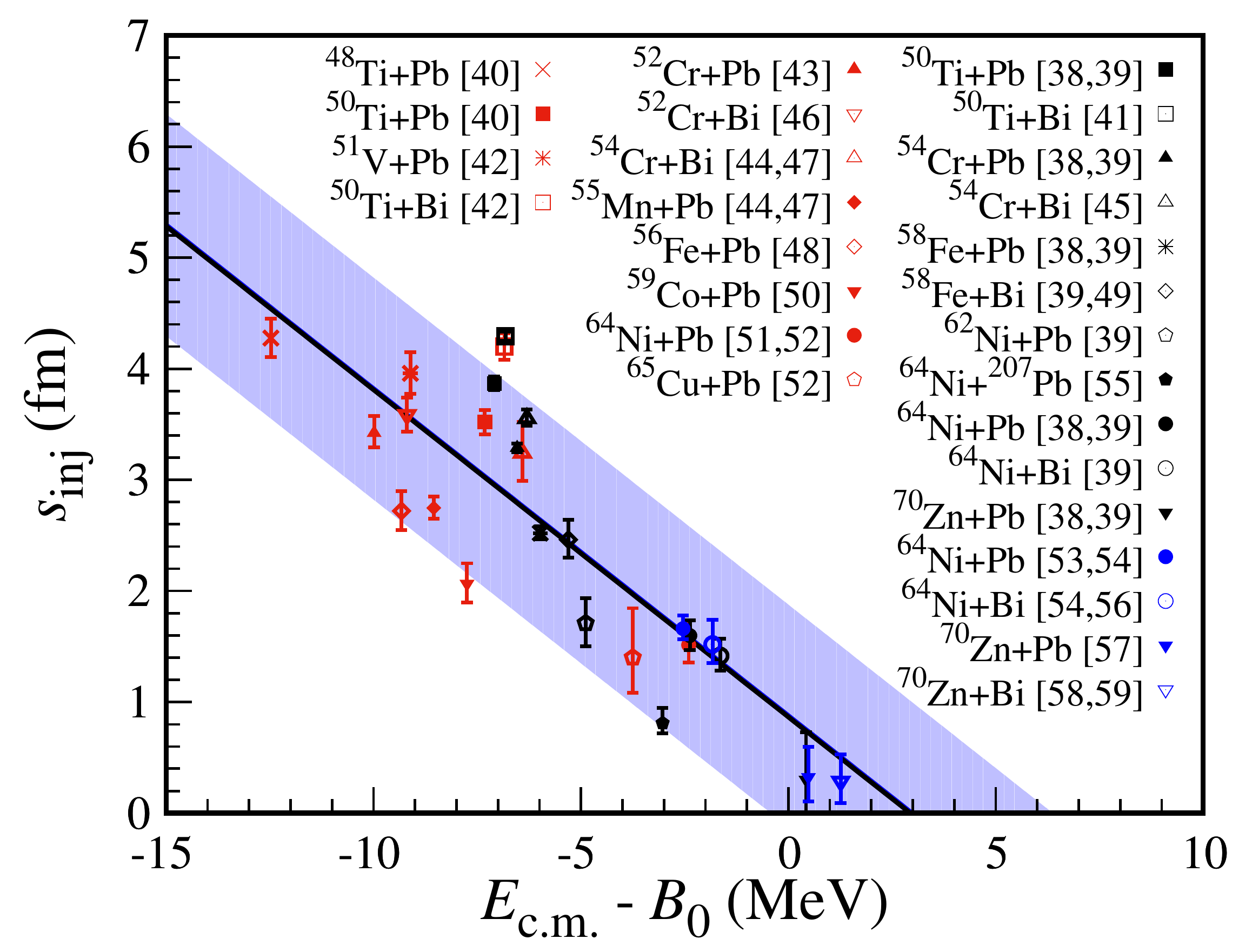}}
\caption{The ``injection point'' systematics obtained for the set of 1n cold fusion reactions \cite{HOFMANN200493,Hofmann_1998,PhysRevC.78.024605,ISI:000172568800008,PhysRevC.78.034604,PhysRevC.79.027602,PhysRevC.78.024606,ISI:A1989U697600008,PhysRevLett.100.022501,PhysRevC.73.014611,PhysRevC.79.011602,ISI:A1997XX26900004,PhysRevC.79.027605,PhysRevC.67.064609,PhysRevLett.93.212702,ISI:000223717300009,MORITA2004101,ISI:000167664500002,doi:10.1143/JPSJ.73.1738,doi:10.1143/JPSJ.76.043201,doi:10.1143/JPSJ.73.2593,New113} using the new nuclear data tables~\cite{Jach2021}. If not shown explicitly, targets were $^{208}$Pb or $^{209}$Bi. The color of the points indicates the laboratory where the reaction was studied: LBNL (red), GSI (black), RIKEN (blue). See text for details.}\label{fig:sinj_cold}
\end{figure}

\subsection{Fusion cross section}\label{sec:fusion}

The next step, the merging of the interacting system and reaching the compound nucleus (CN) configuration, is the least studied and the most difficult to describe part of the formation of superheavy nuclei.

In the FBD model, we assume that just after overcoming the entrance channel barrier, the neck formed between the colliding nuclei grows much faster than the changes in the remaining collective degrees of freedom, i.e., system elongation and its mass asymmetry \cite{FBD-Acta, FBD-05}. This fast neck zipping locates the colliding system at a certain point, which we call the ``injection point''. This point is located in the asymmetric fusion-fission valley of the compound nucleus in three-dimensional (asymmetry, neck parameter, and elongation) potential energy surface and marks the beginning of the diffusion process. At this moment, the available kinetic energy that remains after passing the entrance channel barrier is already transformed into internal degrees of freedom in the over-damped regime.

The shape parametrization used in the model to describe the interacting system is that of two spheres joined smoothly by a third quadratic surface and is adapted from Ref.~\cite{Blocki} and presented in Fig.~\ref{fig:geo}. The elongation of the system is defined as $L = 2(R_1+R_2) + s$, where $R_1$ and $R_2$ are the radii of the two spheres, and $s$ is the distance between their surfaces, which can be negative in case of compact shapes (see Refs.~\cite{Blocki} and~\cite{FBD-11} for more details). The distance separating the surfaces of the two colliding heavy ions when the fusion starts will be denoted as the ``injection point'' distance $s_{\rm inj}$. This distance is the adjustable parameter of the model and is used to calculate the fusion probability. The method of estimating this key model parameter is described later in the text. The elongation of the system corresponding to the ``injection point'' distance $s_{\rm inj}$ will be denoted by $L_{\rm inj}$.

In order to fuse, the system must overcome the saddle separating the ``injection point'' from the compound nucleus configuration. In the diffusion approach, this happens by thermal fluctuations in the shape degrees of freedom. The fusion probability, $P_{\rm fus}(l)$, may be derived by solving the Smoluchowski diffusion equation. (Note that the fusion probability is, in general, an $l$-dependent quantity.) Let us denote the elongation of the system at the macroscopic saddle by $L_{\rm sp}$. When $L_{\rm inj} > L_{\rm sp}$ the barrier separating the ``injection point'' from the compound nucleus configuration is at the front and the system has to climb uphill to overcome the saddle. In the $L_{\rm inj} < L_{\rm sp}$ case the ``injection point'' configuration is more compact than the saddle configuration, and the system is already behind the barrier. In the latter case, the barrier prevents the system from re-separation by reducing the outgoing flux of particles. Assuming that the internal barrier has height $H(l)$ and the form of an inverted parabola one gets \cite{FBD-Acta}
\begin{equation}
\label{eq:Smoluchowski}
  P_{\rm fus}(l) = \frac{1}{2} \left\{
\begin{array}{lr}
   1 + { \rm erf}\sqrt{\frac{H(l)}{T}} & : L_{\rm inj} < L_{\rm sp} \\
   1 - { \rm erf}\sqrt{\frac{H(l)}{T}} & : L_{\rm inj} \ge L_{\rm sp} \\
  \end{array}.
\right.
\end{equation}
where $T$ is the average temperature during the fusion process (see~\cite{FBD-11} for details).

The energy threshold $H(l)$ opposing fusion in Eq.~\ref{eq:Smoluchowski} is calculated as the difference between the energy of the saddle point $E_{\rm sp}$ and the energy of the combined system at the ``injection point'' $E_{\rm inj}$, corrected by the rotational energies of these configurations,
\begin{equation}\label{H}
H(l) = (E_{\rm sp} - E_{\rm inj}) + (E_{\rm sp}^{\rm rot}(l)-E_{\rm inj}^{\rm rot}(l)).
\end{equation}
Energies $E_{\rm sp}$ and $E_{\rm inj}$ are calculated using algebraic expressions listed in Section C in Ref.~\cite{FBD-11} approximating the potential energy surfaces obtained by B\l{}ocki and \'Swi\c{a}tecki~\cite{Blocki}. These surfaces take into account the most important collective variables describing the fusion process, such as mass asymmetry, the neck variable, and the system elongation. Rotational energies at the injection point $E_{\rm inj}^{\rm rot}(l)$ and the saddle point $E_{\rm sp}^{\rm rot}(l)$ are calculated assuming the rigid-body moments of inertia for the particular shapes~\cite{FBD-11}.

In this review, we present new parametrizations of the ``injection point'' distance for both cold and hot fusion reactions. The expression for the ``injection point'' distance $s_{\rm inj}$ can be derived from the experimental data by fitting Eq.~\ref{factorize} to the maxima of the measured evaporation residue cross sections. One expression is used for cold fusion reactions and the other for hot fusion reactions, each of these expressions contains two fitted parameters, see Eqs.~\ref{sinj1} and~\ref{sinj2}, and Ref.~\cite{FBD-11} for the fitting procedure. The ``experimental'' $s_{\rm inj}$ values are obtained in a model - dependent way, assuming particular theoretical values of the ground state masses, fission barrier heights, and other relevant properties of SHN, such as deformation parameters and shell corrections. In this work, all necessary input data were taken from the new nuclear data tables of SHN by Jachimowicz \emph{et al.}~\cite{Jach2021}.

\begin{figure}[t]
\center{\includegraphics[width=0.8\columnwidth,angle=-90]{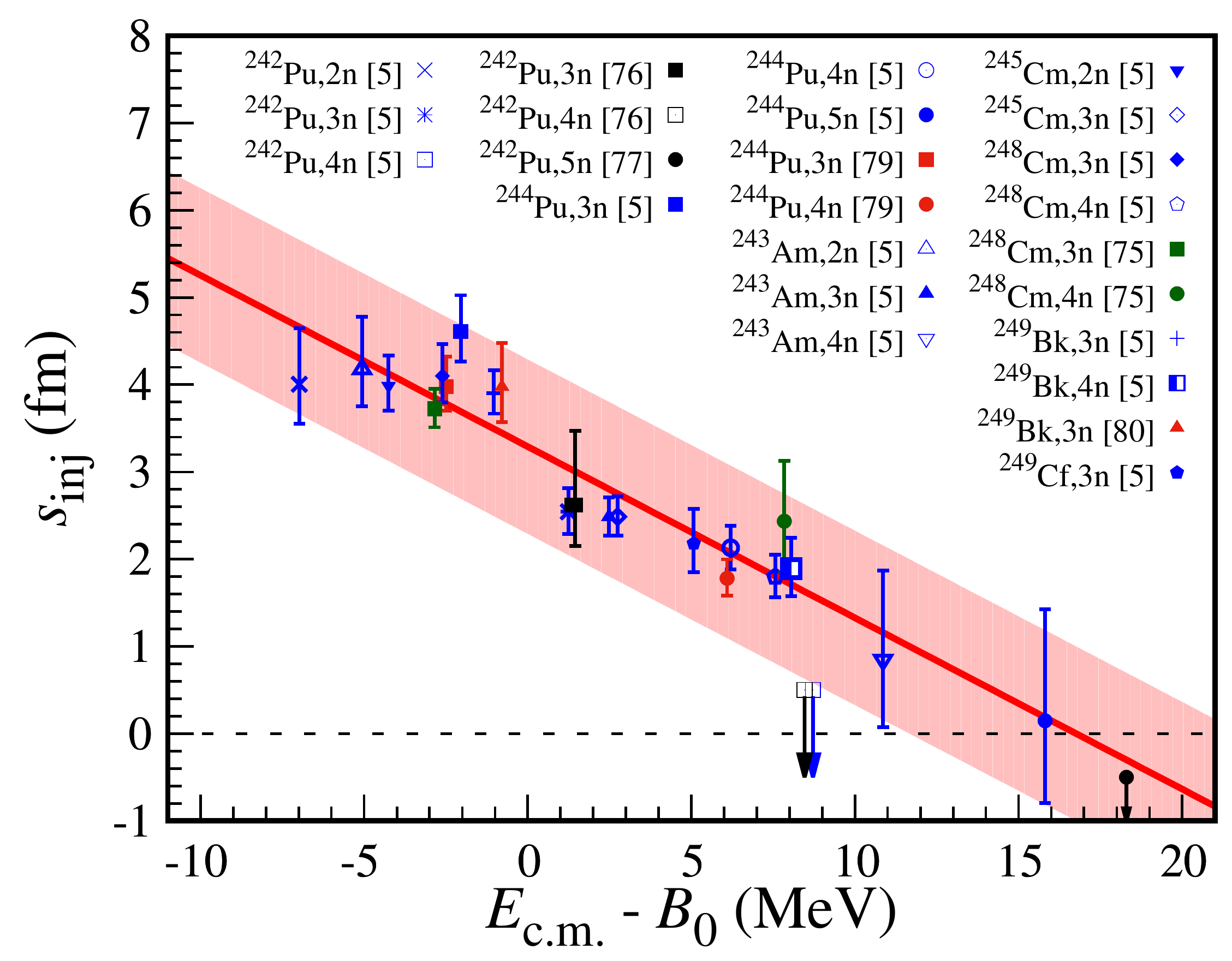}}
\caption{The ``injection point'' systematics obtained for the set of 2n-5n hot fusion reactions with a $^{48}$Ca projectile studied in Refs. \cite{DUBNA-114,DUBNA-116_1,DUBNA-116_2,DUBNA-114+116,DUBNA-112.114.116,DUBNA-116.118,DUBNA-113+115,DUBNA-117,DUBNA-117_2,DUBNA-117+118,DUBNA-113+115+117,DUBNA-115,DUBNA-113.115.117,DUBNA-117_3,GSI-116,LBNL-114,LBNL-114_2,TASCA-114_1,TASCA-114_2,TASCA-117}. Target nucleus and reaction channel are given in the legend for each reaction. The color of the points indicates the experimental setup used to study a given reaction: DGFRS (blue), SHIP (dark-green), TASCA (red), BGS (black). For three reactions only the upper limit of $s_{\rm inj}$ was established.}\label{fig:sinj_hot}
\end{figure}

Fig.~\ref{fig:sinj_cold} shows the new parametrization of the ``injection point distance'' for cold fusion reactions as a function of the excess of the center-of-mass energy $E_{\rm c.m.}$ over the mean barrier $B_{0}$. Each point in the figure represents the value of the $s_{\rm inj}$ distance obtained from the fit to experimentally measured 1n evaporation residue cross sections for 27 cold fusion reactions~\cite{Riken,HOFMANN200493,Hofmann_1998,PhysRevC.78.024605,ISI:000172568800008,PhysRevC.78.034604,PhysRevC.79.027602,PhysRevC.78.024606,ISI:A1989U697600008,PhysRevLett.100.022501,PhysRevC.73.014611,PhysRevC.79.011602,ISI:A1997XX26900004,PhysRevC.79.027605,PhysRevC.67.064609,PhysRevLett.93.212702,ISI:000223717300009,MORITA2004101,ISI:000167664500002,doi:10.1143/JPSJ.73.1738,doi:10.1143/JPSJ.76.043201,doi:10.1143/JPSJ.73.2593,New113}. The systematics can be approximated by a straight line in the form:
\begin{equation}
\label{sinj1}
s_{\rm inj} = 0.878~\textrm{fm} - 0.294\times(E_{\rm c.m.} - B_{0})~\textrm{fm/MeV}.
\end{equation}
A similar (approximately linear) behavior of the $s_{\rm inj}$ distance as a function of $E_{\rm c.m.}-B_0$ was also obtained by solving Langevin-type equations in Ref. \cite{Boilley}.

The new parametrization for hot fusion reactions is shown in Fig.~\ref{fig:sinj_hot}. In this case, $s_{\rm inj}$ values were obtained from 24 evaporation residue cross sections (2n-5n) for $^{48}$Ca reactions incident on various actinide targets measured with DGFRS  \cite{DUBNA-114,DUBNA-116_1,DUBNA-116_2,DUBNA-114+116,DUBNA-112.114.116,DUBNA-116.118,DUBNA-113+115,DUBNA-117,DUBNA-117_2,DUBNA-117+118,DUBNA-113+115+117,DUBNA-115,DUBNA-113.115.117,DUBNA-117_3}, SHIP \cite{GSI-116}, BGS \cite{LBNL-114,LBNL-114_2}, and TASCA \cite{TASCA-114_1,TASCA-114_2,TASCA-117}. This systematics can also be approximated by a straight line:
\begin{equation}\label{sinj2}
s_{\rm inj} = 3.291~\textrm{fm} - 0.196\times(E_{\rm c.m.} - B_{0})~\textrm{fm/MeV}.
\end{equation}

The shaded areas in Figs.~\ref{fig:sinj_cold} and~\ref{fig:sinj_hot} correspond to the $s_{\rm inj}$ error corridors, which were estimated to be $\pm 1$ fm. This will later be used to estimate the uncertainties in the calculated fusion probabilities. The parametrizations given by Eqs.~\ref{sinj1} and ~\ref{sinj2} can be used to predict the fusion probabilities and evaporation-residue cross sections for non-studied colliding systems. However, they should be used for interpolation rather than extrapolation beyond the range of studied values of $ E_{\rm c.m.} - B_{0}$. Negative values of the $s_{\rm inj}$ distance would correspond to a large overlap of the density distributions at the first reaction stage, which is not realistic in low energy nuclear collisions. Therefore, for cold fusion reactions in collisions at energies higher than a few MeV above $B_{0}$, we assume $s_{\rm inj}=0$ (allowing a deviation in the range of the error corridor given). For the hot fusion reactions this limit can be extended to about 15 MeV above $B_{0}$.

The $s_{\rm inj}$ distance determines the relative position between the ``injection point'' and the saddle point, which has to be overcome by the interacting system in order to fuse. The closer the distance the lower the internal fusion barrier (see Eq.~\ref{H}).
In the case of the hot fusion reactions, the saddle is always ``symmetric'' (which means that it is located along the symmetric fission valley and might be associated with the CN fission saddle point). In cold fusion reactions, this is not always the case. For lighter systems, such as $^{48}$Ca + $^{208}$Pb or $^{50}$Ti + $^{208}$Pb, the potential energy surface topology shows two distinct saddles.  In addition to the ``symmetric'' saddle, there is an ``asymmetric'' one which is usually located closer to the ``injection point''. Which saddle has to be overcome depends on the incident energy and angular momentum.

\begin{figure}[t!!]
\centering
\includegraphics[width=0.7\columnwidth,angle=-90]{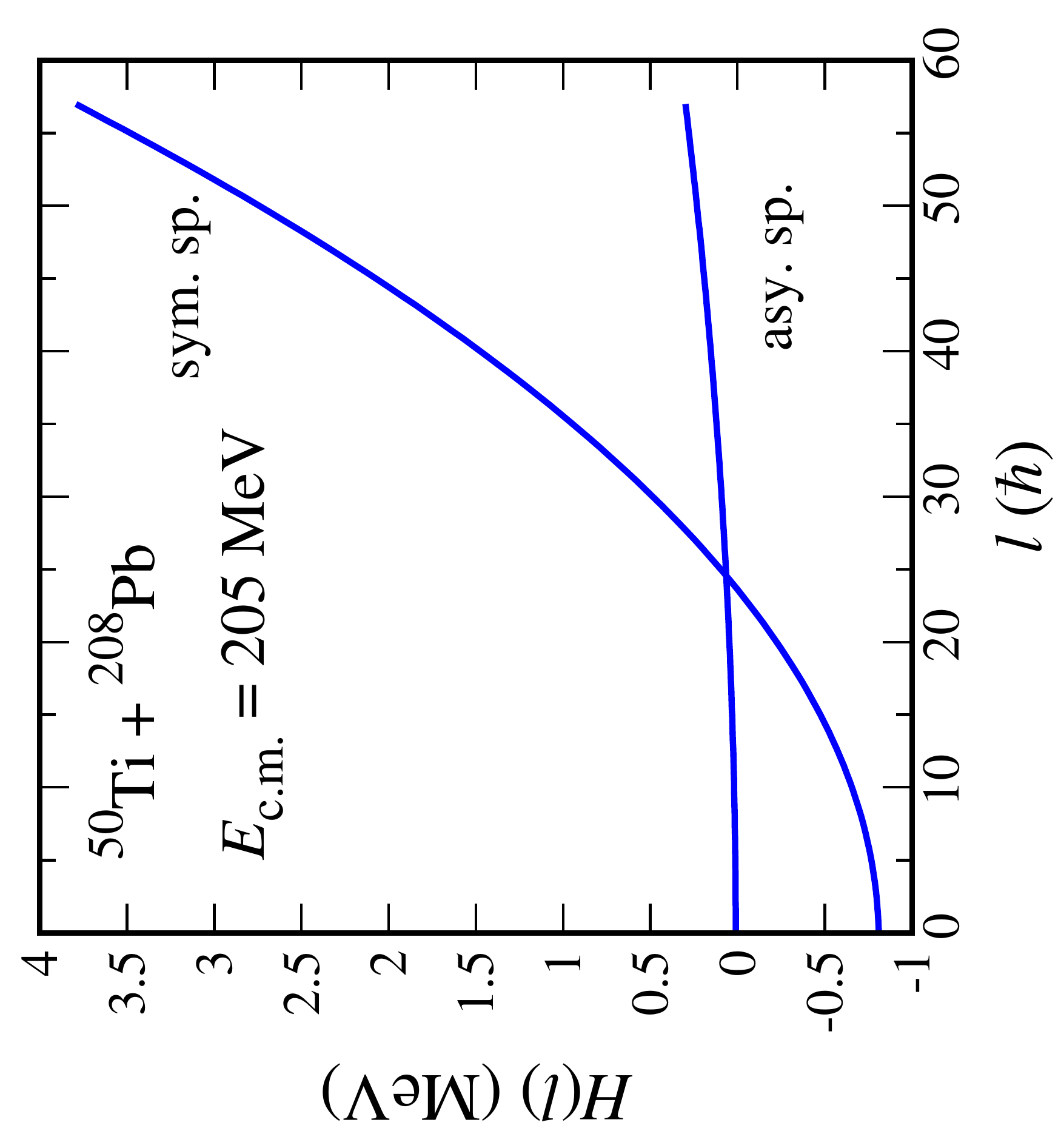}
\caption{The heights $H(l)$ of the internal fusion barriers for the ``symmetric'' and ``asymmetric'' saddles as a function of the angular momentum $l$. Calculations for the $^{50}$Ti + $^{208}$Pb reaction at a center-of-mass energy $E_{\rm c.m.}=205$ MeV. See text for details.}\label{fig:H}
\end{figure}

As an example, the heights $H(l)$ of the internal fusion barriers for ``symmetric'' and ``asymmetric'' saddles are shown in Fig.~\ref{fig:H} for the $^{50}$Ti + $^{208}$Pb reaction. Calculations were done using Eq.~\ref{H} at a center-of-mass energy equal to 205 MeV, thus above $B_0$, when the $s_{\rm inj}=0$ limit is already reached. At lower partial waves, the ``asymmetric'' saddle dominates. After passing this saddle, the system slides down towards the CN configuration (the ``symmetric'' saddle is lower in energy and does not hinder the process). Since the ``symmetric'' saddle is much more compact, at $l \approx 25$, it starts to dominate in the process. As the ``injection point'' is
outside the ``symmetric" saddle, the system must now climb uphill in order to fuse. As one can see, the influence of the rotational energy becomes essential to the fusion hindrance as $H(l)$ quickly increases with increasing $l$.

\begin{figure}[t!!]
\centering
\includegraphics[width=0.85\columnwidth,angle=-90]{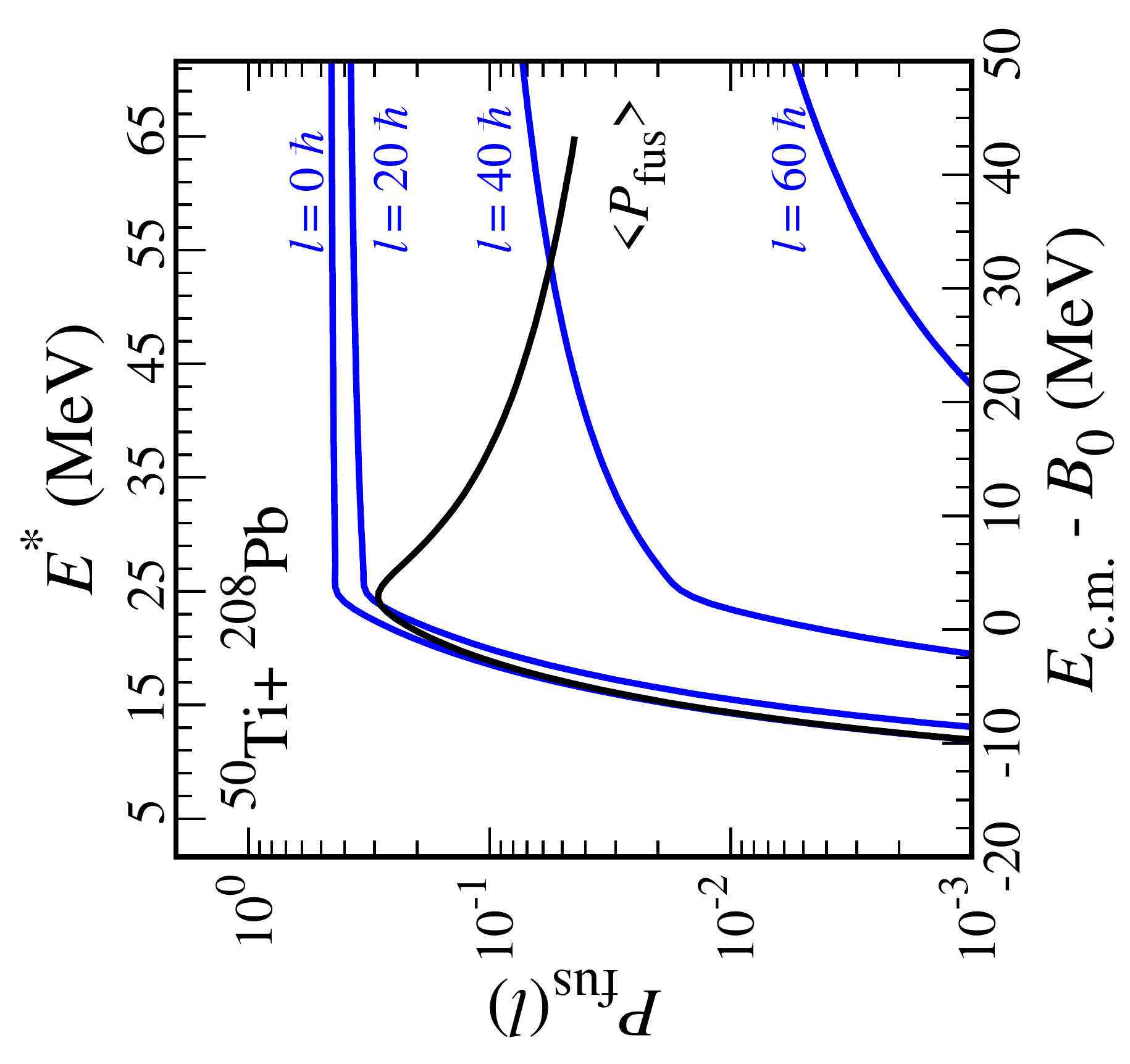}
\caption{The fusion probabilities $P_{\rm fus}(l)$ for the $^{50}$Ti + $^{208}$Pb reaction calculated for selected values of angular momentum $l$ (blue lines) as a function of $ E_{\rm c.m.} - B_{0}$ (lower $x$-scale) or excitation energy $E^*$ (upper $x$-scale). The black line shows the averaged fusion probability $<P_{\rm fus}>$ given by Eq.~\ref{fuseq}.}\label{pfusl}
\end{figure}

Examples of the fusion probability calculations for the $^{50}$Ti + $^{208}$Pb reaction are shown in Fig.~\ref{pfusl}. Blue lines show the fusion probabilities $P_{\rm fus}(l)$ for selected values of angular momentum ($l=0$, 20, 40, and 60) as a function of $ E_{\rm c.m.} - B_{0}$. Let us start the discussion by analyzing central collisions. For $l=0$ the height of the barrier is simply the difference between the ``asymmetric'' saddle point energy and the energy of the combined system of the projectile and target nuclei separated by the distance $s_{\rm inj}$ (see Eq.~\ref{H} and Fig.~\ref{fig:H}). As the available energy increases, the ``injection point distance'' decreases (see Fig.~\ref{fig:sinj_cold}), leading to a lowering of the barrier height. This leads to the rapid growth of the fusion probability, which is eventually stopped when the $s_{\rm inj} = 0$ limit is reached.

For higher partial waves ($l>25$), the more compact, ``symmetric'' saddle becomes dominant (see Fig.~\ref{fig:H}). As the height of the internal barrier $H(l)$ increases with increasing $l$, fusion becomes less and less likely. A steady increase of the fusion probability observed at higher energies is due to the heating up of the interacting system as more of the interaction energy is dissipated (see Eq.~\ref{eq:Smoluchowski}).

It is impossible to measure the fusion probability for a given $l$-value, therefore we define a quantity
\begin{equation}
\label{fuseq}
 <P_{\rm fus}> =\frac {1}{(l_{\rm max}+1)^2}\sum_{l = 0}^{l_{\rm max}}(2l+1)\times P_{\rm fus}(l),
\end{equation}
which is the fusion probability ``averaged'' over all angular momenta contributing to the fusion cross section at a given energy. This formula can be compared with the experimental data, and we will refer to it as the averaged fusion probability.

The averaged fusion probability $<P_{\rm fus}>$ for the $^{50}$Ti + $^{208}$Pb reaction discussed above is shown in Fig.~\ref{pfusl} by the black line.
The dependence of $<P_{\rm fus}>$ on the energy (when $E_{\rm c.m.}>B_{0}$) is based on two opposite effects and may be briefly described as follows: the higher the center-of-mass energy, the more partial waves contribute to the process but, higher $l$-values lead to higher fusion barriers, and consequently smaller partial fusion probabilities $P_{\rm fus}(l)$. Therefore, Eq.~\ref{fuseq} automatically takes into account the physical effect of a limiting angular momentum for fusion process.

The averaged fusion probability in the latter part of this article is used to calculate the fusion cross section which can then be compared with the experimental results:
\begin{equation}
\centering
\label{PCNeq}
 \sigma_{\rm fus} = \pi \lambdabar^2 \sum_{l = 0}^{l_{\rm max}}(2l+1)
 T(l)P_{\rm fus}(l) = \sigma_{\rm cap}\times <P_{\rm fus}>.
\end{equation}

\subsection{Survival probability}\label{sec:surv}

\begin{figure}[t!!]
\centering
\includegraphics[width=1.\columnwidth,angle=-90]{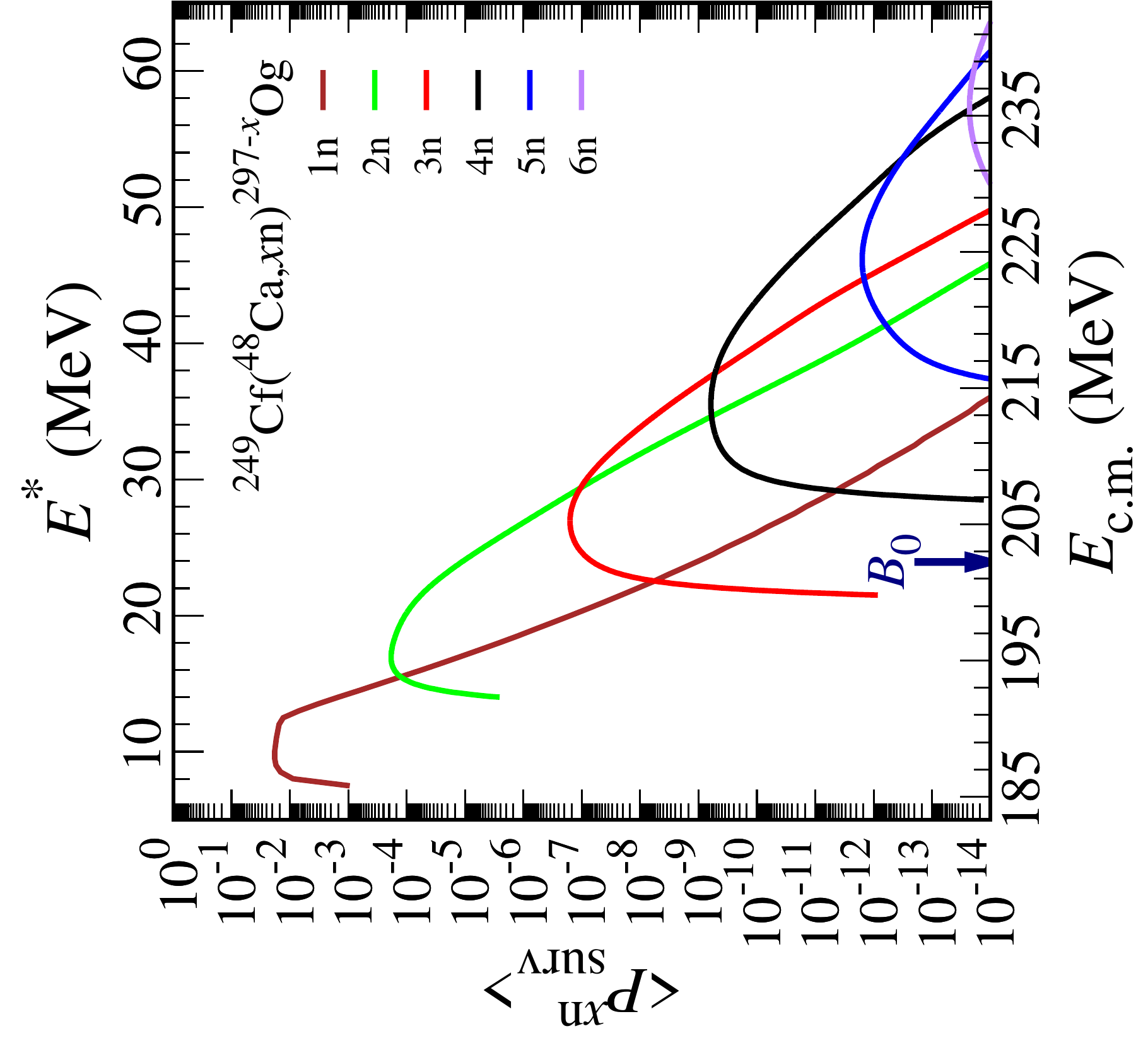}
\caption{Averaged survival probability $<P_{\rm surv}^{x{\rm n}}>$ for the excited $^{297}$Og$^*$ compound nucleus. Neutron evaporation channels from 1n to 6n are shown by colored lines. The lower $x$-axis shows the center-of-mass energy available in the collision, while the upper $x$-axis shows the corresponding excitation energy of $^{297}$Og$^*$. The arrow indicates the value of the mean entrance channel barrier $B_0$.} \label{Psurv_Og}
\end{figure}

The last term in Eq.~\ref{factorize} represents the survival probability $P_{\rm surv}^{x{\rm n}}(l)$, i.e., the probability that the excited fusion product will avoid fission during the deexcitation process and reach the ground state of the final nucleus. The deexcitation can occur through the emission of neutrons, light-charged particles, and gamma rays. However, in the case of excited SHN, usually, only the competition between fission and neutron emission plays a role. The emission of light-charged particles, despite lower separation energies than those for neutrons~\cite{Jach2021}, is hindered by the necessity to overcome the Coulomb barrier~\cite{PRC-ap}. In turn, gamma-ray emission mainly occurs at low excitation energies, below fission thresholds, and does not significantly affect the survival probability. So far, there have been no reports on the observation of light-charged particle emission in the $^{48}$Ca reactions incident on actinide targets. Recently revised data shows that the p channel was populated in the $^{50}$Ti + $^{209}$Bi reaction~\cite{TiBip}, which is the only known case for cold fusion reactions so far. However, channels such as p$x$n and $\alpha x$n might get more attention in the future with the availability of more intense beam currents. The possibility of production of new SHN in these channels is discussed in~\cite{PRC-ap,PhysRevC.94.044606,HONG2020135760,HONG201742}.

Results presented in this study were obtained assuming competition between fission and neutron emission only, using formulas and methods described in~\cite{FBD-11,FBD-APP-xn}. Below we briefly summarize how the survival probability is calculated within the FBD model. One modification regarding the calculation of neutron kinetic energies was introduced and is described later.

The survival probability $P_{\rm surv}^{x{\rm n}}(l)$ is calculated using the standard statistical methods by applying the Weisskopf formula for the neutron emission width

\begin{equation}
\centering\label{Gn}
\Gamma_{\rm n} = \frac{gm_{\rm n}\sigma_{\rm n}}{\pi^2\hbar^2\rho_{\rm G.S.}}\int_0^{X_{\rm n}}\rho_{\rm n}(X_{\rm n}-\epsilon_{\rm n})\epsilon_{\rm n} {\rm d}\epsilon_{\rm n},
\end{equation}
and the conventional expression of transition-state theory for the fission width
\begin{equation}
\centering\label{Gf}
\Gamma_{\rm f} = \frac{1}{2\pi\rho_{\rm G.S.}}\int_0^{X_{\rm f}}\rho_{\rm f}(X_{\rm f}-\epsilon_{\rm f}){\rm d}\epsilon_{\rm f},
\end{equation}
where $\epsilon_{\rm n}$ and $\epsilon_{\rm f}$ are the kinetic energies taken away by the neutron and the two fission fragments, respectively. The integral upper bounds, $X_{\rm n} $ and $X_{\rm f} $, are the maximum available excitation energies of the system after neutron emission or at the fission saddle point. These energies were calculated also taking into account the differences in the rotational energies between the CN and the appropriate final shapes (the daughter nucleus after neutron emission or the fission saddle point). In the formula for the neutron decay width $g$, $m_{\rm n}$, and $\sigma_{\rm n}$ stand for the neutron spin degeneracy, neutron mass, and the cross section for neutron capture in the inverse process. In both formulas, $\rho_{\rm G.S.}$ is the primary compound nucleus level density calculated at its excitation energy, while $\rho_{\rm n}$ is the level density after neutron emission and $\rho_{\rm f}$
at the saddle-point configuration. To evaluate the integrals in Eqs.~\ref{Gn} and~\ref{Gf} we use the Fermi-gas-model level densities $\propto \exp{2\sqrt{aU}}$, where $U$ is the effective excitation energy of the nucleus corrected for rotational and pairing energies. The level density parameters $a_{\rm n}$ and $a_{\rm f}$ characterizing the neutron emission and fission saddle point configurations are calculated using shape-dependent formulas proposed by Reisdorf~\cite{Reisdorf}, with shell effects accounted for by the Ignatyuk formula with standard damping energy~\cite{Ignatiuk}. See~\cite{FBD-11} for details.

\begin{figure*}[h!!!]
\centering
\includegraphics[width=\textwidth,angle=-90,scale=0.40]{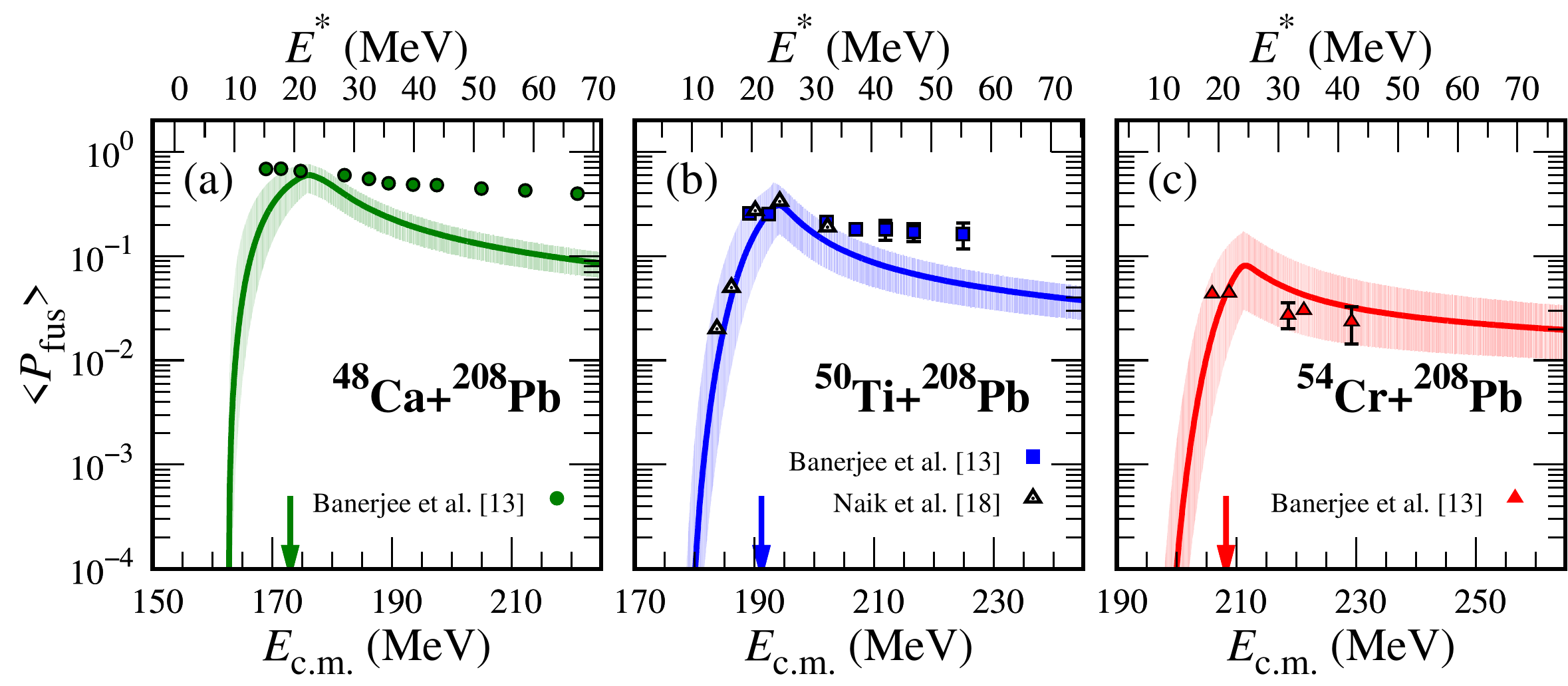}
\caption{The averaged fusion probability $<P_{\rm fus}>$ as a function of the center-of-mass energy for three cold fusion reactions induced by $^{48}$Ca, $^{50}$Ti and $^{54}$Cr projectiles on a $^{208}$Pb target. Experimental data are taken from~\cite{HindePRL,Naik}. The FBD model calculations (Eq.~\ref{fuseq}) are shown by solid lines together with error corridors. Arrows show the positions of the entrance channel barriers $B_0$.}\label{fig:PCN}
\end{figure*}

In cold fusion reactions, when only one neutron is emitted in the deexcitation process, the survival probability is simply given by the ratio of the neutron decay width $\Gamma_{\rm n}$ and the total neutron plus fission width $\Gamma_{\rm tot} = \Gamma_{\rm n}+\Gamma_{\rm f}$, multiplied by the probability $P_<$ that the excitation energy (after the emission of the neutron) is less than the threshold for next chance fission or next neutron emission, whichever is lower (see  Eq. 32 in ~\cite{FBD-11}). In hot fusion reactions, more neutrons can be emitted due to the higher excitation energy of the compound nucleus, and the survival probability is given by the standard expression
\begin{equation}
\centering\label{Psurvxn}
P_{\rm surv}^{x{\rm n}}(l) = \prod_{i=1}^{x}\Big(\frac{\Gamma_{\rm n}}{\Gamma_{\rm n}+\Gamma_{\rm f}}\Big)_{i} \times P_<,
\end{equation}
where $x$ indicates the number of emitted particles. The expression in parentheses is calculated for each step of the evaporation cascade using appropriate decay widths, while $P_<$ applies only for a specified final channel. Calculations take into account the change in excitation energy due to neutron emission while spin reduction is not considered ($s$-wave emission).

In this paper, we modify the method of calculating the neutron kinetic energy $\epsilon_{\rm k}$ carried away at each evaporation step. In our previous papers, we assumed that each neutron carries away an energy equal to the expected value $\bar{\epsilon_{\rm k}}$ resulting from the shape of the neutron evaporation spectrum (which is proportional to the expression $\rho_{\rm n}(X_{\rm n}-\epsilon_{\rm n})\epsilon_{\rm n}$ under the integral in Eq.~\ref{Gn}). Such a simplification allowed us to calculate evaporation residue cross sections without using Monte Carlo methods~\cite{FBD-APP-xn}.

The exact spectrum can be well approximated by the standard Maxwell-type neutron evaporation spectrum proportional to $\epsilon_{\rm k} \exp(-\epsilon_{\rm k} / T)$, where $T$ is the temperature of the transition state for neutron emission (at a given excitation energy). In this paper, neutron kinetic energies are randomly selected from Maxwell-type distribution, and the final survival probability is obtained using the Monte Carlo method. This approach better describes the overall shape of the experimentally observed excitation functions. The evaporation residue cross section maxima obtained with the Monte Carlo method are in agreement with the values obtained using the simplified method.

Decay widths are sensitive to the input data used, such as ground state masses, fission saddle point masses (fission barrier heights), shell corrections, and nuclear deformations. The dependence of the level density parameters on the excitation energy may also play a significant role in the competition of different evaporation channels, as recently demonstrated in Refs.~\cite{level1,level2}. Undoubtedly, the heights of the fission barriers are decisive for the probability of survival. For instance, a change of 1 MeV in the height of the fission barrier at some stage of the evaporation cascade may result in a more than one order of magnitude change in the survival probability for consecutive channels~\cite{PRC-hot}. Therefore, it is crucial in the calculations to use a coherent set of nuclear data which reliably describes the properties of SHN.

In this study all necessary input data were taken from \cite{Jach2021}, where calculations of basic nuclear properties for 1305 heavy and superheavy nuclei with $Z=98-126$ and $N=134-192$ were performed using the microscopic-macroscopic Warsaw method with a deformed Woods-Saxon single-particle potential~\cite{dudek} and the Yukawa-plus-exponential macroscopic energy~\cite{krappe} taken as the smooth part. Ground-state shapes and masses were found by minimization over seven axially-symmetric deformations. A search for the fission saddle points was performed using the ``imaginary water flow'' method with five- (for nonaxial shapes) and seven-dimensional (for reflection-asymmetric shapes) deformation spaces. For systems with odd numbers of protons, neutrons, or both, a standard BCS method with blocking was used.

An example of the survival probability calculations for the excited $^{297}$Og$^*$ nucleus formed in the fusion of $^{48}$Ca and $^{249}$Cf nuclei is shown in Fig.~\ref{Psurv_Og}. The survival probability for each channel was averaged over all angular momenta contributing to the fusion cross section at a given energy in the same way as the fusion probability in Eq.~\ref{fuseq}. In the region of SHN, the fission barrier heights are comparable to the neutron separation energies. In fact, for most of the superheavy nuclei considered in this article, fission barrier heights are lower than neutron separation energies, which highly reduces the survival probability. This can be seen in the discussed example, where each further emitted neutron reduces the chance of surviving fission by at least two orders of magnitude.

\begin{figure}[t]
\center{\includegraphics[width=0.8\columnwidth,angle=-90]{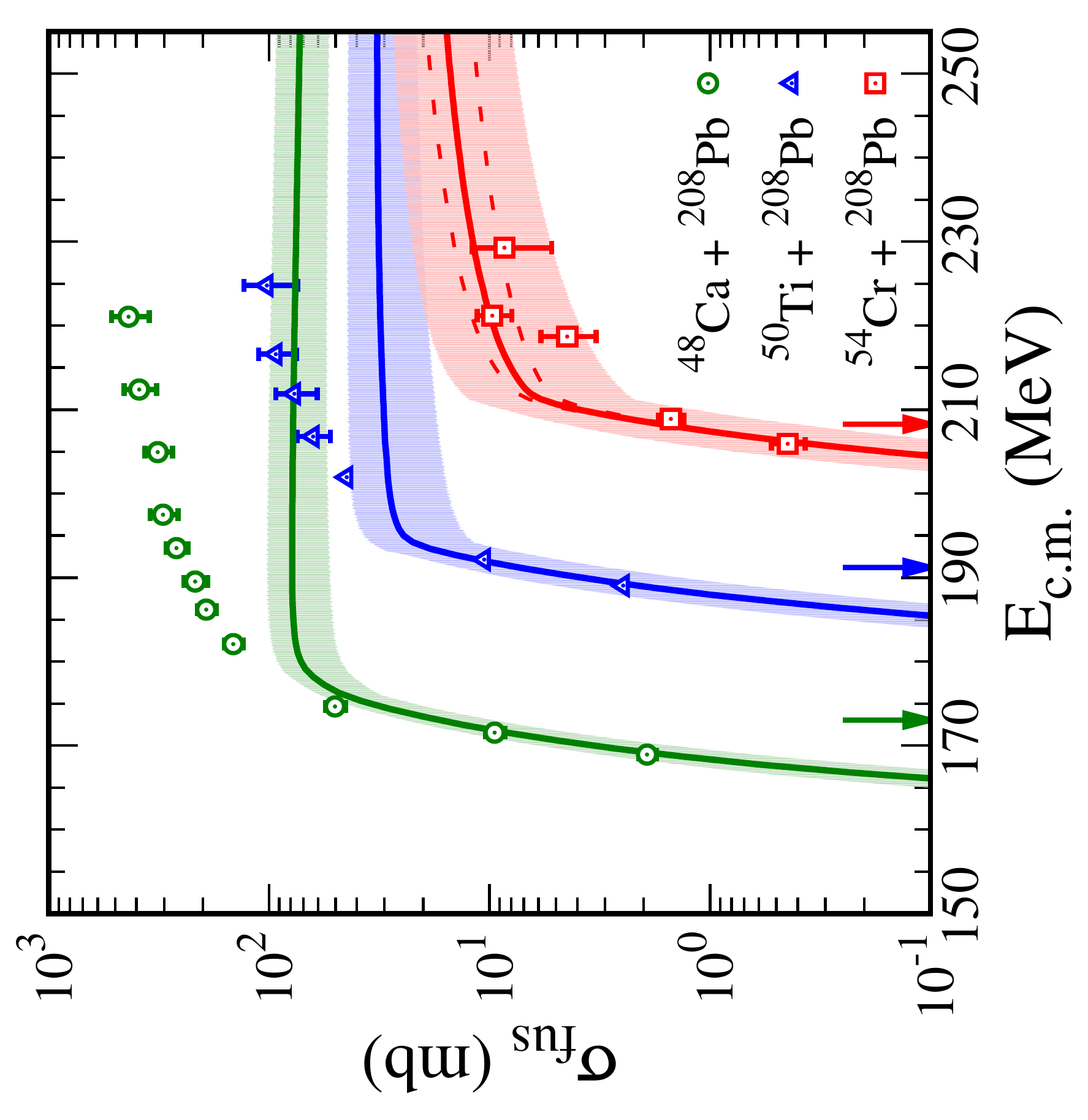}}
\caption{Compound nucleus formation cross sections, $\sigma_{\rm fus}$, for the $^{48}$Ca +$^{208}$Pb, $^{50}$Ti +$^{208}$Pb, and $^{54}$Ca +$^{208}$Pb reactions. Points were derived from the experimental data presented in Ref.~\cite{HindePRL}. Solid lines show the predictions of the FBD model. Dashed lines show calculations for two extreme orientations of spherical target and $^{54}$Cr projectile in the entrance channel. The arrows indicate the values of the mean entrance channel barrier, $B_{0}$, for each reaction. See text for details. Adapted from~\cite{PRC-Pfus}.}\label{fig:fis}
\end{figure}

\section{Results and discussion}\label{sec:results}

\begin{figure}[t]
\centering
\includegraphics[width=1.3\columnwidth,angle=-90]{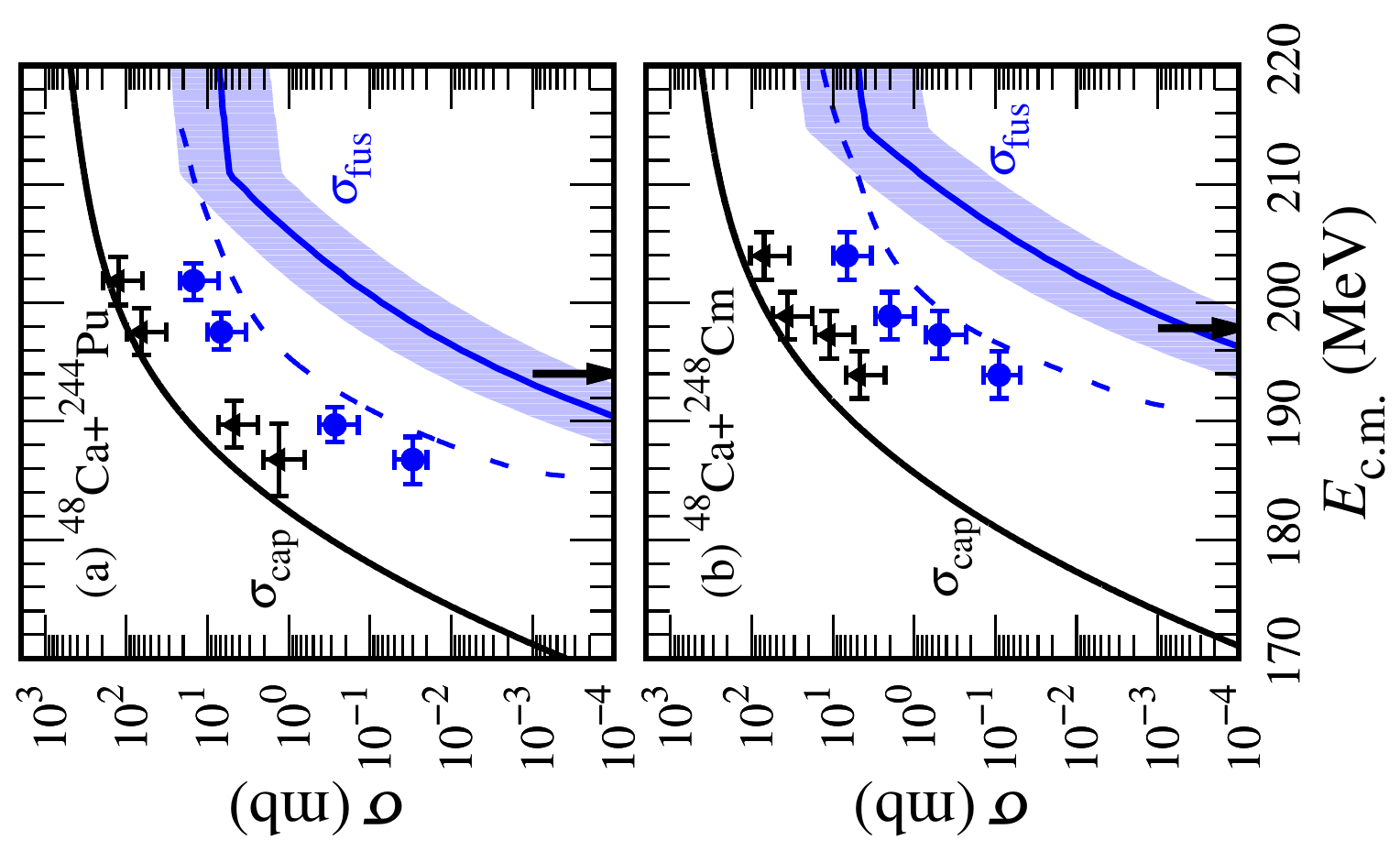}
\caption{The experimentally measured fissionlike cross sections (black triangles), cross sections for mass-symmetric fission $\sigma_{\rm sym}$ (within an $A_{\rm CN}/2 \pm 20$ mass gate; blue circles), and the component of these mass-symmetric fission events attributed to fusion-fission (blue-dashed lines) for $^{48}$Ca + $^{244}$Pu (panel (a)) and $^{48}$Ca + $^{248}$Cm (panel (b))~\cite{Kozlulin_CaUPuCm}. Black and blue solid lines represent capture cross sections $\sigma_{\rm cap}$ and fusion cross sections $\sigma_{\rm fus}$ (with error corridors) calculated within the FBD model.}\label{fig:Ca+PuCm}
\end{figure}

As presented in the previous sections, the FBD model can describe different stages of the process leading to the formation of SHN. In this section, we compare the model predictions with selected experimental data on fusion probabilities, fusion cross sections, and evaporation residue cross sections.
These comparisons need to be taken with caution. In addition to the different methods of experimental data analysis applied by distinct groups, such as background subtraction, mass gates and detection efficiencies, some problems arise from the fact that the fusion probability can not be directly measured. However, some quantitative and qualitative comparisons are possible.

We will begin the discussion with the fusion probabilities for cold synthesis reactions. Unfortunately, the amount of experimental data on this topic is rather limited. Recently, the symmetric-peaked fission cross sections $\sigma_{\rm sym}$ were measured for $^{48}$Ca, $^{50}$Ti, and $^{54}$Cr projectiles incident on a $^{208}$Pb target~\cite{HindePRL}. The measurements allowed upper limits on the CN formation probabilities $P_{\rm sym}$ (which can be compared with the calculated fusion probabilities) to be estimated at energies around and above the interaction barrier $B_0$ for all three reactions. The $P_{\rm sym}$ probabilities were derived in Ref.~\cite{HindePRL} in a model-dependent way by dividing measured $\sigma_{\rm sym}$ cross sections by scaled measured total fission-like cross sections. Scaling factors were estimated using the CCFULL model based on the coupled channels formalism~\cite{Hagino}.

The comparison of the $P_{\rm sym}$ values from Ref.~\cite{HindePRL} with the averaged fusion probabilities $<P_{\rm fus}>$ (Eq.~\ref{fuseq}) is show in Fig.~\ref{fig:PCN}. The FBD model calculations are shown with the error corridors resulting from the uncertainty in the ``injection point'' systematics (see Fig.~\ref{sinj1}). Data from Ref.~\cite{HindePRL} cover energies around and above the interaction barrier $B_0$ only. However, for the $^{50}$Ti + $^{208}$Pb reaction, additional data for energies up to 10 MeV below $B_0$ are also available~\cite{Naik} (open triangles in Fig.~\ref{fig:PCN}b).

For $^{48}$Ca and $^{50}$Ti induced reactions on a $^{208}$Pb target, the data points from Ref.~\cite{HindePRL} are clearly above the model calculations,
but as mentioned previously, these data represent the upper limit to the fusion probabilities. For the $^{50}$Ti + $^{208}$Pb reaction
the steep fall in the fusion probability at the sub-barrier energies is exactly reproduced ($B_0$ values are indicated by the arrows in Fig.~\ref{fig:PCN}). For the $^{54}$Cr + $^{208}$Pb reaction the calculations are in agreement with the data within the error corridor. In this case the model predictions slightly surpass the experimental data. Experimental fusion probabilities for this reaction were extracted using more restricted mass gates applied to symmetric fission fragments and a different method of background subtraction than for the reactions with $^{48}$Ca and $^{50}$Ti projectiles.

In Fig.~\ref{fig:fis} we show the calculated fusion cross sections (Eq.~\ref{PCNeq}) together with the experimentally measured symmetric-peaked cross sections $\sigma_{\rm sym}$ from Ref.~\cite{HindePRL} for the same three cold fusion reactions. (The $\sigma_{\rm sym}$ values are not given explicitly in~\cite{HindePRL}. However, they can be easily derived using data presented in the paper and corresponding supplementary material. See Ref.~\cite{PRC-Pfus} for details.). As can be seen in Fig.~\ref{fig:fis}, the model works best in the most important energy region, i.e., around the mean entrance channel barrier energies, thus close to the optimal bombardment energies for 1n reaction channels. At higher center-of-mass energies, the experimentally measured fusion cross sections exceed the calculated values by several times. Even if the $\sigma_{\rm sym}$ values represent the upper limits to the fusion cross sections, such discrepancies are not expected. To obtain the fusion cross section, one has to subtract the background from symmetric and asymmetric quasi-fission events, which, as the authors of~\cite{HindePRL} pointed out, is not straightforward. One can also expect some uncertainties in the model calculations. The parabolic approximation of the shape of the internal fusion barrier that we use allows the solution of the Smoluchowski diffusion equation in the analytical form only if the temperature is sufficiently low and the thermal fluctuations are small. These conditions are not met at high excitation energies. The calculated values of the fusion cross sections at excitation energies above 40 MeV should be treated as extrapolations. More detailed analysis of the $^{48}$Ca, $^{50}$Ti, and $^{54}$Cr reactions with a $^{208}$Pb target done within the FBD model framework can be found in Ref~\cite{PRC-Pfus}.

Fusion probabilities for hot fusion reactions were extracted from several experiments (see for instance Refs.~\cite{KOZULIN2010227,Kozlulin_CaUPuCm,PhysRevC.83.064613,PhysRevC.94.054613}). These measurements were all analyzed using similar methods. We will discuss here the results for two reactions only, namely $^{48}$Ca + $^{244}$Pu and $^{48}$Ca + $^{248}$Cm, reported in Ref.~\cite{Kozlulin_CaUPuCm}. The experimentally measured fissionlike cross sections, the cross sections for mass-symmetric fission $\sigma_{\rm sym}$ (within an $A_{\rm CN}/2 \pm 20$ mass gate), and the component of these mass-symmetric fission events attributed to fusion-fission for these two reactions are shown in Fig.~\ref{fig:Ca+PuCm} by black triangles, blue circles, and blue-dashed lines respectively. The experimental values of $\sigma_{\rm sym}$ comprise quasifission and fusion-fission events (there was no background subtraction), while the estimated fusion-fission component is equivalent to the upper limit for the fusion cross section for the respective reaction (see Ref~\cite{Kozlulin_CaUPuCm} for more details). The FBD model calculations are also shown in the figure. Black solid lines are the capture cross sections calculated using Eq.~\ref{capture}, and blue solid lines with error corridors are the fusion cross sections obtained with Eq.~\ref{PCNeq}. The shapes of the fissionlike excitation functions are quite well reproduced within the model. However, the calculated fusion cross sections are well below the experimentally estimated upper limits (blue-dashed lines on Fig.~\ref{fig:Ca+PuCm}).

\begin{figure}[t!!]
\center{\includegraphics[width=1.3\columnwidth,angle=-90]{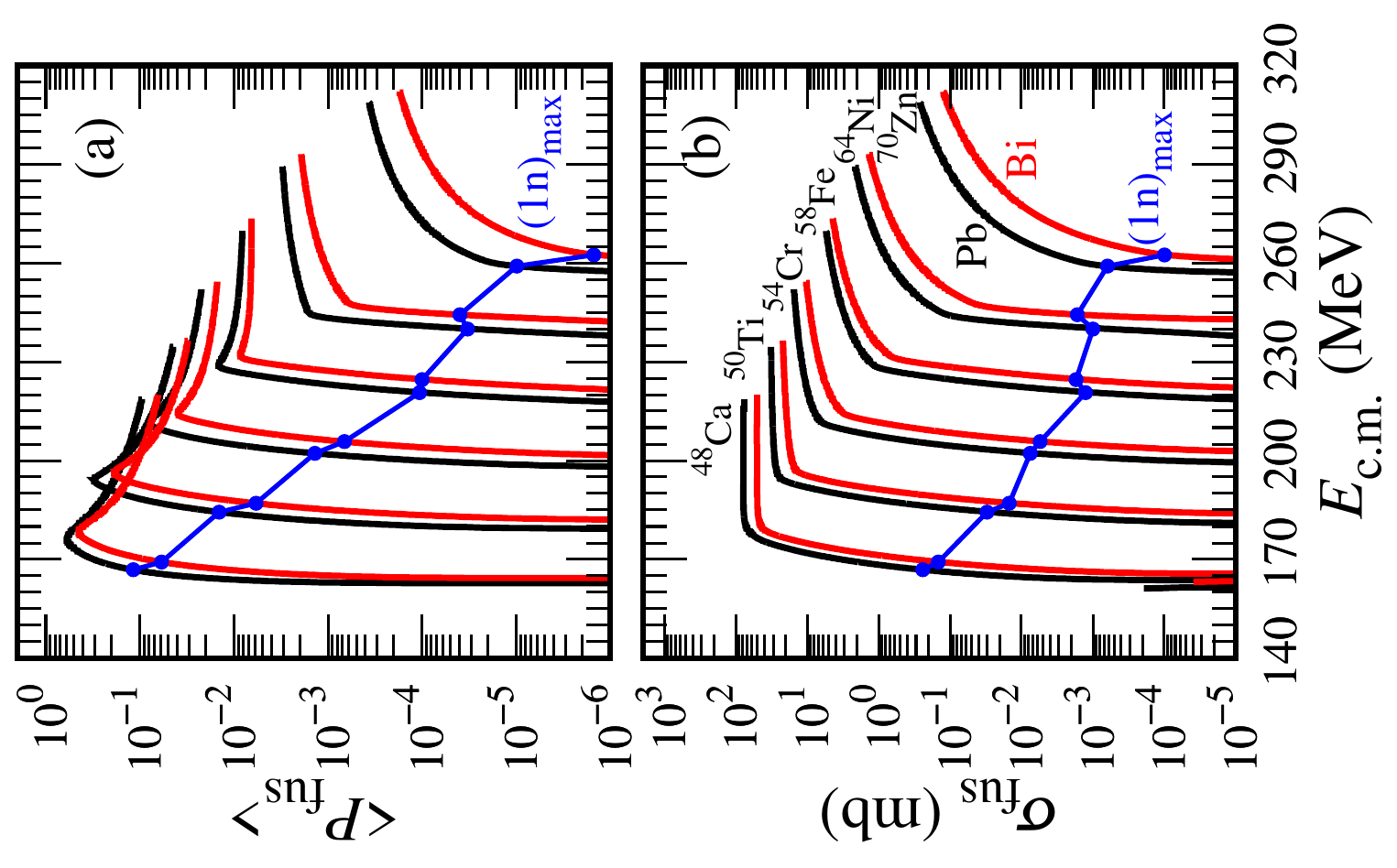}}
\caption{Panel (a): averaged fusion probability $<P_{\rm fus}>$ as a function of the center-of-mass energy for $^{48}$Ca, $^{50}$Ti, $^{54}$Cr, $^{56}$Fe, $^{64}$Ni and $^{70}$Zn projectiles on $^{208}$Pb (black lines) and $^{209}$Bi (red lines) targets. The blue line, marked (1n)$_{\textrm{max}}$, joins $<P_{\rm fus}>$ values at the maxima of the calculated $1n$ evaporation residue cross sections. Panel (b): Corresponding fusion cross sections, $\sigma_{\rm fus}$. The blue line joins $\sigma_{\rm fus}$ values at the maxima of the calculated 1n evaporation residue cross sections.}\label{fig:pfus_cold}
\end{figure}

\begin{figure}[t!!]
 \center{\includegraphics[width=1.3\columnwidth,angle=-90]{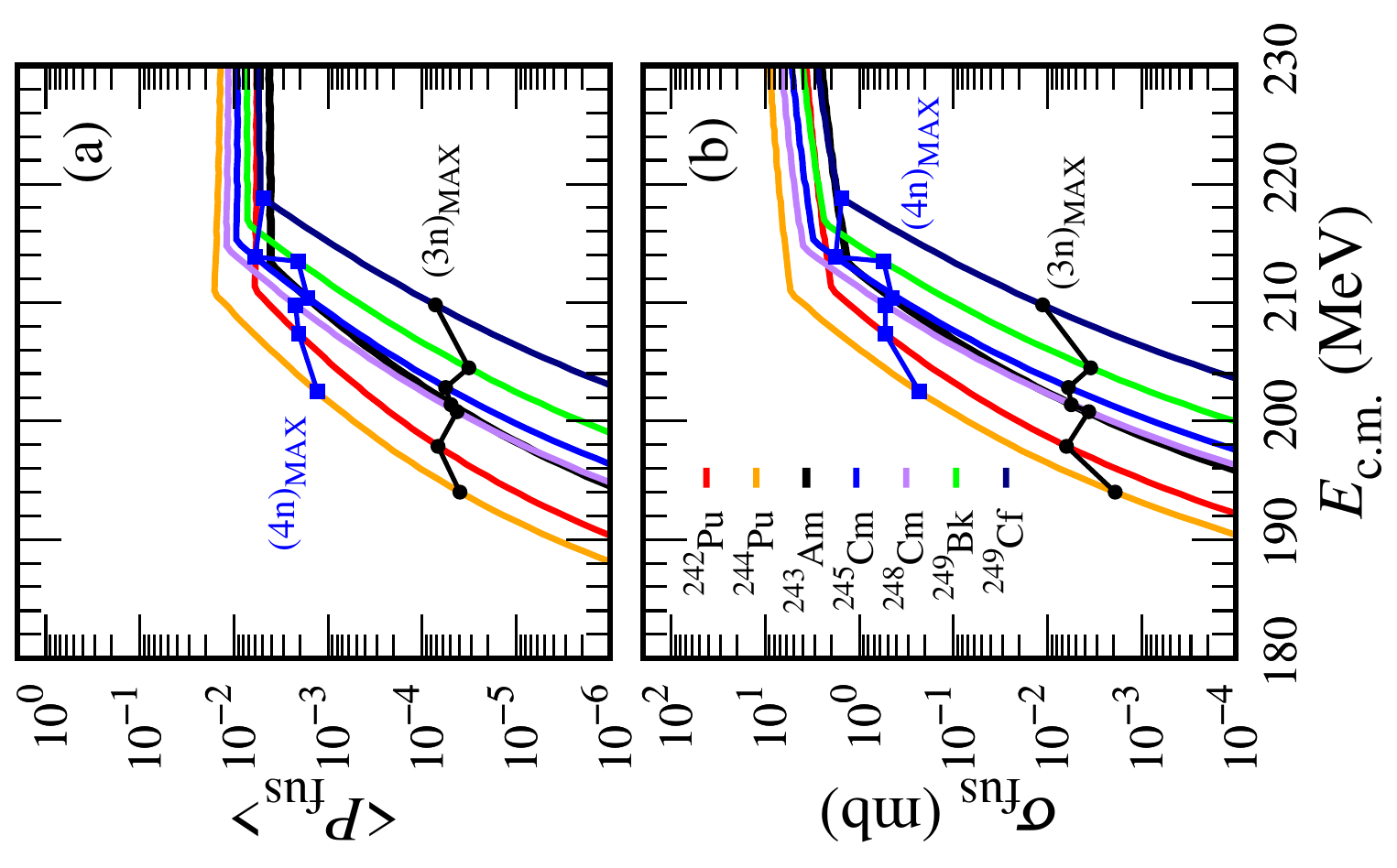}}
\caption{Panel (a): averaged fusion probability $<P_{\rm fus}>$ as a function of the center-of-mass energy for $^{48}$Ca induced reactions on indicated actinide targets. The black and blue lines, marked (3n)$_{\textrm{max}}$ and (4n)$_{\textrm{max}}$, join $<P_{\rm fus}>$ values at the maxima of the calculated 3n and 4n evaporation residue cross sections, respectively. Panel (b): Corresponding fusion cross sections, $\sigma_{\rm fus}$. The black and blue lines join $\sigma_{\rm fus}$ values at the maxima of the calculated 3n and 4n evaporation residue cross sections, respectively.}\label{fig:pfus_hot}
\end{figure}

Figs.~\ref{fig:pfus_cold} and~\ref{fig:pfus_hot} show the summary  of the averaged fusion probabilities and fusion cross section calculations for cold and hot synthesis reactions. In the top panel of Fig.~\ref{fig:pfus_cold} we present averaged fusion probabilities $<P_{\rm fus}>$ (see Eq.~\ref{pfusl}) calculated for a set of cold fusion reactions induced by projectiles ranging from $^{48}$Ca to $^{70}$Zn on $^{208}$Pb and $^{209}$Bi targets.
The target nuclei are distinguished by the line color. Black lines correspond to reactions incident on $^{208}$Pb and red lines on $^{209}$Bi.
The blue line connects points corresponding to the theoretical average fusion probabilities for the maxima of the 1n evaporation residue excitation functions. The $<P_{\rm fus}>$ at the optimum bombarding energy for the 1n channel drops five orders of magnitude with the change of the projectile nucleus from $^{48}$Ca to $^{70}$Zn.

This dependence confirms the known fact that an increase in the symmetry of the colliding system is not conducive to the success of the synthesis. This is a topological effect related to the disadvantageous position of the sticking configuration to the fusion saddle that must be overcome. The greater the initial symmetry, the deeper the system is under the fusion barrier and the less chance it has to overcome it.

Figure~\ref{fig:pfus_cold}b shows the fusion cross sections $\sigma_{\rm fus}$ (see Eq.~\ref{PCNeq}) for cold fusion reactions. The behavior of the fusion excitation functions is defined by the rapid growth of the fusion probability up to a few MeV above $B_0$.  At energies above the mean entrance channel barrier, the increase of the capture cross section is compensated by the decrease in fusion probability which results in a steady and slow growth of the fusion cross section. As mentioned before, this effect is associated with the suppression of contributions from higher partial waves to the cross section at higher bombarding energies.

In Fig.~\ref{fig:pfus_hot} we present calculated values of $<P_{\rm fus}>$ and the corresponding $\sigma_{\rm fus}$ cross sections for a set of hot fusion reactions of the $^{48}$Ca projectile incident on various actinide targets, from $^{242}$Pu to $^{249}$Cf. This set represents the reactions in which superheavy elements with atomic numbers $114 \le Z \le 118$ were discovered. The top panel shows averaged fusion probabilities $<P_{\rm fus}>$. The black and blue lines connect points showing $<P_{\rm fus}>$ values corresponding to the maxima of the theoretical evaporation residue excitation functions for 3n and 4n channels. Trends of these lines are different from the trend observed for the 1n cold fusion reactions. Values of fusion probabilities for a given evaporation channel are almost constant. The averaged fusion probability for the 3n channel is of the order of $5\times10^{-3}$ and for the 4n of the order of $2\times10^{-2}$.

The corresponding $\sigma_{\rm fus} $ cross sections are shown in Fig.~\ref{fig:pfus_hot}b. The entrance channel barriers $B_0$ and the barrier distribution widths $\omega$ for the discussed reactions are similar (see Figs.~\ref{fig:CaU_CaCm} and~\ref{fig:Ca+PuCm} for comparison) resulting in similar values of the capture cross sections. Therefore, the $\sigma_{\rm fus}$ cross sections reflect, to a large extent, the behavior of the fusion probabilities. The fusion cross sections approach the level of 10 mb as the energy increases.

\begin{figure*}[ht!]
\centering
\includegraphics[angle=0,scale=0.6]{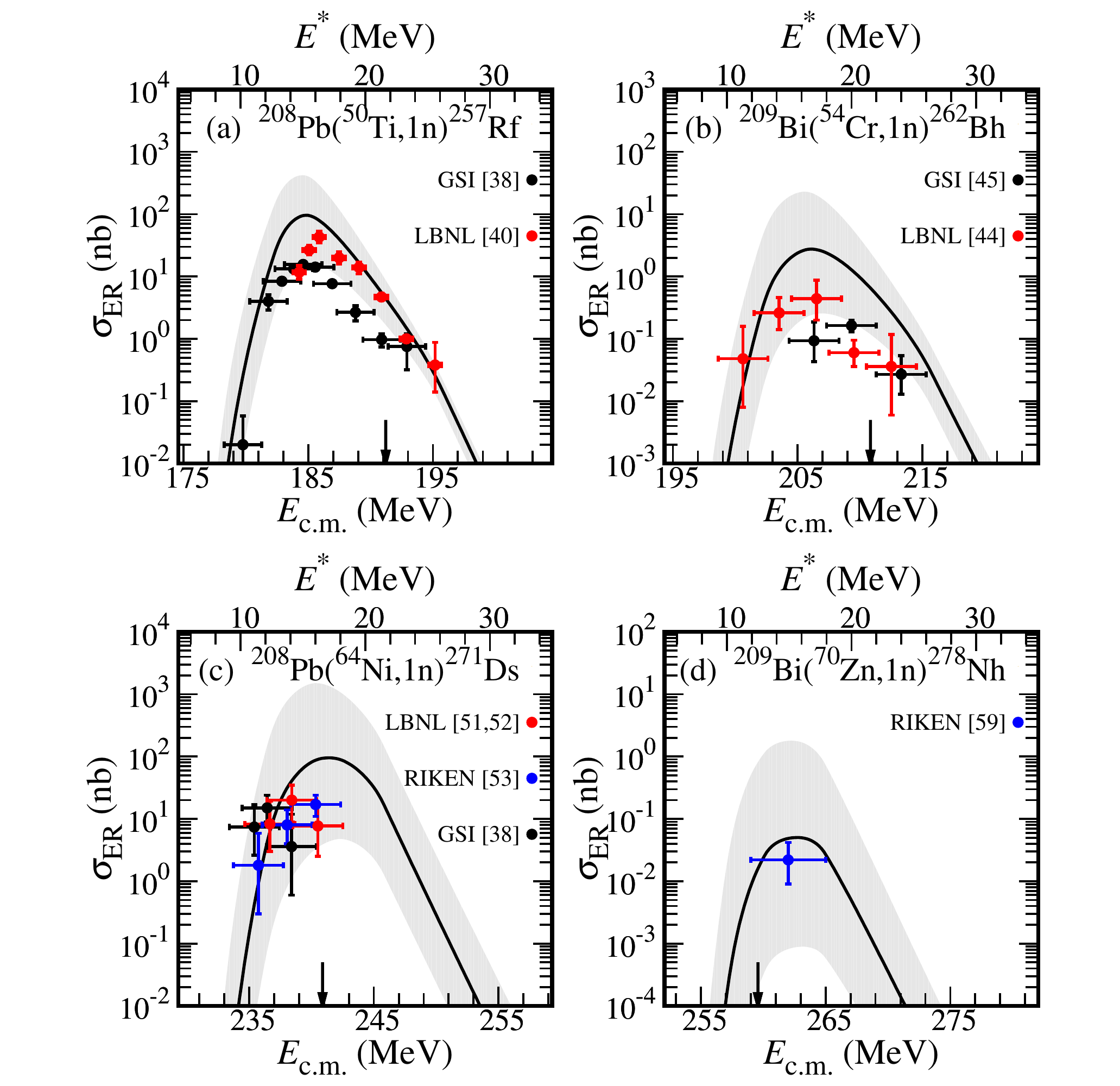}
\caption{Evaporation residue cross sections $\sigma_{\rm ER}$ for cold fusion reactions leading to the formation of the SHN with atomic numbers 104~\cite{HOFMANN200493,PhysRevC.78.024605}, 107~\cite{ISI:A1989U697600008,PhysRevC.78.024606}, 110~\cite{HOFMANN200493,PhysRevC.67.064609,PhysRevLett.93.212702,ISI:000223717300009}, and 113~\cite{New113}. The FBD model calculations are shown as solid black lines together with corresponding error corridors.}\label{fig:coldER}
\end{figure*}

\begin{figure*}[ht!]
\centering
\includegraphics[angle=-90,scale=0.7]{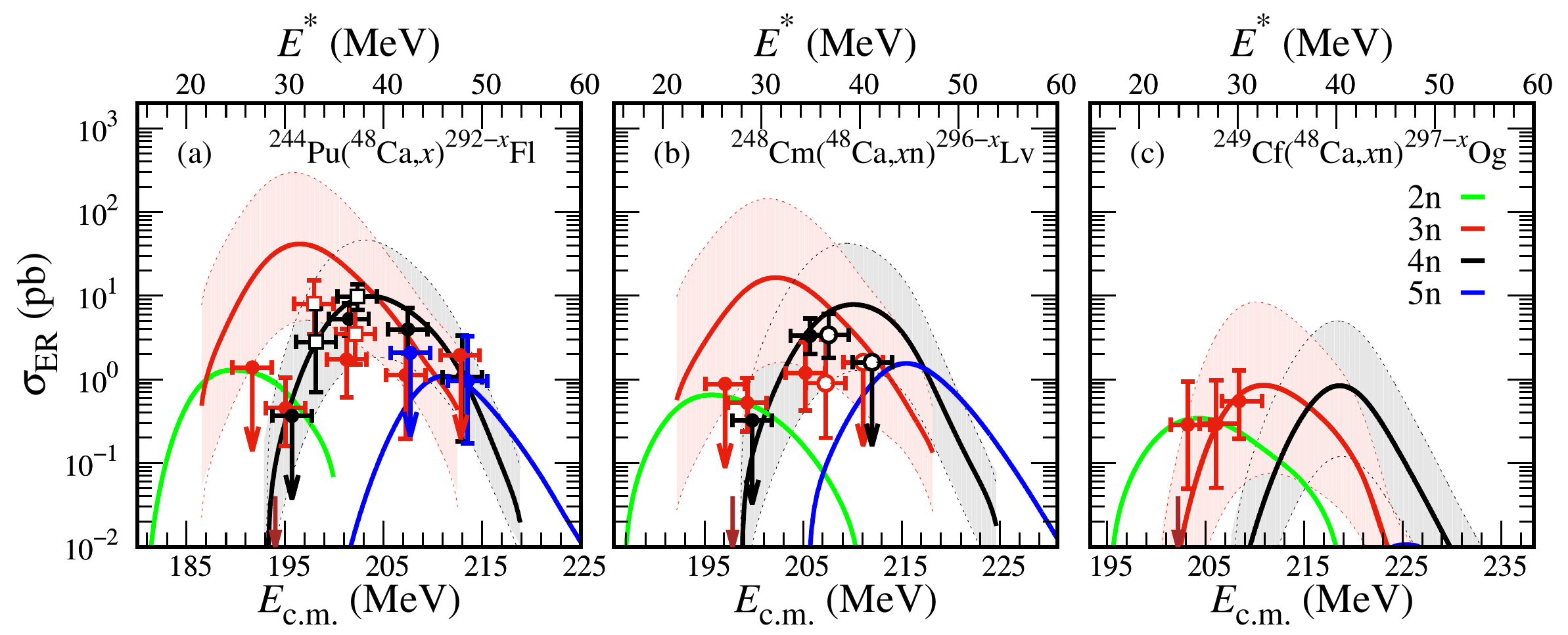}
\caption{Evaporation residue cross sections $\sigma_{\rm ER}$ for $^{48}$Ca induced reactions on $^{244}$Pu, $^{248}$Cm and $^{249}$Cf targets measured at DGFRS~\cite{Oganessian_2015} (full circles), TASCA~\cite{TASCA-114_2} (open squares) and SHIP~\cite{GSI-116} (open circles). Reaction channels are shown in different colors (see panel (c) for description). Solid lines represent the FBD model calculations of 2n, 3n, 4n and 5n reaction channels. Shaded areas limited by dotted colored lines correspond to the error corridors for the 3n and 4n channels.}\label{fig:hotER}
\end{figure*}

Finally, in Fig.~\ref{fig:coldER} and~\ref{fig:hotER}, we present a comparison of the experimentally measured evaporation residue cross sections $\sigma_{\rm ER}$ for selected cold and hot fusion reactions with the calculations done within the FBD model (see Eq.~\ref{factorize}) using the new $s_{\rm inj}$ systematics. The model calculations for a given reaction were corrected by taking into account the target thickness and respective projectile energy losses (see Ref.~\cite{FBD-11}). Shaded areas in the figures represent the $\sigma_{\rm ER}$ error corridors resulting from the uncertainties in the $s_{\rm inj}$ parametrizations (see Fig.~\ref{fig:sinj_cold} and~\ref{fig:sinj_hot} for comparison).

Figure~\ref{fig:coldER} shows evaporation residue cross sections $\sigma_{\rm ER}$ for four 1n-type cold fusion reactions leading to the formation of the SHN with atomic numbers 104 \cite{HOFMANN200493,Hofmann_1998,PhysRevC.78.024605}, 107 \cite{PhysRevC.78.024606,ISI:A1989U697600008}, 110 \cite{HOFMANN200493,Hofmann_1998,PhysRevC.67.064609,PhysRevLett.93.212702,ISI:000223717300009,MORITA2004101}, and 113 \cite{doi:10.1143/JPSJ.73.2593,New113}, respectively. The color of the points indicates the laboratory where the reaction was studied: red (LBNL), black (GSI), and blue (RIKEN). Model calculations are shown as solid black lines. The uncertainties in our calculations
are usually within one order of magnitude in both directions, which is comparable with the differences in the experimentally measured $\sigma_{\rm ER}$ values at similar energies in different laboratories. Both the shapes of the excitation functions and values of the evaporation residue cross sections are reasonably well reproduced within the model, including a six order of magnitude drop in cross section with the increase of the CN atomic number from 104 to 113. This mainly results from the systematic decrease in $<P_{\rm fus}>$ as the system symmetry in the entrance channel increases (see Fig. \ref{fig:pfus_cold} for comparison).

Shapes of the excitation functions and the $\sigma_{\rm ER}$ values result from the interplay between $\sigma_{\rm fus}$ and $<P_{\rm surv}>$. This is especially important in hot fusion reactions when one observes a fast decrease of $<P_{\rm surv}>$ in the subsequent steps of the neutron evaporation cascade.
Let us take the $^{297}$Og$^*$ nucleus formed in the complete fusion of $^{48}$Ca and $^{249}$Cf nuclei as an example (see Fig.~\ref{Psurv_Og}).
In this particular case, the decrease in the survival probability for the 4n channel (black curve) with respect to the 3n channel (red curve) is compensated by the increase in the fusion probability $<P_{\rm fus}>$ (dark blue curve in Fig.~\ref{Psurv_Og}), which finally leads to close values of the evaporation residue cross sections. Excitation functions for this reaction and two other hot fusion reactions leading to the formation of various isotopes of Fl and Lv, are shown in Fig.~\ref{fig:hotER}. As can be seen, the model calculations (solid lines) reproduce the experimental data taken from Refs.~\cite{Oganessian_2015,TASCA-114_2,GSI-116} reasonably well including a slow decrease in the value of the evaporation residue cross sections for 2n-5n channels as the CN atomic number increases from 114 to 118. The possibility of SHN synthesis in the 6n-9n evaporation channels was discussed by Hong \emph{et al.} in Ref.~\cite{Hong}.

In the case of the $x$n-type hot fusion reactions, the range of partial waves contributing to the total cross section is wider than for 1n cold fusion reactions. Therefore, one would expect a greater role of angular momentum in the compound nucleus formation for this type of reaction. The angular momentum gained by the system during the first step of the merging process modifies the whole potential energy landscape.
This effect is more significant as the shape of the nuclear system becomes more compact and the corresponding moment of inertia decreases, increasing the value of the rotational energy. Therefore, high angular momentum adversely affects the probability of synthesis success.
The fusion barrier that the system has to overcome increases and the fission barrier lowers, increasing the chance of compound nucleus disintegration.

\section{Summary and prospects}\label{sec:summary}

In this paper we summarized recent developments in the FBD model. The model is based on the assumption of three independent steps: capture, fusion, and deexcitation of the compound nucleus. The fusion step, the least studied and most difficult part in the description of the synthesis process, was analyzed with the diffusion approach. Each reaction stage was described in detail and compared with selected experimental data for both cold and hot fusion reactions.

Despite its simplicity, the FBD model allows for surprisingly accurate reproduction of the evaporation residue cross sections and allows predictions to be made for as yet not studied reactions. By using the Fusion-by-Diffusion approach one can intuitively understand the very complex phenomenon of synthesis of superheavy nuclei. However, there are still some elements we plan to improve or modify. The following changes are, in our opinion, important for a better understanding of SHN formation:
\begin{itemize}
\item Inclusion of the shell corrections in the fusion step, not only in the ground state (as it is now) but also in the full range of a  multidimensional potential energy surface (PES).
\item Fully account for the centrifugal barrier by adding it at each PES point.
\item Accounting for higher-order deformations of the colliding nuclei and considering various possible projectile-target configurations in the entrance channel.
\item Elimination of the $s_{\rm inj}$ parameter by determining it with more basic rules.
\end{itemize}

\begin{acknowledgements}
M.K. was co-financed by the International Research Project COPIGAL.
\end{acknowledgements}

\bibliography{fbd_EPJ_bibfile}

\providecommand{\noopsort}[1]{}\providecommand{\singleletter}[1]{#1}%
\begin{thebibliography}{96}
\providecommand{\natexlab}[1]{#1}
\providecommand{\url}[1]{\texttt{#1}}
\expandafter\ifx\csname urlstyle\endcsname\relax
  \providecommand{\doi}[1]{doi: #1}\else
  \providecommand{\doi}{doi: \begingroup \urlstyle{rm}\Url}\fi

\bibitem[Hofmann(2011)]{Hofmann}
S.~Hofmann.
\newblock Synthesis of superheavy elements by cold fusion.
\newblock \emph{Radiochimica Acta}, 99\penalty0 (7-8):\penalty0 405--428, 2011.
\newblock \doi{doi:10.1524/ract.2011.1854}.
\newblock URL \url{https://doi.org/10.1524/ract.2011.1854}.

\bibitem[Düllmann et~al.(2022)Düllmann, Block, Heßberger, Khuyagbaatar,
  Kindler, Kratz, Lommel, Münzenberg, Pershina, Renisch, Schädel, and
  Yakushev]{GSI-5}
Christoph~E. Düllmann, Michael Block, Fritz~P. Heßberger, Jadambaa
  Khuyagbaatar, Birgit Kindler, Jens~V. Kratz, Bettina Lommel, Gottfried
  Münzenberg, Valeria Pershina, Dennis Renisch, Matthias Schädel, and
  Alexander Yakushev.
\newblock Five decades of gsi superheavy element discoveries and chemical
  investigation.
\newblock \emph{Radiochimica Acta}, 110\penalty0 (6-9):\penalty0 417--439,
  2022.
\newblock \doi{doi:10.1515/ract-2022-0015}.
\newblock URL \url{https://doi.org/10.1515/ract-2022-0015}.

\bibitem[Morita et~al.(2018)Morita, Morimoto, Kaji, Habaa, and Kudo]{Riken}
Kosuke Morita, Kouji Morimoto, Daiya Kaji, Hiromitsu Habaa, and Hisaaki Kudo.
\newblock Discovery of new element, nihonium, and perspectives (plenary).
\newblock \emph{Progress in Nuclear Science and Technology}, 5:\penalty0 8--13,
  2018.
\newblock \doi{http://dx.doi.org/10.15669/pnst.5.8}.
\newblock URL \url{http://dx.doi.org/10.15669/pnst.5.8}.

\bibitem[Oganessian(2011)]{Oganessian+2011+429+439}
Yu.~Ts. Oganessian.
\newblock Synthesis of the heaviest elements in 48ca-induced reactions.
\newblock \emph{Radiochimica Acta}, 99\penalty0 (7-8):\penalty0 429--439, 2011.
\newblock \doi{doi:10.1524/ract.2011.1860}.
\newblock URL \url{https://doi.org/10.1524/ract.2011.1860}.

\bibitem[Oganessian and Utyonkov(2015)]{Oganessian_2015}
Yu~Ts Oganessian and V~K Utyonkov.
\newblock Super-heavy element research.
\newblock \emph{Reports on Progress in Physics}, 78\penalty0 (3):\penalty0
  036301, feb 2015.
\newblock \doi{10.1088/0034-4885/78/3/036301}.
\newblock URL \url{https://doi.org/10.1088/0034-4885/78/3/036301}.

\bibitem[Hofmann et~al.(2019)Hofmann, Ackermann, Antalic, Comas, Heinz,
  Heredia, Hessberger, Khuyagbaatar, Kindler, Kojouharov, Leino, Lommel, Mann,
  Nishio, Popeko, Saro, Uusitalo, Venhart, and A.V.]{NiU}
S.~Hofmann, D.~Ackermann, S.~Antalic, V.F. Comas, S.~Heinz, J.A. Heredia, F.P.
  Hessberger, J.~Khuyagbaatar, B.~Kindler, I.~Kojouharov, M.~Leino, B.~Lommel,
  R.~Mann, K.~Nishio, A.G. Popeko, S.~Saro, J.~Uusitalo, M.~Venhart, and
  Yeremin A.V.
\newblock Probing shell effects at z=120 and n=184.
\newblock \emph{GSI Scientific Report 2009-1 (GSI, 2019)}, page 131, 2019.
\newblock URL \url{http://repository.gsi.de/record/53523}.

\bibitem[Oganessian et~al.(2009)Oganessian, Utyonkov, Lobanov, Abdullin,
  Polyakov, Sagaidak, Shirokovsky, Tsyganov, Voinov, Mezentsev, Subbotin,
  Sukhov, Subotic, Zagrebaev, Dmitriev, Henderson, Moody, Kenneally, Landrum,
  Shaughnessy, Stoyer, Stoyer, and Wilk]{PhysRevC.79.024603}
Yu.~Ts. Oganessian, V.~K. Utyonkov, Yu.~V. Lobanov, F.~Sh. Abdullin, A.~N.
  Polyakov, R.~N. Sagaidak, I.~V. Shirokovsky, Yu.~S. Tsyganov, A.~A. Voinov,
  A.~N. Mezentsev, V.~G. Subbotin, A.~M. Sukhov, K.~Subotic, V.~I. Zagrebaev,
  S.~N. Dmitriev, R.~A. Henderson, K.~J. Moody, J.~M. Kenneally, J.~H. Landrum,
  D.~A. Shaughnessy, M.~A. Stoyer, N.~J. Stoyer, and P.~A. Wilk.
\newblock Attempt to produce element 120 in the
  $^{244}\mathrm{Pu}+^{58}\mathrm{Fe}$ reaction.
\newblock \emph{Phys. Rev. C}, 79:\penalty0 024603, Feb 2009.
\newblock \doi{10.1103/PhysRevC.79.024603}.
\newblock URL \url{https://link.aps.org/doi/10.1103/PhysRevC.79.024603}.

\bibitem[Hofmann et~al.(2016)Hofmann, Heinz, Mann, Maurer, Münzenberg,
  Antalic, Barth, Burkhard, Dahl, Eberhardt, Grzywacz, Hamilton, Henderson,
  Kenneally, Kindler, Kojouharov, Lang, Lommel, Miernik, Miller, Moody, Morita,
  Nishio, Popeko, Roberto, Runke, Rykaczewski, Saro, Scheidenberger, Schött,
  Shaughnessy, Stoyer, Thörle-Pospiech, Tinschert, Trautmann, Uusitalo, and
  Yeremin]{CrCm}
S.~Hofmann, S.~Heinz, R.~Mann, J.~Maurer, G.~Münzenberg, S.~Antalic, W.~Barth,
  H.~G. Burkhard, L.~Dahl, K.~Eberhardt, R.~Grzywacz, J.~H. Hamilton, R.~A.
  Henderson, J.~M. Kenneally, B.~Kindler, I.~Kojouharov, R.~Lang, B.~Lommel,
  K.~Miernik, D.~Miller, K.~J. Moody, K.~Morita, K.~Nishio, A.~G. Popeko, J.~B.
  Roberto, J.~Runke, K.~P. Rykaczewski, S.~Saro, C.~Scheidenberger, H.~J.
  Schött, D.~A. Shaughnessy, M.~A. Stoyer, P.~Thörle-Pospiech, K.~Tinschert,
  N.~Trautmann, J.~Uusitalo, and A.~V. Yeremin.
\newblock Review of even element super-heavy nuclei and search for element 120.
\newblock \emph{The European Physical Journal A}, 52:\penalty0 180, 2016.
\newblock \doi{10.1140/epja/i2016-16180-4}.
\newblock URL \url{https://doi.org/10.1140/epja/i2016-16180-4}.

\bibitem[Khuyagbaatar et~al.(2020)Khuyagbaatar, Yakushev, D\"ullmann,
  Ackermann, Andersson, Asai, Block, Boll, Brand, Cox, Dasgupta, Derkx,
  Di~Nitto, Eberhardt, Even, Evers, Fahlander, Forsberg, Gates, Gharibyan,
  Golubev, Gregorich, Hamilton, Hartmann, Herzberg, He\ss{}berger, Hinde,
  Hoffmann, Hollinger, H\"ubner, J\"ager, Kindler, Kratz, Krier, Kurz,
  Laatiaoui, Lahiri, Lang, Lommel, Maiti, Miernik, Minami, Mistry, Mokry,
  Nitsche, Omtvedt, Pang, Papadakis, Renisch, Roberto, Rudolph, Runke,
  Rykaczewski, Sarmiento, Sch\"adel, Schausten, Semchenkov, Shaughnessy,
  Steinegger, Steiner, Tereshatov, Th\"orle-Pospiech, Tinschert, Torres
  De~Heidenreich, Trautmann, T\"urler, Uusitalo, Wegrzecki, Wiehl, Van~Cleve,
  and Yakusheva]{PhysRevC.102.064602}
J.~Khuyagbaatar, A.~Yakushev, Ch.~E. D\"ullmann, D.~Ackermann, L.-L. Andersson,
  M.~Asai, M.~Block, R.~A. Boll, H.~Brand, D.~M. Cox, M.~Dasgupta, X.~Derkx,
  A.~Di~Nitto, K.~Eberhardt, J.~Even, M.~Evers, C.~Fahlander, U.~Forsberg,
  J.~M. Gates, N.~Gharibyan, P.~Golubev, K.~E. Gregorich, J.~H. Hamilton,
  W.~Hartmann, R.-D. Herzberg, F.~P. He\ss{}berger, D.~J. Hinde, J.~Hoffmann,
  R.~Hollinger, A.~H\"ubner, E.~J\"ager, B.~Kindler, J.~V. Kratz, J.~Krier,
  N.~Kurz, M.~Laatiaoui, S.~Lahiri, R.~Lang, B.~Lommel, M.~Maiti, K.~Miernik,
  S.~Minami, A.~K. Mistry, C.~Mokry, H.~Nitsche, J.~P. Omtvedt, G.~K. Pang,
  P.~Papadakis, D.~Renisch, J.~B. Roberto, D.~Rudolph, J.~Runke, K.~P.
  Rykaczewski, L.~G. Sarmiento, M.~Sch\"adel, B.~Schausten, A.~Semchenkov,
  D.~A. Shaughnessy, P.~Steinegger, J.~Steiner, E.~E. Tereshatov,
  P.~Th\"orle-Pospiech, K.~Tinschert, T.~Torres De~Heidenreich, N.~Trautmann,
  A.~T\"urler, J.~Uusitalo, M.~Wegrzecki, N.~Wiehl, S.~M. Van~Cleve, and
  V.~Yakusheva.
\newblock Search for elements 119 and 120.
\newblock \emph{Phys. Rev. C}, 102:\penalty0 064602, Dec 2020.
\newblock \doi{10.1103/PhysRevC.102.064602}.
\newblock URL \url{https://link.aps.org/doi/10.1103/PhysRevC.102.064602}.

\bibitem[Bao(2020)]{Bao}
Xiao~Jun Bao.
\newblock Fusion dynamics of low-energy heavy-ion collisions for production of
  superheavy nuclei.
\newblock \emph{Frontiers in Physics}, 8, 2020.
\newblock ISSN 2296-424X.
\newblock \doi{10.3389/fphy.2020.00014}.
\newblock URL
  \url{https://www.frontiersin.org/article/10.3389/fphy.2020.00014}.

\bibitem[Jachimowicz et~al.(2021)Jachimowicz, Kowal, and Skalski]{Jach2021}
P.~Jachimowicz, M.~Kowal, and J.~Skalski.
\newblock Properties of heaviest nuclei with $98 \le$ z $\le 126$ and $134 \le$
  n $\le 192$.
\newblock \emph{Atomic Data and Nuclear Data Tables}, 138:\penalty0 101393,
  2021.
\newblock ISSN 0092-640X.
\newblock \doi{https://doi.org/10.1016/j.adt.2020.101393}.
\newblock URL
  \url{https://www.sciencedirect.com/science/article/pii/S0092640X20300504}.

\bibitem[Tanaka et~al.(2018)Tanaka, Narikiyo, Morita, Fujita, Kaji, Morimoto,
  Yamaki, Wakabayashi, Tanaka, Takeyama, Yoneda, Haba, Komori, Yanou, Gall,
  Asfari, Faure, Hasebe, Huang, Kanaya, Murakami, Yoshida, Yamaguchi, Tokanai,
  Yoshida, Yamamoto, Yamano, Watanabe, Ishizawa, Asai, Aono, Goto, Katori, and
  Hagino]{Tanaka_cold}
Taiki Tanaka, Yoshihiro Narikiyo, Kosuke Morita, Kunihiro Fujita, Daiya Kaji,
  Kouji Morimoto, Sayaka Yamaki, Yasuo Wakabayashi, Kengo Tanaka, Mirei
  Takeyama, Akira Yoneda, Hiromitsu Haba, Yukiko Komori, Shinya Yanou,
  Beno\^{\i}t Jean-Paul Gall, Zouhair Asfari, Hugo Faure, Hiroo Hasebe, Minghui
  Huang, Jumpei Kanaya, Masashi Murakami, Atsushi Yoshida, Takayuki Yamaguchi,
  Fuyuki Tokanai, Tomomi Yoshida, Shoya Yamamoto, Yuki Yamano, Kenyu Watanabe,
  Satoshi Ishizawa, Masato Asai, Ryuji Aono, Shin-ichi Goto, Kenji Katori, and
  Kouichi Hagino.
\newblock Determination of fusion barrier distributions from quasielastic
  scattering cross sections towards superheavy nuclei synthesis.
\newblock \emph{Journal of the Physical Society of Japan}, 87\penalty0
  (1):\penalty0 014201, 2018.
\newblock \doi{10.7566/JPSJ.87.014201}.
\newblock URL \url{https://doi.org/10.7566/JPSJ.87.014201}.

\bibitem[Banerjee et~al.(2019)Banerjee, Hinde, Dasgupta, Simpson, Jeung,
  Simenel, Swinton-Bland, Williams, Carter, Cook, David, D\"ullmann,
  Khuyagbaatar, Kindler, Lommel, Prasad, Sengupta, Smith, Vo-Phuoc, Walshe, and
  Yakushev]{HindePRL}
K.~Banerjee, D.~J. Hinde, M.~Dasgupta, E.~C. Simpson, D.~Y. Jeung, C.~Simenel,
  B.~M.~A. Swinton-Bland, E.~Williams, I.~P. Carter, K.~J. Cook, H.~M. David,
  Ch.~E. D\"ullmann, J.~Khuyagbaatar, B.~Kindler, B.~Lommel, E.~Prasad,
  C.~Sengupta, J.~F. Smith, K.~Vo-Phuoc, J.~Walshe, and A.~Yakushev.
\newblock Mechanisms suppressing superheavy element yields in cold fusion
  reactions.
\newblock \emph{Phys. Rev. Lett.}, 122:\penalty0 232503, Jun 2019.
\newblock \doi{10.1103/PhysRevLett.122.232503}.
\newblock URL \url{https://link.aps.org/doi/10.1103/PhysRevLett.122.232503}.

\bibitem[Bock et~al.(1982)Bock, Chu, Dakowski, Gobbi, Grosse, Olmi, Sann,
  Schwalm, Lynen, Müller, Bjørnholm, Esbensen, Wölfli, and Morenzoni]{BOCK}
R.~Bock, Y.T. Chu, M.~Dakowski, A.~Gobbi, E.~Grosse, A.~Olmi, H.~Sann,
  D.~Schwalm, U.~Lynen, W.~Müller, S.~Bjørnholm, H.~Esbensen, W.~Wölfli, and
  E.~Morenzoni.
\newblock Dynamics of the fusion process.
\newblock \emph{Nuclear Physics A}, 388\penalty0 (2):\penalty0 334--380, 1982.
\newblock ISSN 0375-9474.
\newblock \doi{https://doi.org/10.1016/0375-9474(82)90420-1}.
\newblock URL
  \url{https://www.sciencedirect.com/science/article/pii/0375947482904201}.

\bibitem[Pacheco et~al.(1992)Pacheco, Fern\'andez~Niello, DiGregorio, di~Tada,
  Testoni, Chan, Ch\'avez, Gazes, Plagnol, and Stokstad]{Pacheco}
A.~J. Pacheco, J.~O. Fern\'andez~Niello, D.~E. DiGregorio, M.~di~Tada, J.~E.
  Testoni, Y.~Chan, E.~Ch\'avez, S.~Gazes, E.~Plagnol, and R.~G. Stokstad.
\newblock Capture reactions in the $^{40,48}\mathrm{Ca}$${+}^{197}$au and
  $^{40,48}\mathrm{Ca}$${+}^{208}$pb systems.
\newblock \emph{Phys. Rev. C}, 45:\penalty0 2861--2869, Jun 1992.
\newblock \doi{10.1103/PhysRevC.45.2861}.
\newblock URL \url{https://link.aps.org/doi/10.1103/PhysRevC.45.2861}.

\bibitem[Prokhorova et~al.(2008)Prokhorova, Bogachev, Itkis, Itkis, Knyazheva,
  Kondratiev, Kozulin, Krupa, Oganessian, Pokrovsky, Pashkevich, and
  Rusanov]{PROKH}
E.V. Prokhorova, A.A. Bogachev, M.G. Itkis, I.M. Itkis, G.N. Knyazheva, N.A.
  Kondratiev, E.M. Kozulin, L.~Krupa, Yu.Ts. Oganessian, I.V. Pokrovsky, V.V.
  Pashkevich, and A.Ya. Rusanov.
\newblock The fusion–fission and quasi-fission processes in the reaction 48ca
  + 208pb at energies near the coulomb barrier.
\newblock \emph{Nuclear Physics A}, 802\penalty0 (1):\penalty0 45--66, 2008.
\newblock ISSN 0375-9474.
\newblock \doi{https://doi.org/10.1016/j.nuclphysa.2008.01.016}.
\newblock URL
  \url{https://www.sciencedirect.com/science/article/pii/S037594740800050X}.

\bibitem[Clerc et~al.(1984)Clerc, Keller, Sahm, Schmidt, Schulte, and
  Vermeulen]{CLERC}
H.-G. Clerc, J.G. Keller, C.-C. Sahm, K.-H. Schmidt, H.~Schulte, and
  D.~Vermeulen.
\newblock Fusion-fission and neutron-evaporation-residue cross-sections in
  40ar- and 50ti-induced fusion reactions.
\newblock \emph{Nuclear Physics A}, 419\penalty0 (3):\penalty0 571--588, 1984.
\newblock ISSN 0375-9474.
\newblock \doi{https://doi.org/10.1016/0375-9474(84)90634-1}.
\newblock URL
  \url{https://www.sciencedirect.com/science/article/pii/0375947484906341}.

\bibitem[Naik et~al.(2007)Naik, Loveland, Sprunger, Vinodkumar, Peterson,
  Jiang, Zhu, Tang, Moore, and Chowdhury]{Naik}
R.~S. Naik, W.~Loveland, P.~H. Sprunger, A.~M. Vinodkumar, D.~Peterson, C.~L.
  Jiang, S.~Zhu, X.~Tang, E.~F. Moore, and P.~Chowdhury.
\newblock Measurement of the fusion probability ${P}_{\mathrm{cn}}$ for the
  reaction of $^{50}\mathrm{Ti}$ with $^{208}\mathrm{Pb}$.
\newblock \emph{Phys. Rev. C}, 76:\penalty0 054604, Nov 2007.
\newblock \doi{10.1103/PhysRevC.76.054604}.
\newblock URL \url{https://link.aps.org/doi/10.1103/PhysRevC.76.054604}.

\bibitem[Tanaka et~al.(2020)Tanaka, Morita, Morimoto, Kaji, Haba, Boll, Brewer,
  Van~Cleve, Dean, Ishizawa, Ito, Komori, Nishio, Niwase, Rasco, Roberto,
  Rykaczewski, Sakai, Stracener, and Hagino]{Tanaka_hot}
T.~Tanaka, K.~Morita, K.~Morimoto, D.~Kaji, H.~Haba, R.~A. Boll, N.~T. Brewer,
  S.~Van~Cleve, D.~J. Dean, S.~Ishizawa, Y.~Ito, Y.~Komori, K.~Nishio,
  T.~Niwase, B.~C. Rasco, J.~B. Roberto, K.~P. Rykaczewski, H.~Sakai, D.~W.
  Stracener, and K.~Hagino.
\newblock Study of quasielastic barrier distributions as a step towards the
  synthesis of superheavy elements with hot fusion reactions.
\newblock \emph{Phys. Rev. Lett.}, 124:\penalty0 052502, Feb 2020.
\newblock \doi{10.1103/PhysRevLett.124.052502}.
\newblock URL \url{https://link.aps.org/doi/10.1103/PhysRevLett.124.052502}.

\bibitem[Kozulin et~al.(2010)Kozulin, Knyazheva, Itkis, Itkis, Bogachev, Krupa,
  Loktev, Smirnov, Zagrebaev, Äystö, Trzaska, Rubchenya, Vardaci, Stefanini,
  Cinausero, Corradi, Fioretto, Mason, Prete, Silvestri, Beghini, Montagnoli,
  Scarlassara, Hanappe, Khlebnikov, Kliman, Brondi, {Di Nitto}, Moro, Gelli,
  and Szilner]{KOZULIN2010227}
E.M. Kozulin, G.N. Knyazheva, I.M. Itkis, M.G. Itkis, A.A. Bogachev, L.~Krupa,
  T.A. Loktev, S.V. Smirnov, V.I. Zagrebaev, J.~Äystö, W.H. Trzaska, V.A.
  Rubchenya, E.~Vardaci, A.M. Stefanini, M.~Cinausero, L.~Corradi, E.~Fioretto,
  P.~Mason, G.F. Prete, R.~Silvestri, S.~Beghini, G.~Montagnoli,
  F.~Scarlassara, F.~Hanappe, S.V. Khlebnikov, J.~Kliman, A.~Brondi, A.~{Di
  Nitto}, R.~Moro, N.~Gelli, and S.~Szilner.
\newblock Investigation of the reaction 64ni+238u being an option of
  synthesizing element 120.
\newblock \emph{Physics Letters B}, 686\penalty0 (4):\penalty0 227--232, 2010.
\newblock ISSN 0370-2693.
\newblock \doi{https://doi.org/10.1016/j.physletb.2010.02.041}.
\newblock URL
  \url{https://www.sciencedirect.com/science/article/pii/S0370269310002285}.

\bibitem[Tōke et~al.(1985)Tōke, Bock, Dai, Gobbi, Gralla, Hildenbrand,
  Kuzminski, Müller, Olmi, Stelzer, Back, and Bjørnholm]{TOKE1985327}
J.~Tōke, R.~Bock, G.X. Dai, A.~Gobbi, S.~Gralla, K.D. Hildenbrand,
  J.~Kuzminski, W.F.J. Müller, A.~Olmi, H.~Stelzer, B.B. Back, and
  S.~Bjørnholm.
\newblock Quasi-fission — the mass-drift mode in heavy-ion reactions.
\newblock \emph{Nuclear Physics A}, 440\penalty0 (2):\penalty0 327--365, 1985.
\newblock ISSN 0375-9474.
\newblock \doi{https://doi.org/10.1016/0375-9474(85)90344-6}.
\newblock URL
  \url{https://www.sciencedirect.com/science/article/pii/0375947485903446}.

\bibitem[Nishio et~al.(2012)Nishio, Mitsuoka, Nishinaka, Makii, Wakabayashi,
  Ikezoe, Hirose, Ohtsuki, Aritomo, and Hofmann]{Nishio_CaU}
K.~Nishio, S.~Mitsuoka, I.~Nishinaka, H.~Makii, Y.~Wakabayashi, H.~Ikezoe,
  K.~Hirose, T.~Ohtsuki, Y.~Aritomo, and S.~Hofmann.
\newblock Fusion probabilities in the reactions ${}^{40,48}$ca+${}^{238}$u at
  energies around the coulomb barrier.
\newblock \emph{Phys. Rev. C}, 86:\penalty0 034608, Sep 2012.
\newblock \doi{10.1103/PhysRevC.86.034608}.
\newblock URL \url{https://link.aps.org/doi/10.1103/PhysRevC.86.034608}.

\bibitem[Kozulin et~al.(2014)Kozulin, Knyazheva, Itkis, Itkis, Bogachev,
  Chernysheva, Krupa, Hanappe, Dorvaux, Stuttg\'e, Trzaska, Schmitt, and
  Chubarian]{Kozlulin_CaUPuCm}
E.~M. Kozulin, G.~N. Knyazheva, I.~M. Itkis, M.~G. Itkis, A.~A. Bogachev, E.~V.
  Chernysheva, L.~Krupa, F.~Hanappe, O.~Dorvaux, L.~Stuttg\'e, W.~H. Trzaska,
  C.~Schmitt, and G.~Chubarian.
\newblock Fusion-fission and quasifission of superheavy systems with
  $z=110--116$ formed in $^{48}\mathrm{Ca}$-induced reactions.
\newblock \emph{Phys. Rev. C}, 90:\penalty0 054608, Nov 2014.
\newblock \doi{10.1103/PhysRevC.90.054608}.
\newblock URL \url{https://link.aps.org/doi/10.1103/PhysRevC.90.054608}.

\bibitem[Swiatecki et~al.(2003)Swiatecki, Siwek-Wilczynska, and
  Wilczynski]{FBD-Acta}
W.~J. Swiatecki, K.~Siwek-Wilczynska, and J.~Wilczynski.
\newblock Fusion by diffusion.
\newblock \emph{ACTA PHYSICA POLONICA B}, 34\penalty0 (4):\penalty0 2049--2071,
  APR 2003.
\newblock ISSN 0587-4254.
\newblock 37th Zakopane Meeting of Physics on Trends in Nuclear Physics,
  ZAKOPANE, POLAND, SEP 03-10, 2002.

\bibitem[Swiatecki et~al.(2005)Swiatecki, Siwek-Wilczynska, and
  Wilczynski]{FBD-05}
W.~J. Swiatecki, K.~Siwek-Wilczynska, and J.~Wilczynski.
\newblock Fusion by diffusion. ii. synthesis of transfermium elements in cold
  fusion reactions.
\newblock \emph{Phys. Rev. C}, 71:\penalty0 014602, Jan 2005.
\newblock \doi{10.1103/PhysRevC.71.014602}.
\newblock URL \url{https://link.aps.org/doi/10.1103/PhysRevC.71.014602}.

\bibitem[Cap et~al.(2011)Cap, Siwek-Wilczy\ifmmode~\acute{n}\else
  \'{n}\fi{}ska, and Wilczy\ifmmode~\acute{n}\else \'{n}\fi{}ski]{FBD-11}
T.~Cap, K.~Siwek-Wilczy\ifmmode~\acute{n}\else \'{n}\fi{}ska, and
  J.~Wilczy\ifmmode~\acute{n}\else \'{n}\fi{}ski.
\newblock Nucleus-nucleus cold fusion reactions analyzed with the $l$-dependent
  ``fusion by diffusion'' model.
\newblock \emph{Phys. Rev. C}, 83:\penalty0 054602, May 2011.
\newblock \doi{10.1103/PhysRevC.83.054602}.
\newblock URL \url{https://link.aps.org/doi/10.1103/PhysRevC.83.054602}.

\bibitem[Cap et~al.(2012)Cap, Siwek-Wilczynska, Skwira-Chalot, and
  Wilczynski]{FBD-APP-xn}
T.~Cap, K.~Siwek-Wilczynska, I.~Skwira-Chalot, and J.~Wilczynski.
\newblock Extension of the fusion by diffusion model for description of the
  synthesis of superheavy nuclei in (fusion, xn) reactions.
\newblock \emph{Acta Physica Polonica. Series B: Elementary Particle Physics,
  Nuclear Physics, Statistical Physics, Theory of Relativity, Field Theory},
  43:\penalty0 297--303, 2012.
\newblock \doi{10.5506/APhysPolB.43.297}.

\bibitem[Siwek-Wilczy\ifmmode~\acute{n}\else \'{n}\fi{}ska
  et~al.(2012)Siwek-Wilczy\ifmmode~\acute{n}\else \'{n}\fi{}ska, Cap, Kowal,
  Sobiczewski, and Wilczy\ifmmode~\acute{n}\else \'{n}\fi{}ski]{PRC-hot}
K.~Siwek-Wilczy\ifmmode~\acute{n}\else \'{n}\fi{}ska, T.~Cap, M.~Kowal,
  A.~Sobiczewski, and J.~Wilczy\ifmmode~\acute{n}\else \'{n}\fi{}ski.
\newblock Predictions of the fusion-by-diffusion model for the synthesis cross
  sections of $z=114$--120 elements based on macroscopic-microscopic fission
  barriers.
\newblock \emph{Phys. Rev. C}, 86:\penalty0 014611, Jul 2012.
\newblock \doi{10.1103/PhysRevC.86.014611}.
\newblock URL \url{https://link.aps.org/doi/10.1103/PhysRevC.86.014611}.

\bibitem[Cap et~al.(2013)Cap, Siwek-Wilczy\ifmmode~\acute{n}\else
  \'{n}\fi{}ska, Kowal, and Wilczy\ifmmode~\acute{n}\else \'{n}\fi{}ski]{PRCOg}
T.~Cap, K.~Siwek-Wilczy\ifmmode~\acute{n}\else \'{n}\fi{}ska, M.~Kowal, and
  J.~Wilczy\ifmmode~\acute{n}\else \'{n}\fi{}ski.
\newblock Calculations of the cross sections for the synthesis of new
  ${}^{293--296}$118 isotopes in ${}^{249--252}$cf(${}^{48}$ca,$xn$) reactions.
\newblock \emph{Phys. Rev. C}, 88:\penalty0 037603, Sep 2013.
\newblock \doi{10.1103/PhysRevC.88.037603}.
\newblock URL \url{https://link.aps.org/doi/10.1103/PhysRevC.88.037603}.

\bibitem[Siwek-Wilczy\ifmmode~\acute{n}\else \'{n}\fi{}ska
  et~al.(2019)Siwek-Wilczy\ifmmode~\acute{n}\else \'{n}\fi{}ska, Cap, and
  Kowal]{PRC-ap}
K.~Siwek-Wilczy\ifmmode~\acute{n}\else \'{n}\fi{}ska, T.~Cap, and M.~Kowal.
\newblock Exploring the production of new superheavy nuclei with proton and
  $\ensuremath{\alpha}$-particle evaporation channels.
\newblock \emph{Phys. Rev. C}, 99:\penalty0 054603, May 2019.
\newblock \doi{10.1103/PhysRevC.99.054603}.
\newblock URL \url{https://link.aps.org/doi/10.1103/PhysRevC.99.054603}.

\bibitem[Siwek-Wilczy\'nska and Wilczy\'nski(2004)]{KSW04}
K.~Siwek-Wilczy\'nska and J.~Wilczy\'nski.
\newblock Empirical nucleus-nucleus potential deduced from fusion excitation
  functions.
\newblock \emph{Phys. Rev. C}, 69:\penalty0 024611, Feb 2004.
\newblock \doi{10.1103/PhysRevC.69.024611}.
\newblock URL \url{https://link.aps.org/doi/10.1103/PhysRevC.69.024611}.

\bibitem[Ntshangase et~al.(2007)Ntshangase, Rowley, Bark, Förtsch, Lawrie,
  Lawrie, Lindsay, Lipoglavsek, Maliage, Mudau, Mullins, Ndwandwe, Neveling,
  Sletten, Smit, and Theron]{NTSHANGASE200727}
S.S. Ntshangase, N.~Rowley, R.A. Bark, S.V. Förtsch, J.J. Lawrie, E.A. Lawrie,
  R.~Lindsay, M.~Lipoglavsek, S.M. Maliage, L.J. Mudau, S.M. Mullins, O.M.
  Ndwandwe, R.~Neveling, G.~Sletten, F.D. Smit, and C.~Theron.
\newblock Barrier distribution for a ‘superheavy’ nucleus–nucleus
  collision.
\newblock \emph{Physics Letters B}, 651\penalty0 (1):\penalty0 27--32, 2007.
\newblock ISSN 0370-2693.
\newblock \doi{https://doi.org/10.1016/j.physletb.2007.05.039}.
\newblock URL
  \url{https://www.sciencedirect.com/science/article/pii/S0370269307006314}.

\bibitem[Dasgupta et~al.(1998)Dasgupta, Hinde, Rowley, and
  Stefanini]{doi:10.1146/annurev.nucl.48.1.401}
M.~Dasgupta, D.~J. Hinde, N.~Rowley, and A.~M. Stefanini.
\newblock Measuring barriers to fusion.
\newblock \emph{Annual Review of Nuclear and Particle Science}, 48\penalty0
  (1):\penalty0 401--461, 1998.
\newblock \doi{10.1146/annurev.nucl.48.1.401}.
\newblock URL \url{https://doi.org/10.1146/annurev.nucl.48.1.401}.

\bibitem[Back et~al.(2014)Back, Esbensen, Jiang, and Rehm]{RevModPhys.86.317}
B.~B. Back, H.~Esbensen, C.~L. Jiang, and K.~E. Rehm.
\newblock Recent developments in heavy-ion fusion reactions.
\newblock \emph{Rev. Mod. Phys.}, 86:\penalty0 317--360, Mar 2014.
\newblock \doi{10.1103/RevModPhys.86.317}.
\newblock URL \url{https://link.aps.org/doi/10.1103/RevModPhys.86.317}.

\bibitem[Balantekin and Takigawa(1998)]{RevModPhys7077}
A.~B. Balantekin and N.~Takigawa.
\newblock Quantum tunneling in nuclear fusion.
\newblock \emph{Rev. Mod. Phys.}, 70:\penalty0 77--100, 1998.
\newblock \doi{10.1103/RevModPhys.70.77}.
\newblock URL \url{https://link.aps.org/doi/10.1103/RevModPhys.70.77}.

\bibitem[Rowley et~al.(1991)Rowley, Satchler, and Stelson]{ROWLEY199125}
N.~Rowley, G.R. Satchler, and P.H. Stelson.
\newblock On the “distribution of barriers” interpretation of heavy-ion
  fusion.
\newblock \emph{Physics Letters B}, 254\penalty0 (1):\penalty0 25--29, 1991.
\newblock ISSN 0370-2693.
\newblock \doi{https://doi.org/10.1016/0370-2693(91)90389-8}.
\newblock URL
  \url{https://www.sciencedirect.com/science/article/pii/0370269391903898}.

\bibitem[Timmers et~al.(1995)Timmers, Leigh, Dasgupta, Hinde, Lemmon, Mein,
  Morton, Newton, and Rowley]{TIMMERS1995190}
H.~Timmers, J.R. Leigh, M.~Dasgupta, D.J. Hinde, R.C. Lemmon, J.C. Mein, C.R.
  Morton, J.O. Newton, and N.~Rowley.
\newblock Probing fusion barrier distributions with quasi-elastic scattering.
\newblock \emph{Nuclear Physics A}, 584\penalty0 (1):\penalty0 190--204, 1995.
\newblock ISSN 0375-9474.
\newblock \doi{https://doi.org/10.1016/0375-9474(94)00521-N}.
\newblock URL
  \url{https://www.sciencedirect.com/science/article/pii/037594749400521N}.

\bibitem[B\l{}ocki and \'Swi\c{a}tecki(1982)]{Blocki}
J.~B\l{}ocki and W.~J. \'Swi\c{a}tecki.
\newblock Nuclear-deformation energiesaccording to a liquid-drop model with a
  sharp surface.
\newblock \emph{Lawrence Berkeley Laboratory preprint LBL-12811, May 1982
  (unpublished)}, 1982.
\newblock URL \url{https://www.osti.gov/biblio/6632591}.

\bibitem[Hofmann et~al.(2004)Hofmann, Heßberger, Ackermann, Antalic, Cagarda,
  Kindler, Kuusiniemi, Leino, Lommel, Malyshev, Mann, Mu¨nzenberg, Popeko,
  S´aro, Streicher, and Yeremin]{HOFMANN200493}
S.~Hofmann, F.P. Heßberger, D.~Ackermann, S.~Antalic, P.~Cagarda, B.~Kindler,
  P.~Kuusiniemi, M.~Leino, B.~Lommel, O.N. Malyshev, R.~Mann, G.~Mu¨nzenberg,
  A.G. Popeko, S.~S´aro, B.~Streicher, and A.V. Yeremin.
\newblock Properties of heavy nuclei measured at the gsi ship.
\newblock \emph{Nuclear Physics A}, 734:\penalty0 93--100, 2004.
\newblock ISSN 0375-9474.
\newblock \doi{https://doi.org/10.1016/j.nuclphysa.2004.01.018}.
\newblock URL
  \url{https://www.sciencedirect.com/science/article/pii/S0375947404000235}.

\bibitem[Hofmann(1998)]{Hofmann_1998}
S~Hofmann.
\newblock New elements - approaching $z=114$.
\newblock \emph{Reports on Progress in Physics}, 61\penalty0 (6):\penalty0
  639--689, jun 1998.
\newblock \doi{10.1088/0034-4885/61/6/002}.
\newblock URL \url{https://doi.org/10.1088/0034-4885/61/6/002}.

\bibitem[Dragojevi\ifmmode~\acute{c}\else \'{c}\fi{}
  et~al.(2008)Dragojevi\ifmmode~\acute{c}\else \'{c}\fi{}, Gregorich,
  D\"ullmann, Garcia, Gates, Nelson, Stavsetra, Sudowe, and
  Nitsche]{PhysRevC.78.024605}
I.~Dragojevi\ifmmode~\acute{c}\else \'{c}\fi{}, K.~E. Gregorich, Ch.~E.
  D\"ullmann, M.~A. Garcia, J.~M. Gates, S.~L. Nelson, L.~Stavsetra, R.~Sudowe,
  and H.~Nitsche.
\newblock Influence of projectile neutron number in the
  ${}^{208}\mathrm{Pb}({}^{48}\mathrm{Ti}, n){}^{255}\mathrm{Rf}$ and
  ${}^{208}\mathrm{Pb}({}^{50}\mathrm{Ti}, n){}^{257}\mathrm{Rf}$ reactions.
\newblock \emph{Phys. Rev. C}, 78:\penalty0 024605, Aug 2008.
\newblock \doi{10.1103/PhysRevC.78.024605}.
\newblock URL \url{https://link.aps.org/doi/10.1103/PhysRevC.78.024605}.

\bibitem[Hessberger et~al.(2001)Hessberger, Hofmann, Ackermann, Ninov, Leino,
  Munzenberg, Saro, Lavrentev, Popeko, Yeremin, and
  Stodel]{ISI:000172568800008}
FP~Hessberger, S~Hofmann, D~Ackermann, V~Ninov, M~Leino, G~Munzenberg, S~Saro,
  A~Lavrentev, AG~Popeko, AV~Yeremin, and C~Stodel.
\newblock Decay properties of neutron-deficient isotopes $^{256,257}$db,
  $^{255}$rf, $^{252,253}$lr.
\newblock \emph{EUROPEAN PHYSICAL JOURNAL A}, 12\penalty0 (1):\penalty0 57--67,
  SEP 2001.
\newblock ISSN 1434-6001.
\newblock \doi{10.1007/s100500170039}.

\bibitem[Gates et~al.(2008)Gates, Nelson, Gregorich,
  Dragojevi\ifmmode~\acute{c}\else \'{c}\fi{}, D\"ullmann, Ellison, Folden~III,
  Garcia, Stavsetra, Sudowe, Hoffman, and Nitsche]{PhysRevC.78.034604}
J.~M. Gates, S.~L. Nelson, K.~E. Gregorich, I.~Dragojevi\ifmmode~\acute{c}\else
  \'{c}\fi{}, Ch.~E. D\"ullmann, P.~A. Ellison, C.~M. Folden~III, M.~A. Garcia,
  L.~Stavsetra, R.~Sudowe, D.~C. Hoffman, and H.~Nitsche.
\newblock Comparison of reactions for the production of
  $^{258,257}\mathrm{Db}$: ${}^{208}\mathrm{Pb}({}^{51}\mathrm{V},\mathit{xn})$
  and ${}^{209}\mathrm{Bi}({}^{50}\mathrm{Ti},\mathit{xn})$.
\newblock \emph{Phys. Rev. C}, 78:\penalty0 034604, Sep 2008.
\newblock \doi{10.1103/PhysRevC.78.034604}.
\newblock URL \url{https://link.aps.org/doi/10.1103/PhysRevC.78.034604}.

\bibitem[Folden~III et~al.(2009)Folden~III, Dragojevi\ifmmode~\acute{c}\else
  \'{c}\fi{}, D\"ullmann, Eichler, Garcia, Gates, Nelson, Sudowe, Gregorich,
  Hoffman, and Nitsche]{PhysRevC.79.027602}
C.~M. Folden~III, I.~Dragojevi\ifmmode~\acute{c}\else \'{c}\fi{}, Ch.~E.
  D\"ullmann, R.~Eichler, M.~A. Garcia, J.~M. Gates, S.~L. Nelson, R.~Sudowe,
  K.~E. Gregorich, D.~C. Hoffman, and H.~Nitsche.
\newblock Measurement of the
  $^{208}\mathrm{Pb}(^{52}\mathrm{Cr},n)^{259}\mathrm{Sg}$ excitation function.
\newblock \emph{Phys. Rev. C}, 79:\penalty0 027602, Feb 2009.
\newblock \doi{10.1103/PhysRevC.79.027602}.
\newblock URL \url{https://link.aps.org/doi/10.1103/PhysRevC.79.027602}.

\bibitem[Nelson et~al.(2008{\natexlab{a}})Nelson, Folden~III, Gregorich,
  Dragojevi\ifmmode~\acute{c}\else \'{c}\fi{}, D\"ullmann, Eichler, Garcia,
  Gates, Sudowe, and Nitsche]{PhysRevC.78.024606}
S.~L. Nelson, C.~M. Folden~III, K.~E. Gregorich,
  I.~Dragojevi\ifmmode~\acute{c}\else \'{c}\fi{}, Ch.~E. D\"ullmann,
  R.~Eichler, M.~A. Garcia, J.~M. Gates, R.~Sudowe, and H.~Nitsche.
\newblock Comparison of complementary reactions for the production of
  $^{261,262}\mathrm{Bh}$.
\newblock \emph{Phys. Rev. C}, 78:\penalty0 024606, Aug 2008{\natexlab{a}}.
\newblock \doi{10.1103/PhysRevC.78.024606}.
\newblock URL \url{https://link.aps.org/doi/10.1103/PhysRevC.78.024606}.

\bibitem[MUNZENBERG et~al.(1989)MUNZENBERG, ARMBRUSTER, HOFMANN, HESSBERGER,
  FOLGER, KELLER, NINOV, POPPENSIEKER, QUINT, REISDORF, SCHMIDT, SCHNEIDER,
  SCHOTT, SUMMERER, ZYCHOR, LEINO, ACKERMANN, GOLLERTHAN, HANELT, MORAWEK,
  VERMEULEN, FUJITA, and SCHWAB]{ISI:A1989U697600008}
G~MUNZENBERG, P~ARMBRUSTER, S~HOFMANN, FP~HESSBERGER, H~FOLGER, JG~KELLER,
  V~NINOV, K~POPPENSIEKER, AB~QUINT, W~REISDORF, KH~SCHMIDT, JRH SCHNEIDER,
  HJ~SCHOTT, K~SUMMERER, I~ZYCHOR, ME~LEINO, D~ACKERMANN, U~GOLLERTHAN,
  E~HANELT, W~MORAWEK, D~VERMEULEN, Y~FUJITA, and T~SCHWAB.
\newblock Element 107.
\newblock \emph{ZEITSCHRIFT FUR PHYSIK A-HADRONS AND NUCLEI}, 333\penalty0
  (2):\penalty0 163--175, 1989.
\newblock ISSN 0939-7922.
\newblock \doi{10.1007/BF01565147}.

\bibitem[Nelson et~al.(2008{\natexlab{b}})Nelson, Gregorich,
  Dragojevi\ifmmode~\acute{c}\else \'{c}\fi{}, Garcia, Gates, Sudowe, and
  Nitsche]{PhysRevLett.100.022501}
S.~L. Nelson, K.~E. Gregorich, I.~Dragojevi\ifmmode~\acute{c}\else \'{c}\fi{},
  M.~A. Garcia, J.~M. Gates, R.~Sudowe, and H.~Nitsche.
\newblock Lightest isotope of bh produced via the
  $^{209}\mathrm{Bi}(^{52}\mathrm{Cr},n)^{260}\mathrm{Bh}$ reaction.
\newblock \emph{Phys. Rev. Lett.}, 100:\penalty0 022501, Jan
  2008{\natexlab{b}}.
\newblock \doi{10.1103/PhysRevLett.100.022501}.
\newblock URL \url{https://link.aps.org/doi/10.1103/PhysRevLett.100.022501}.

\bibitem[Folden~III et~al.(2006)Folden~III, Nelson, D\"ullmann, Schwantes,
  Sudowe, Zielinski, Gregorich, Nitsche, and Hoffman]{PhysRevC.73.014611}
C.~M. Folden~III, S.~L. Nelson, Ch.~E. D\"ullmann, J.~M. Schwantes, R.~Sudowe,
  P.~M. Zielinski, K.~E. Gregorich, H.~Nitsche, and D.~C. Hoffman.
\newblock Excitation function for the production of $^{262}\mathrm{Bh}$
  ($z=107$) in the odd-z-projectile reaction
  $^{208}\mathrm{Pb}$($^{55}\mathrm{Mn}$, $n$).
\newblock \emph{Phys. Rev. C}, 73:\penalty0 014611, Jan 2006.
\newblock \doi{10.1103/PhysRevC.73.014611}.
\newblock URL \url{https://link.aps.org/doi/10.1103/PhysRevC.73.014611}.

\bibitem[Dragojevi\ifmmode~\acute{c}\else \'{c}\fi{}
  et~al.(2009)Dragojevi\ifmmode~\acute{c}\else \'{c}\fi{}, Gregorich,
  D\"ullmann, Dvorak, Ellison, Gates, Nelson, Stavsetra, and
  Nitsche]{PhysRevC.79.011602}
I.~Dragojevi\ifmmode~\acute{c}\else \'{c}\fi{}, K.~E. Gregorich, Ch.~E.
  D\"ullmann, J.~Dvorak, P.~A. Ellison, J.~M. Gates, S.~L. Nelson,
  L.~Stavsetra, and H.~Nitsche.
\newblock New isotope $^{263}\mathrm{Hs}$.
\newblock \emph{Phys. Rev. C}, 79:\penalty0 011602, Jan 2009.
\newblock \doi{10.1103/PhysRevC.79.011602}.
\newblock URL \url{https://link.aps.org/doi/10.1103/PhysRevC.79.011602}.

\bibitem[Hofmann et~al.(1997)Hofmann, Hessberger, Ninov, Armbruster,
  Munzenberg, Stodel, Popeko, Yeremin, Saro, and Leino]{ISI:A1997XX26900004}
S~Hofmann, FP~Hessberger, V~Ninov, P~Armbruster, G~Munzenberg, C~Stodel,
  AG~Popeko, AV~Yeremin, S~Saro, and M~Leino.
\newblock Excitation function for the production of (265)108 and (266)109.
\newblock \emph{ZEITSCHRIFT FUR PHYSIK A-HADRONS AND NUCLEI}, 358\penalty0
  (4):\penalty0 377--378, SEP 1997.
\newblock ISSN 0939-7922.
\newblock \doi{10.1007/s002180050343}.

\bibitem[Nelson et~al.(2009)Nelson, Gregorich, Dragojevi\ifmmode~\acute{c}\else
  \'{c}\fi{}, Dvo\ifmmode~\check{r}\else \v{r}\fi{}\'ak, Ellison, Garcia,
  Gates, Stavsetra, Ali, and Nitsche]{PhysRevC.79.027605}
S.~L. Nelson, K.~E. Gregorich, I.~Dragojevi\ifmmode~\acute{c}\else \'{c}\fi{},
  J.~Dvo\ifmmode~\check{r}\else \v{r}\fi{}\'ak, P.~A. Ellison, M.~A. Garcia,
  J.~M. Gates, L.~Stavsetra, M.~N. Ali, and H.~Nitsche.
\newblock Comparison of complementary reactions in the production of mt.
\newblock \emph{Phys. Rev. C}, 79:\penalty0 027605, Feb 2009.
\newblock \doi{10.1103/PhysRevC.79.027605}.
\newblock URL \url{https://link.aps.org/doi/10.1103/PhysRevC.79.027605}.

\bibitem[Ginter et~al.(2003)Ginter, Gregorich, Loveland, Lee, Kirbach, Sudowe,
  Folden, Patin, Seward, Wilk, Zielinski, Aleklett, Eichler, Nitsche, and
  Hoffman]{PhysRevC.67.064609}
T.~N. Ginter, K.~E. Gregorich, W.~Loveland, D.~M. Lee, U.~W. Kirbach,
  R.~Sudowe, C.~M. Folden, J.~B. Patin, N.~Seward, P.~A. Wilk, P.~M. Zielinski,
  K.~Aleklett, R.~Eichler, H.~Nitsche, and D.~C. Hoffman.
\newblock Confirmation of production of element 110 by the
  ${}^{208}\mathrm{Pb}{(}^{64}\mathrm{Ni},n)$ reaction.
\newblock \emph{Phys. Rev. C}, 67:\penalty0 064609, Jun 2003.
\newblock \doi{10.1103/PhysRevC.67.064609}.
\newblock URL \url{https://link.aps.org/doi/10.1103/PhysRevC.67.064609}.

\bibitem[Folden et~al.(2004)Folden, Gregorich, D\"ullmann, Mahmud, Pang,
  Schwantes, Sudowe, Zielinski, Nitsche, and Hoffman]{PhysRevLett.93.212702}
C.~M. Folden, K.~E. Gregorich, Ch.~E. D\"ullmann, H.~Mahmud, G.~K. Pang, J.~M.
  Schwantes, R.~Sudowe, P.~M. Zielinski, H.~Nitsche, and D.~C. Hoffman.
\newblock Development of an odd-$z$-projectile reaction for heavy element
  synthesis:
  $^{208}\mathrm{P}\mathrm{b}(^{64}\mathrm{N}\mathrm{i},n)^{271}\mathrm{D}\mathrm{s}$
  and $^{208}\mathrm{P}\mathrm{b}(^{65}\mathrm{C}\mathrm{u},n)^{272}111$.
\newblock \emph{Phys. Rev. Lett.}, 93:\penalty0 212702, Nov 2004.
\newblock \doi{10.1103/PhysRevLett.93.212702}.
\newblock URL \url{https://link.aps.org/doi/10.1103/PhysRevLett.93.212702}.

\bibitem[Morita et~al.(2004{\natexlab{a}})Morita, Morimoto, Kaji, Haba,
  Ideguchi, Kanungo, Katori, Koura, Kudo, Ohnishi, Ozawa, Suda, Sueki,
  Tanihata, Xu, Yeremin, Yoneda, Yoshida, Zhao, and Zheng]{ISI:000223717300009}
K~Morita, K~Morimoto, D~Kaji, H~Haba, E~Ideguchi, R~Kanungo, K~Katori, H~Koura,
  H~Kudo, T~Ohnishi, A~Ozawa, T~Suda, K~Sueki, I~Tanihata, H~Xu, AV~Yeremin,
  A~Yoneda, A~Yoshida, YL~Zhao, and T~Zheng.
\newblock Production and decay of the isotope (271)ds (z=110).
\newblock \emph{EUROPEAN PHYSICAL JOURNAL A}, 21\penalty0 (2):\penalty0
  257--263, AUG 2004{\natexlab{a}}.
\newblock ISSN 1434-6001.
\newblock \doi{10.1140/epja/i2003-10205-1}.

\bibitem[Morita et~al.(2004{\natexlab{b}})Morita, Morimoto, Kaji, Goto, Haba,
  Ideguchi, Kanungo, Katori, Koura, Kudo, Ohnishi, Ozawa, Peter, Suda, Sueki,
  Tanihata, Tokanai, Xu, Yeremin, Yoneda, Yoshida, Zhao, and
  Zheng]{MORITA2004101}
K.~Morita, K.K. Morimoto, D.~Kaji, S.~Goto, H.~Haba, E.~Ideguchi, R.~Kanungo,
  K.~Katori, H.~Koura, H.~Kudo, T.~Ohnishi, A.~Ozawa, J.C. Peter, T.~Suda,
  K.~Sueki, I.~Tanihata, F.~Tokanai, H.~Xu, A.V. Yeremin, A.~Yoneda,
  A.~Yoshida, Y.-L. Zhao, and T.~Zheng.
\newblock Status of heavy element research using garis at riken.
\newblock \emph{Nuclear Physics A}, 734:\penalty0 101--108, 2004{\natexlab{b}}.
\newblock ISSN 0375-9474.
\newblock \doi{https://doi.org/10.1016/j.nuclphysa.2004.01.019}.
\newblock URL
  \url{https://www.sciencedirect.com/science/article/pii/S0375947404000247}.

\bibitem[Hofmann et~al.(2001)Hofmann, Hessberger, Ackermann, Antalic, Cagarda,
  Cwiok, Kindler, Kojouharova, Lommel, Mann, Munzenberg, Popeko, Saro, Schott,
  and Yeremin]{ISI:000167664500002}
S~Hofmann, FP~Hessberger, D~Ackermann, S~Antalic, P~Cagarda, S~Cwiok,
  B~Kindler, J~Kojouharova, B~Lommel, R~Mann, G~Munzenberg, AG~Popeko, S~Saro,
  HJ~Schott, and AV~Yeremin.
\newblock The new isotope (270)110 and its decay products (266)hs and (262)sg.
\newblock \emph{EUROPEAN PHYSICAL JOURNAL A}, 10\penalty0 (1):\penalty0 5--10,
  JAN 2001.
\newblock ISSN 1434-6001.
\newblock \doi{10.1007/s100500170137}.

\bibitem[Morita et~al.(2004{\natexlab{c}})Morita, Morimoto, Kaji, Haba,
  Ideguchi, C.~Peter, Kanungo, Katori, Koura, Kudo, Ohnishi, Ozawa, Suda,
  Sueki, Tanihata, Xu, V.~Yeremin, Yoneda, Yoshida, Zhao, Zheng, Goto, and
  Tokanai]{doi:10.1143/JPSJ.73.1738}
K.~Morita, K.~Morimoto, D.~Kaji, H.~Haba, E.~Ideguchi, J.~C.~Peter, R.~Kanungo,
  K.~Katori, H.~Koura, H.~Kudo, T.~Ohnishi, A.~Ozawa, T.~Suda, K.~Sueki,
  I.~Tanihata, H.~Xu, A.~V.~Yeremin, A.~Yoneda, A.~Yoshida, Y.-L. Zhao,
  T.~Zheng, S.~Goto, and F.~Tokanai.
\newblock Production and decay properties of 272111 and its daughter nuclei.
\newblock \emph{Journal of the Physical Society of Japan}, 73\penalty0
  (7):\penalty0 1738--1744, 2004{\natexlab{c}}.
\newblock \doi{10.1143/JPSJ.73.1738}.
\newblock URL \url{https://doi.org/10.1143/JPSJ.73.1738}.

\bibitem[Morita et~al.(2007)Morita, Morimoto, Kaji, Akiyama, Goto, Haba,
  Ideguchi, Katori, Koura, Kudo, Ohnishi, Ozawa, Suda, Sueki, Tokanai,
  Yamaguchi, Yoneda, and Yoshida]{doi:10.1143/JPSJ.76.043201}
Kosuke Morita, Kouji Morimoto, Daiya Kaji, Takahiro Akiyama, Sin-ichi Goto,
  Hiromitsu Haba, Eiji Ideguchi, Kenji Katori, Hiroyuki Koura, Hisaaki Kudo,
  Tetsuya Ohnishi, Akira Ozawa, Toshimi Suda, Keisuke Sueki, Fuyuki Tokanai,
  Takayuki Yamaguchi, Akira Yoneda, and Atsushi Yoshida.
\newblock Experiment on synthesis of an isotope 277112 by 208pb+70zn reaction.
\newblock \emph{Journal of the Physical Society of Japan}, 76\penalty0
  (4):\penalty0 043201, 2007.
\newblock \doi{10.1143/JPSJ.76.043201}.
\newblock URL \url{https://doi.org/10.1143/JPSJ.76.043201}.

\bibitem[Morita et~al.(2004{\natexlab{d}})Morita, Morimoto, Kaji, Akiyama,
  Goto, Haba, Ideguchi, Kanungo, Katori, Koura, Kudo, Ohnishi, Ozawa, Suda,
  Sueki, Xu, Yamaguchi, Yoneda, Yoshida, and Zhao]{doi:10.1143/JPSJ.73.2593}
Kosuke Morita, Kouji Morimoto, Daiya Kaji, Takahiro Akiyama, Sin-ichi Goto,
  Hiromitsu Haba, Eiji Ideguchi, Rituparna Kanungo, Kenji Katori, Hiroyuki
  Koura, Hisaaki Kudo, Tetsuya Ohnishi, Akira Ozawa, Toshimi Suda, Keisuke
  Sueki, HuShan Xu, Takayuki Yamaguchi, Akira Yoneda, Atsushi Yoshida, and
  YuLiang Zhao.
\newblock Experiment on the synthesis of element 113 in the reaction
  209bi(70zn,n)278113.
\newblock \emph{Journal of the Physical Society of Japan}, 73\penalty0
  (10):\penalty0 2593--2596, 2004{\natexlab{d}}.
\newblock \doi{10.1143/JPSJ.73.2593}.
\newblock URL \url{https://doi.org/10.1143/JPSJ.73.2593}.

\bibitem[Morita et~al.(2012)Morita, Morimoto, Kaji, Haba, Ozeki, Kudou, Sumita,
  Wakabayashi, Yoneda, Tanaka, Yamaki, Sakai, Akiyama, Goto, Hasebe, Huang,
  Huang, Ideguchi, Kasamatsu, Katori, Kariya, Kikunaga, Koura, Kudo, Mashiko,
  Mayama, Mitsuok~a, Moriya, Murakami, Murayama, Namai, Ozawa, Sato, Sueki,
  Takeyama, Tokanai, Yamaguchi, and Yoshida]{New113}
Kosuke Morita, Kouji Morimoto, Daiya Kaji, Hiromitsu Haba, Kazutaka Ozeki, Yuki
  Kudou, Takayuki Sumita, Yasuo Wakabayashi, Aki~ra Yoneda, Kengo Tanaka,
  Sayaka Yamaki, Ryutaro Sakai, Takahiro Akiyama, Shin-ichi Goto, Hiroo Hasebe,
  Minghui Huang, Tianheng Huang, Eiji Ideguchi, Yoshitaka Kasamatsu, Kenji
  Katori, Yoshiki Kariya, Hidetoshi Kikunaga, Hiroyuki Koura, Hisaaki Kudo,
  Akihiro Mashiko, Keita Mayama, Shin-ichi Mitsuok~a, Toru Moriya, Masashi
  Murakami, Hirohumi Murayama, Saori Namai, Akira Ozawa, Nozomi Sato, Keisuke
  Sueki, Mirei Takeyama, Fuyuki Tokanai, Takayuki Yamaguchi, and Atsushi
  Yoshida.
\newblock New result in the production and decay of an isotope, 278113, of the
  113th element.
\newblock \emph{Journal of the Physical Society of Japan}, 81\penalty0
  (10):\penalty0 103201, 2012.
\newblock \doi{10.1143/JPSJ.81.103201}.
\newblock URL \url{https://doi.org/10.1143/JPSJ.81.103201}.

\bibitem[Oganessian et~al.(2000{\natexlab{a}})Oganessian, Utyonkov, Lobanov,
  Abdullin, Polyakov, Shirokovsky, Tsyganov, Gulbekian, Bogomolov, Gikal,
  Mezentsev, Iliev, Subbotin, Sukhov, Ivanov, Buklanov, Subotic, Itkis, Moody,
  Wild, Stoyer, Stoyer, and Lougheed]{DUBNA-114}
Yu.~Ts. Oganessian, V.~K. Utyonkov, Yu.~V. Lobanov, F.~Sh. Abdullin, A.~N.
  Polyakov, I.~V. Shirokovsky, Yu.~S. Tsyganov, G.~G. Gulbekian, S.~L.
  Bogomolov, B.~N. Gikal, A.~N. Mezentsev, S.~Iliev, V.~G. Subbotin, A.~M.
  Sukhov, O.~V. Ivanov, G.~V. Buklanov, K.~Subotic, M.~G. Itkis, K.~J. Moody,
  J.~F. Wild, N.~J. Stoyer, M.~A. Stoyer, and R.~W. Lougheed.
\newblock Synthesis of superheavy nuclei in the
  ${}^{48}\mathrm{Ca}{+}^{244}\mathrm{Pu}$ reaction: ${}^{288}114$.
\newblock \emph{Phys. Rev. C}, 62:\penalty0 041604, Sep 2000{\natexlab{a}}.
\newblock \doi{10.1103/PhysRevC.62.041604}.
\newblock URL \url{https://link.aps.org/doi/10.1103/PhysRevC.62.041604}.

\bibitem[Oganessian et~al.(2000{\natexlab{b}})Oganessian, Utyonkov, Lobanov,
  Abdullin, Polyakov, Shirokovsky, Tsyganov, Gulbekian, Bogomolov, Gikal,
  Mezentsev, Iliev, Subbotin, Sukhov, Ivanov, Buklanov, Subotic, Itkis, Moody,
  Wild, Stoyer, Stoyer, Lougheed, Laue, Karelin, and Tatarinov]{DUBNA-116_1}
Yu.~Ts. Oganessian, V.~K. Utyonkov, Yu.~V. Lobanov, F.~Sh. Abdullin, A.~N.
  Polyakov, I.~V. Shirokovsky, Yu.~S. Tsyganov, G.~G. Gulbekian, S.~L.
  Bogomolov, B.~N. Gikal, A.~N. Mezentsev, S.~Iliev, V.~G. Subbotin, A.~M.
  Sukhov, O.~V. Ivanov, G.~V. Buklanov, K.~Subotic, M.~G. Itkis, K.~J. Moody,
  J.~F. Wild, N.~J. Stoyer, M.~A. Stoyer, R.~W. Lougheed, C.~A. Laue, Ye.~A.
  Karelin, and A.~N. Tatarinov.
\newblock Observation of the decay of ${}^{292}116$.
\newblock \emph{Phys. Rev. C}, 63:\penalty0 011301, Dec 2000{\natexlab{b}}.
\newblock \doi{10.1103/PhysRevC.63.011301}.
\newblock URL \url{https://link.aps.org/doi/10.1103/PhysRevC.63.011301}.

\bibitem[Oganessian et~al.(2001)Oganessian, Utyonkov, and Moody]{DUBNA-116_2}
Yu.~Ts. Oganessian, V.~K. Utyonkov, and K.~J. Moody.
\newblock Synthesis of 292116 in the 248cm + 48ca reaction.
\newblock \emph{Physics of Atomic Nuclei}, 64:\penalty0 1349--1355, 2001.
\newblock \doi{10.1134/1.1398925}.
\newblock URL \url{https://doi.org/10.1134/1.1398925}.

\bibitem[Oganessian et~al.(2004{\natexlab{a}})Oganessian, Utyonkov, Lobanov,
  Abdullin, Polyakov, Shirokovsky, Tsyganov, Gulbekian, Bogomolov, Gikal,
  Mezentsev, Iliev, Subbotin, Sukhov, Voinov, Buklanov, Subotic, Zagrebaev,
  Itkis, Patin, Moody, Wild, Stoyer, Stoyer, Shaughnessy, Kenneally, and
  Lougheed]{DUBNA-114+116}
Yu.~Ts. Oganessian, V.~K. Utyonkov, Yu.~V. Lobanov, F.~Sh. Abdullin, A.~N.
  Polyakov, I.~V. Shirokovsky, Yu.~S. Tsyganov, G.~G. Gulbekian, S.~L.
  Bogomolov, B.~N. Gikal, A.~N. Mezentsev, S.~Iliev, V.~G. Subbotin, A.~M.
  Sukhov, A.~A. Voinov, G.~V. Buklanov, K.~Subotic, V.~I. Zagrebaev, M.~G.
  Itkis, J.~B. Patin, K.~J. Moody, J.~F. Wild, M.~A. Stoyer, N.~J. Stoyer,
  D.~A. Shaughnessy, J.~M. Kenneally, and R.~W. Lougheed.
\newblock Measurements of cross sections for the fusion-evaporation reactions
  $^{244}\mathrm{Pu}{(^{48}\mathrm{Ca},xn)}^{292\ensuremath{-}x}114$ and
  $^{245}\mathrm{Cm}{(^{48}\mathrm{Ca},xn)}^{293\ensuremath{-}x}116$.
\newblock \emph{Phys. Rev. C}, 69:\penalty0 054607, May 2004{\natexlab{a}}.
\newblock \doi{10.1103/PhysRevC.69.054607}.
\newblock URL \url{https://link.aps.org/doi/10.1103/PhysRevC.69.054607}.

\bibitem[Oganessian et~al.(2004{\natexlab{b}})Oganessian, Utyonkov, Lobanov,
  Abdullin, Polyakov, Shirokovsky, Tsyganov, Gulbekian, Bogomolov, Gikal,
  Mezentsev, Iliev, Subbotin, Sukhov, Voinov, Buklanov, Subotic, Zagrebaev,
  Itkis, Patin, Moody, Wild, Stoyer, Stoyer, Shaughnessy, Kenneally, Wilk,
  Lougheed, Il'kaev, and Vesnovskii]{DUBNA-112.114.116}
Yu.~Ts. Oganessian, V.~K. Utyonkov, Yu.~V. Lobanov, F.~Sh. Abdullin, A.~N.
  Polyakov, I.~V. Shirokovsky, Yu.~S. Tsyganov, G.~G. Gulbekian, S.~L.
  Bogomolov, B.~N. Gikal, A.~N. Mezentsev, S.~Iliev, V.~G. Subbotin, A.~M.
  Sukhov, A.~A. Voinov, G.~V. Buklanov, K.~Subotic, V.~I. Zagrebaev, M.~G.
  Itkis, J.~B. Patin, K.~J. Moody, J.~F. Wild, M.~A. Stoyer, N.~J. Stoyer,
  D.~A. Shaughnessy, J.~M. Kenneally, P.~A. Wilk, R.~W. Lougheed, R.~I.
  Il'kaev, and S.~P. Vesnovskii.
\newblock Measurements of cross sections and decay properties of the isotopes
  of elements 112, 114, and 116 produced in the fusion reactions
  $^{233,238}\mathrm{U}$, $^{242}\mathrm{Pu}$, and
  $^{248}\mathrm{Cm}+^{48}\mathrm{Ca}$.
\newblock \emph{Phys. Rev. C}, 70:\penalty0 064609, Dec 2004{\natexlab{b}}.
\newblock \doi{10.1103/PhysRevC.70.064609}.
\newblock URL \url{https://link.aps.org/doi/10.1103/PhysRevC.70.064609}.

\bibitem[Oganessian et~al.(2006)Oganessian, Utyonkov, Lobanov, Abdullin,
  Polyakov, Sagaidak, Shirokovsky, Tsyganov, Voinov, Gulbekian, Bogomolov,
  Gikal, Mezentsev, Iliev, Subbotin, Sukhov, Subotic, Zagrebaev, Vostokin,
  Itkis, Moody, Patin, Shaughnessy, Stoyer, Stoyer, Wilk, Kenneally, Landrum,
  Wild, and Lougheed]{DUBNA-116.118}
Yu.~Ts. Oganessian, V.~K. Utyonkov, Yu.~V. Lobanov, F.~Sh. Abdullin, A.~N.
  Polyakov, R.~N. Sagaidak, I.~V. Shirokovsky, Yu.~S. Tsyganov, A.~A. Voinov,
  G.~G. Gulbekian, S.~L. Bogomolov, B.~N. Gikal, A.~N. Mezentsev, S.~Iliev,
  V.~G. Subbotin, A.~M. Sukhov, K.~Subotic, V.~I. Zagrebaev, G.~K. Vostokin,
  M.~G. Itkis, K.~J. Moody, J.~B. Patin, D.~A. Shaughnessy, M.~A. Stoyer, N.~J.
  Stoyer, P.~A. Wilk, J.~M. Kenneally, J.~H. Landrum, J.~F. Wild, and R.~W.
  Lougheed.
\newblock Synthesis of the isotopes of elements 118 and 116 in the
  $^{249}\mathrm{Cf}$ and $^{245}\mathrm{Cm}+^{48}\mathrm{Ca}$ fusion
  reactions.
\newblock \emph{Phys. Rev. C}, 74:\penalty0 044602, Oct 2006.
\newblock \doi{10.1103/PhysRevC.74.044602}.
\newblock URL \url{https://link.aps.org/doi/10.1103/PhysRevC.74.044602}.

\bibitem[Oganessian et~al.(2005)Oganessian, Utyonkov, Dmitriev, Lobanov, Itkis,
  Polyakov, Tsyganov, Mezentsev, Yeremin, Voinov, Sokol, Gulbekian, Bogomolov,
  Iliev, Subbotin, Sukhov, Buklanov, Shishkin, Chepygin, Vostokin, Aksenov,
  Hussonnois, Subotic, Zagrebaev, Moody, Patin, Wild, Stoyer, Stoyer,
  Shaughnessy, Kenneally, Wilk, Lougheed, G\"aggeler, Schumann, Bruchertseifer,
  and Eichler]{DUBNA-113+115}
Yu.~Ts. Oganessian, V.~K. Utyonkov, S.~N. Dmitriev, Yu.~V. Lobanov, M.~G.
  Itkis, A.~N. Polyakov, Yu.~S. Tsyganov, A.~N. Mezentsev, A.~V. Yeremin, A.~A.
  Voinov, E.~A. Sokol, G.~G. Gulbekian, S.~L. Bogomolov, S.~Iliev, V.~G.
  Subbotin, A.~M. Sukhov, G.~V. Buklanov, S.~V. Shishkin, V.~I. Chepygin, G.~K.
  Vostokin, N.~V. Aksenov, M.~Hussonnois, K.~Subotic, V.~I. Zagrebaev, K.~J.
  Moody, J.~B. Patin, J.~F. Wild, M.~A. Stoyer, N.~J. Stoyer, D.~A.
  Shaughnessy, J.~M. Kenneally, P.~A. Wilk, R.~W. Lougheed, H.~W. G\"aggeler,
  D.~Schumann, H.~Bruchertseifer, and R.~Eichler.
\newblock Synthesis of elements 115 and 113 in the reaction
  $^{243}\mathrm{Am}+^{48}\mathrm{Ca}$.
\newblock \emph{Phys. Rev. C}, 72:\penalty0 034611, Sep 2005.
\newblock \doi{10.1103/PhysRevC.72.034611}.
\newblock URL \url{https://link.aps.org/doi/10.1103/PhysRevC.72.034611}.

\bibitem[Oganessian et~al.(2011)Oganessian, Abdullin, Bailey, Benker, Bennett,
  Dmitriev, Ezold, Hamilton, Henderson, Itkis, Lobanov, Mezentsev, Moody,
  Nelson, Polyakov, Porter, Ramayya, Riley, Roberto, Ryabinin, Rykaczewski,
  Sagaidak, Shaughnessy, Shirokovsky, Stoyer, Subbotin, Sudowe, Sukhov, Taylor,
  Tsyganov, Utyonkov, Voinov, Vostokin, and Wilk]{DUBNA-117}
Yu.~Ts. Oganessian, F.~Sh. Abdullin, P.~D. Bailey, D.~E. Benker, M.~E. Bennett,
  S.~N. Dmitriev, J.~G. Ezold, J.~H. Hamilton, R.~A. Henderson, M.~G. Itkis,
  Yu.~V. Lobanov, A.~N. Mezentsev, K.~J. Moody, S.~L. Nelson, A.~N. Polyakov,
  C.~E. Porter, A.~V. Ramayya, F.~D. Riley, J.~B. Roberto, M.~A. Ryabinin,
  K.~P. Rykaczewski, R.~N. Sagaidak, D.~A. Shaughnessy, I.~V. Shirokovsky,
  M.~A. Stoyer, V.~G. Subbotin, R.~Sudowe, A.~M. Sukhov, R.~Taylor, Yu.~S.
  Tsyganov, V.~K. Utyonkov, A.~A. Voinov, G.~K. Vostokin, and P.~A. Wilk.
\newblock Eleven new heaviest isotopes of elements $z=105$ to $z=117$
  identified among the products of $^{249}\mathrm{Bk}$$+$$^{48}\mathrm{Ca}$
  reactions.
\newblock \emph{Phys. Rev. C}, 83:\penalty0 054315, May 2011.
\newblock \doi{10.1103/PhysRevC.83.054315}.
\newblock URL \url{https://link.aps.org/doi/10.1103/PhysRevC.83.054315}.

\bibitem[Oganessian et~al.(2013{\natexlab{a}})Oganessian, Abdullin, Alexander,
  Binder, Boll, Dmitriev, Ezold, Felker, Gostic, Grzywacz, Hamilton, Henderson,
  Itkis, Miernik, Miller, Moody, Polyakov, Ramayya, Roberto, Ryabinin,
  Rykaczewski, Sagaidak, Shaughnessy, Shirokovsky, Shumeiko, Stoyer, Stoyer,
  Subbotin, Sukhov, Tsyganov, Utyonkov, Voinov, and Vostokin]{DUBNA-117_2}
Yu.~Ts. Oganessian, F.~Sh. Abdullin, C.~Alexander, J.~Binder, R.~A. Boll, S.~N.
  Dmitriev, J.~Ezold, K.~Felker, J.~M. Gostic, R.~K. Grzywacz, J.~H. Hamilton,
  R.~A. Henderson, M.~G. Itkis, K.~Miernik, D.~Miller, K.~J. Moody, A.~N.
  Polyakov, A.~V. Ramayya, J.~B. Roberto, M.~A. Ryabinin, K.~P. Rykaczewski,
  R.~N. Sagaidak, D.~A. Shaughnessy, I.~V. Shirokovsky, M.~V. Shumeiko, M.~A.
  Stoyer, N.~J. Stoyer, V.~G. Subbotin, A.~M. Sukhov, Yu.~S. Tsyganov, V.~K.
  Utyonkov, A.~A. Voinov, and G.~K. Vostokin.
\newblock Experimental studies of the ${}^{249}$bk + ${}^{48}$ca reaction
  including decay properties and excitation function for isotopes of element
  117, and discovery of the new isotope ${}^{277}$mt.
\newblock \emph{Phys. Rev. C}, 87:\penalty0 054621, May 2013{\natexlab{a}}.
\newblock \doi{10.1103/PhysRevC.87.054621}.
\newblock URL \url{https://link.aps.org/doi/10.1103/PhysRevC.87.054621}.

\bibitem[Oganessian et~al.(2012{\natexlab{a}})Oganessian, Abdullin, Alexander,
  Binder, Boll, Dmitriev, Ezold, Felker, Gostic, Grzywacz, Hamilton, Henderson,
  Itkis, Miernik, Miller, Moody, Polyakov, Ramayya, Roberto, Ryabinin,
  Rykaczewski, Sagaidak, Shaughnessy, Shirokovsky, Shumeiko, Stoyer, Stoyer,
  Subbotin, Sukhov, Tsyganov, Utyonkov, Voinov, and Vostokin]{DUBNA-117+118}
Yu.~Ts. Oganessian, F.~Sh. Abdullin, C.~Alexander, J.~Binder, R.~A. Boll, S.~N.
  Dmitriev, J.~Ezold, K.~Felker, J.~M. Gostic, R.~K. Grzywacz, J.~H. Hamilton,
  R.~A. Henderson, M.~G. Itkis, K.~Miernik, D.~Miller, K.~J. Moody, A.~N.
  Polyakov, A.~V. Ramayya, J.~B. Roberto, M.~A. Ryabinin, K.~P. Rykaczewski,
  R.~N. Sagaidak, D.~A. Shaughnessy, I.~V. Shirokovsky, M.~V. Shumeiko, M.~A.
  Stoyer, N.~J. Stoyer, V.~G. Subbotin, A.~M. Sukhov, Yu.~S. Tsyganov, V.~K.
  Utyonkov, A.~A. Voinov, and G.~K. Vostokin.
\newblock Production and decay of the heaviest nuclei $^{293,294}117$ and
  $^{294}118$.
\newblock \emph{Phys. Rev. Lett.}, 109:\penalty0 162501, Oct
  2012{\natexlab{a}}.
\newblock \doi{10.1103/PhysRevLett.109.162501}.
\newblock URL \url{https://link.aps.org/doi/10.1103/PhysRevLett.109.162501}.

\bibitem[Oganessian et~al.(2013{\natexlab{b}})Oganessian, Abdullin, Dmitriev,
  Gostic, Hamilton, Henderson, Itkis, Moody, Polyakov, Ramayya, Roberto,
  Rykaczewski, Sagaidak, Shaughnessy, Shirokovsky, Stoyer, Stoyer, Subbotin,
  Sukhov, Tsyganov, Utyonkov, Voinov, and Vostokin]{DUBNA-113+115+117}
Yu.~Ts. Oganessian, F.~Sh. Abdullin, S.~N. Dmitriev, J.~M. Gostic, J.~H.
  Hamilton, R.~A. Henderson, M.~G. Itkis, K.~J. Moody, A.~N. Polyakov, A.~V.
  Ramayya, J.~B. Roberto, K.~P. Rykaczewski, R.~N. Sagaidak, D.~A. Shaughnessy,
  I.~V. Shirokovsky, M.~A. Stoyer, N.~J. Stoyer, V.~G. Subbotin, A.~M. Sukhov,
  Yu.~S. Tsyganov, V.~K. Utyonkov, A.~A. Voinov, and G.~K. Vostokin.
\newblock Investigation of the ${}^{243}\mathrm{Am}+{}^{48}\mathrm{Ca}$
  reaction products previously observed in the experiments on elements 113,
  115, and 117.
\newblock \emph{Phys. Rev. C}, 87:\penalty0 014302, Jan 2013{\natexlab{b}}.
\newblock \doi{10.1103/PhysRevC.87.014302}.
\newblock URL \url{https://link.aps.org/doi/10.1103/PhysRevC.87.014302}.

\bibitem[Oganessian et~al.(2004{\natexlab{c}})Oganessian, Utyonkoy, Lobanov,
  Abdullin, Polyakov, Shirokovsky, Tsyganov, Gulbekian, Bogomolov, Mezentsev,
  Iliev, Subbotin, Sukhov, Voinov, Buklanov, Subotic, Zagrebaev, Itkis, Patin,
  Moody, Wild, Stoyer, Stoyer, Shaughnessy, Kenneally, and Lougheed]{DUBNA-115}
Yu.~Ts. Oganessian, V.~K. Utyonkoy, Yu.~V. Lobanov, F.~Sh. Abdullin, A.~N.
  Polyakov, I.~V. Shirokovsky, Yu.~S. Tsyganov, G.~G. Gulbekian, S.~L.
  Bogomolov, A.~N. Mezentsev, S.~Iliev, V.~G. Subbotin, A.~M. Sukhov, A.~A.
  Voinov, G.~V. Buklanov, K.~Subotic, V.~I. Zagrebaev, M.~G. Itkis, J.~B.
  Patin, K.~J. Moody, J.~F. Wild, M.~A. Stoyer, N.~J. Stoyer, D.~A.
  Shaughnessy, J.~M. Kenneally, and R.~W. Lougheed.
\newblock Experiments on the synthesis of element 115 in the reaction
  $^{243}\mathrm{Am}(^{48}\mathrm{Ca},xn)^{291\ensuremath{-}x}115$.
\newblock \emph{Phys. Rev. C}, 69:\penalty0 021601, Feb 2004{\natexlab{c}}.
\newblock \doi{10.1103/PhysRevC.69.021601}.
\newblock URL \url{https://link.aps.org/doi/10.1103/PhysRevC.69.021601}.

\bibitem[Oganessian et~al.(2012{\natexlab{b}})Oganessian, Abdullin, Dmitriev,
  Gostic, Hamilton, Henderson, Itkis, Moody, Polyakov, Ramayya, Roberto,
  Rykaczewski, Sagaidak, Shaughnessy, Shirokovsky, Stoyer, Subbotin, Sukhov,
  Tsyganov, Utyonkov, Voinov, and Vostokin]{DUBNA-113.115.117}
Yu.~Ts. Oganessian, F.~Sh. Abdullin, S.~N. Dmitriev, J.~M. Gostic, J.~H.
  Hamilton, R.~A. Henderson, M.~G. Itkis, K.~J. Moody, A.~N. Polyakov, A.~V.
  Ramayya, J.~B. Roberto, K.~P. Rykaczewski, R.~N. Sagaidak, D.~A. Shaughnessy,
  I.~V. Shirokovsky, M.~A. Stoyer, V.~G. Subbotin, A.~M. Sukhov, Yu.~S.
  Tsyganov, V.~K. Utyonkov, A.~A. Voinov, and G.~K. Vostokin.
\newblock New insights into the $^{243}\mathrm{Am}+^{48}\mathrm{Ca}$ reaction
  products previously observed in the experiments on elements 113, 115, and
  117.
\newblock \emph{Phys. Rev. Lett.}, 108:\penalty0 022502, Jan
  2012{\natexlab{b}}.
\newblock \doi{10.1103/PhysRevLett.108.022502}.
\newblock URL \url{https://link.aps.org/doi/10.1103/PhysRevLett.108.022502}.

\bibitem[Oganessian et~al.(2010)Oganessian, Abdullin, Bailey, Benker, Bennett,
  Dmitriev, Ezold, Hamilton, Henderson, Itkis, Lobanov, Mezentsev, Moody,
  Nelson, Polyakov, Porter, Ramayya, Riley, Roberto, Ryabinin, Rykaczewski,
  Sagaidak, Shaughnessy, Shirokovsky, Stoyer, Subbotin, Sudowe, Sukhov,
  Tsyganov, Utyonkov, Voinov, Vostokin, and Wilk]{DUBNA-117_3}
Yu.~Ts. Oganessian, F.~Sh. Abdullin, P.~D. Bailey, D.~E. Benker, M.~E. Bennett,
  S.~N. Dmitriev, J.~G. Ezold, J.~H. Hamilton, R.~A. Henderson, M.~G. Itkis,
  Yu.~V. Lobanov, A.~N. Mezentsev, K.~J. Moody, S.~L. Nelson, A.~N. Polyakov,
  C.~E. Porter, A.~V. Ramayya, F.~D. Riley, J.~B. Roberto, M.~A. Ryabinin,
  K.~P. Rykaczewski, R.~N. Sagaidak, D.~A. Shaughnessy, I.~V. Shirokovsky,
  M.~A. Stoyer, V.~G. Subbotin, R.~Sudowe, A.~M. Sukhov, Yu.~S. Tsyganov, V.~K.
  Utyonkov, A.~A. Voinov, G.~K. Vostokin, and P.~A. Wilk.
\newblock Synthesis of a new element with atomic number $z=117$.
\newblock \emph{Phys. Rev. Lett.}, 104:\penalty0 142502, Apr 2010.
\newblock \doi{10.1103/PhysRevLett.104.142502}.
\newblock URL \url{https://link.aps.org/doi/10.1103/PhysRevLett.104.142502}.

\bibitem[Hofmann et~al.(2012)Hofmann, Heinz, Mann, Maurer, Khuyagbaatar,
  Ackermann, Antalic, Barth, Block, Burkhard, Comas, Dahl, Eberhardt, Gostic,
  Henderson, Heredia, Heßberger, Kenneally, Kindler, Kojouharov, Kratz, Lang,
  Leino, Lommel, Moody, Münzenberg, Nelson, Nishio, Popeko, Runke, Saro,
  Shaughnessy, Stoyer, Thörle-Pospiech, Tinschert, Trautmann, Uusitalo, Wilk,
  and Yeremin]{GSI-116}
S.~Hofmann, S.~Heinz, R.~Mann, J.~Maurer, J.~Khuyagbaatar, D.~Ackermann,
  S.~Antalic, W.~Barth, M.~Block, H.~G. Burkhard, V.~F. Comas, L.~Dahl,
  K.~Eberhardt, J.~Gostic, R.~A. Henderson, J.~A. Heredia, F.~P. Heßberger,
  J.~M. Kenneally, B.~Kindler, I.~Kojouharov, J.~V. Kratz, R.~Lang, M.~Leino,
  B.~Lommel, K.~J. Moody, G.~Münzenberg, S.~L. Nelson, K.~Nishio, A.~G.
  Popeko, J.~Runke, S.~Saro, D.~A. Shaughnessy, M.~A. Stoyer,
  P.~Thörle-Pospiech, K.~Tinschert, N.~Trautmann, J.~Uusitalo, P.~A. Wilk, and
  A.~V. Yeremin.
\newblock The reaction 48ca + 248cm → 296116* studied at the gsi-ship.
\newblock \emph{The European Physical Journal A}, 48:\penalty0 62, 2012.
\newblock \doi{10.1140/epja/i2012-12062-1}.
\newblock URL \url{https://doi.org/10.1140/epja/i2012-12062-1}.

\bibitem[Stavsetra et~al.(2009)Stavsetra, Gregorich, Dvorak, Ellison,
  Dragojevi\ifmmode~\acute{c}\else \'{c}\fi{}, Garcia, and Nitsche]{LBNL-114}
L.~Stavsetra, K.~E. Gregorich, J.~Dvorak, P.~A. Ellison,
  I.~Dragojevi\ifmmode~\acute{c}\else \'{c}\fi{}, M.~A. Garcia, and H.~Nitsche.
\newblock Independent verification of element 114 production in the
  $^{48}\mathrm{Ca}+^{242}\mathrm{Pu}$ reaction.
\newblock \emph{Phys. Rev. Lett.}, 103:\penalty0 132502, Sep 2009.
\newblock \doi{10.1103/PhysRevLett.103.132502}.
\newblock URL \url{https://link.aps.org/doi/10.1103/PhysRevLett.103.132502}.

\bibitem[Ellison et~al.(2010)Ellison, Gregorich, Berryman, Bleuel, Clark,
  Dragojevi\ifmmode~\acute{c}\else \'{c}\fi{}, Dvorak, Fallon,
  Fineman-Sotomayor, Gates, Gothe, Lee, Loveland, McLaughlin, Paschalis, Petri,
  Qian, Stavsetra, Wiedeking, and Nitsche]{LBNL-114_2}
P.~A. Ellison, K.~E. Gregorich, J.~S. Berryman, D.~L. Bleuel, R.~M. Clark,
  I.~Dragojevi\ifmmode~\acute{c}\else \'{c}\fi{}, J.~Dvorak, P.~Fallon,
  C.~Fineman-Sotomayor, J.~M. Gates, O.~R. Gothe, I.~Y. Lee, W.~D. Loveland,
  J.~P. McLaughlin, S.~Paschalis, M.~Petri, J.~Qian, L.~Stavsetra,
  M.~Wiedeking, and H.~Nitsche.
\newblock New superheavy element isotopes:
  $^{242}\mathrm{Pu}(^{48}\mathrm{Ca},5n)^{285}114$.
\newblock \emph{Phys. Rev. Lett.}, 105:\penalty0 182701, Oct 2010.
\newblock \doi{10.1103/PhysRevLett.105.182701}.
\newblock URL \url{https://link.aps.org/doi/10.1103/PhysRevLett.105.182701}.

\bibitem[D\"ullmann et~al.(2010)D\"ullmann, Sch\"adel, Yakushev, T\"urler,
  Eberhardt, Kratz, Ackermann, Andersson, Block, Br\"uchle, Dvorak, Essel,
  Ellison, Even, Gates, Gorshkov, Graeger, Gregorich, Hartmann, Herzberg,
  He\ss{}berger, Hild, H\"ubner, J\"ager, Khuyagbaatar, Kindler, Krier, Kurz,
  Lahiri, Liebe, Lommel, Maiti, Nitsche, Omtvedt, Parr, Rudolph, Runke,
  Schausten, Schimpf, Semchenkov, Steiner, Th\"orle-Pospiech, Uusitalo,
  Wegrzecki, and Wiehl]{TASCA-114_1}
Ch.~E. D\"ullmann, M.~Sch\"adel, A.~Yakushev, A.~T\"urler, K.~Eberhardt, J.~V.
  Kratz, D.~Ackermann, L.-L. Andersson, M.~Block, W.~Br\"uchle, J.~Dvorak,
  H.~G. Essel, P.~A. Ellison, J.~Even, J.~M. Gates, A.~Gorshkov, R.~Graeger,
  K.~E. Gregorich, W.~Hartmann, R.-D. Herzberg, F.~P. He\ss{}berger, D.~Hild,
  A.~H\"ubner, E.~J\"ager, J.~Khuyagbaatar, B.~Kindler, J.~Krier, N.~Kurz,
  S.~Lahiri, D.~Liebe, B.~Lommel, M.~Maiti, H.~Nitsche, J.~P. Omtvedt, E.~Parr,
  D.~Rudolph, J.~Runke, B.~Schausten, E.~Schimpf, A.~Semchenkov, J.~Steiner,
  P.~Th\"orle-Pospiech, J.~Uusitalo, M.~Wegrzecki, and N.~Wiehl.
\newblock Production and decay of element 114: High cross sections and the new
  nucleus $^{277}\mathrm{Hs}$.
\newblock \emph{Phys. Rev. Lett.}, 104:\penalty0 252701, Jun 2010.
\newblock \doi{10.1103/PhysRevLett.104.252701}.
\newblock URL \url{https://link.aps.org/doi/10.1103/PhysRevLett.104.252701}.

\bibitem[Gates et~al.(2011)Gates, D\"ullmann, Sch\"adel, Yakushev, T\"urler,
  Eberhardt, Kratz, Ackermann, Andersson, Block, Br\"uchle, Dvorak, Essel,
  Ellison, Even, Forsberg, Gellanki, Gorshkov, Graeger, Gregorich, Hartmann,
  Herzberg, He\ss{}berger, Hild, H\"ubner, J\"ager, Khuyagbaatar, Kindler,
  Krier, Kurz, Lahiri, Liebe, Lommel, Maiti, Nitsche, Omtvedt, Parr, Rudolph,
  Runke, Schaffner, Schausten, Schimpf, Semchenkov, Steiner, Th\"orle-Pospiech,
  Uusitalo, Wegrzecki, and Wiehl]{TASCA-114_2}
J.~M. Gates, Ch.~E. D\"ullmann, M.~Sch\"adel, A.~Yakushev, A.~T\"urler,
  K.~Eberhardt, J.~V. Kratz, D.~Ackermann, L.-L. Andersson, M.~Block,
  W.~Br\"uchle, J.~Dvorak, H.~G. Essel, P.~A. Ellison, J.~Even, U.~Forsberg,
  J.~Gellanki, A.~Gorshkov, R.~Graeger, K.~E. Gregorich, W.~Hartmann, R.-D.
  Herzberg, F.~P. He\ss{}berger, D.~Hild, A.~H\"ubner, E.~J\"ager,
  J.~Khuyagbaatar, B.~Kindler, J.~Krier, N.~Kurz, S.~Lahiri, D.~Liebe,
  B.~Lommel, M.~Maiti, H.~Nitsche, J.~P. Omtvedt, E.~Parr, D.~Rudolph,
  J.~Runke, H.~Schaffner, B.~Schausten, E.~Schimpf, A.~Semchenkov, J.~Steiner,
  P.~Th\"orle-Pospiech, J.~Uusitalo, M.~Wegrzecki, and N.~Wiehl.
\newblock First superheavy element experiments at the gsi recoil separator
  tasca: The production and decay of element 114 in the
  $^{244}\mathrm{Pu}$($^{48}\mathrm{Ca}$,3-4$n$) reaction.
\newblock \emph{Phys. Rev. C}, 83:\penalty0 054618, May 2011.
\newblock \doi{10.1103/PhysRevC.83.054618}.
\newblock URL \url{https://link.aps.org/doi/10.1103/PhysRevC.83.054618}.

\bibitem[Khuyagbaatar et~al.(2014)Khuyagbaatar, Yakushev, D\"ullmann,
  Ackermann, Andersson, Asai, Block, Boll, Brand, Cox, Dasgupta, Derkx,
  Di~Nitto, Eberhardt, Even, Evers, Fahlander, Forsberg, Gates, Gharibyan,
  Golubev, Gregorich, Hamilton, Hartmann, Herzberg, He\ss{}berger, Hinde,
  Hoffmann, Hollinger, H\"ubner, J\"ager, Kindler, Kratz, Krier, Kurz,
  Laatiaoui, Lahiri, Lang, Lommel, Maiti, Miernik, Minami, Mistry, Mokry,
  Nitsche, Omtvedt, Pang, Papadakis, Renisch, Roberto, Rudolph, Runke,
  Rykaczewski, Sarmiento, Sch\"adel, Schausten, Semchenkov, Shaughnessy,
  Steinegger, Steiner, Tereshatov, Th\"orle-Pospiech, Tinschert, Torres
  De~Heidenreich, Trautmann, T\"urler, Uusitalo, Ward, Wegrzecki, Wiehl,
  Van~Cleve, and Yakusheva]{TASCA-117}
J.~Khuyagbaatar, A.~Yakushev, Ch.~E. D\"ullmann, D.~Ackermann, L.-L. Andersson,
  M.~Asai, M.~Block, R.~A. Boll, H.~Brand, D.~M. Cox, M.~Dasgupta, X.~Derkx,
  A.~Di~Nitto, K.~Eberhardt, J.~Even, M.~Evers, C.~Fahlander, U.~Forsberg,
  J.~M. Gates, N.~Gharibyan, P.~Golubev, K.~E. Gregorich, J.~H. Hamilton,
  W.~Hartmann, R.-D. Herzberg, F.~P. He\ss{}berger, D.~J. Hinde, J.~Hoffmann,
  R.~Hollinger, A.~H\"ubner, E.~J\"ager, B.~Kindler, J.~V. Kratz, J.~Krier,
  N.~Kurz, M.~Laatiaoui, S.~Lahiri, R.~Lang, B.~Lommel, M.~Maiti, K.~Miernik,
  S.~Minami, A.~Mistry, C.~Mokry, H.~Nitsche, J.~P. Omtvedt, G.~K. Pang,
  P.~Papadakis, D.~Renisch, J.~Roberto, D.~Rudolph, J.~Runke, K.~P.
  Rykaczewski, L.~G. Sarmiento, M.~Sch\"adel, B.~Schausten, A.~Semchenkov,
  D.~A. Shaughnessy, P.~Steinegger, J.~Steiner, E.~E. Tereshatov,
  P.~Th\"orle-Pospiech, K.~Tinschert, T.~Torres De~Heidenreich, N.~Trautmann,
  A.~T\"urler, J.~Uusitalo, D.~E. Ward, M.~Wegrzecki, N.~Wiehl, S.~M.
  Van~Cleve, and V.~Yakusheva.
\newblock $^{48}\mathrm{Ca}+^{249}\mathrm{Bk}$ fusion reaction leading to
  element $z=117$: Long-lived $\ensuremath{\alpha}$-decaying
  $^{270}\mathrm{Db}$ and discovery of $^{266}\mathrm{Lr}$.
\newblock \emph{Phys. Rev. Lett.}, 112:\penalty0 172501, May 2014.
\newblock \doi{10.1103/PhysRevLett.112.172501}.
\newblock URL \url{https://link.aps.org/doi/10.1103/PhysRevLett.112.172501}.

\bibitem[Boilley et~al.(2019)Boilley, Abe, Cauchois, and Shen]{Boilley}
David Boilley, Yasuhisa Abe, Bartholom{\'{e}} Cauchois, and Caiwan Shen.
\newblock Elimination of fast variables and initial slip: a new mechanism for
  fusion hindrance in heavy-ion collisions.
\newblock \emph{Journal of Physics G: Nuclear and Particle Physics},
  46\penalty0 (11):\penalty0 115102, sep 2019.
\newblock \doi{10.1088/1361-6471/ab11ef}.
\newblock URL \url{https://doi.org/10.1088/1361-6471/ab11ef}.

\bibitem[Heßberger(2019)]{TiBip}
F.~P. Heßberger.
\newblock On the synthesis of 258rf via p-deexcitation in the complete fusion
  reaction 50ti + 209bi.
\newblock \emph{The European Physical Journal A}, 55:\penalty0 208, 2019.
\newblock \doi{10.1140/epja/i2019-12912-2}.
\newblock URL \url{https://doi.org/10.1140/epja/i2019-12912-2}.

\bibitem[Hong et~al.(2016)Hong, Adamian, and Antonenko]{PhysRevC.94.044606}
Juhee Hong, G.~G. Adamian, and N.~V. Antonenko.
\newblock Possibilities of production of transfermium nuclei in
  charged-particle evaporation channels.
\newblock \emph{Phys. Rev. C}, 94:\penalty0 044606, Oct 2016.
\newblock \doi{10.1103/PhysRevC.94.044606}.
\newblock URL \url{https://link.aps.org/doi/10.1103/PhysRevC.94.044606}.

\bibitem[Hong et~al.(2020)Hong, Adamian, Antonenko, Jachimowicz, and
  Kowal]{HONG2020135760}
J.~Hong, G.G. Adamian, N.V. Antonenko, P.~Jachimowicz, and M.~Kowal.
\newblock Possibilities of direct production of superheavy nuclei with
  z=112–118 in different evaporation channels.
\newblock \emph{Physics Letters B}, 809:\penalty0 135760, 2020.
\newblock ISSN 0370-2693.
\newblock \doi{https://doi.org/10.1016/j.physletb.2020.135760}.
\newblock URL
  \url{https://www.sciencedirect.com/science/article/pii/S0370269320305633}.

\bibitem[Hong et~al.(2017)Hong, Adamian, and Antonenko]{HONG201742}
Juhee Hong, G.G. Adamian, and N.V. Antonenko.
\newblock Ways to produce new superheavy isotopes with z=111–117 in charged
  particle evaporation channels.
\newblock \emph{Physics Letters B}, 764:\penalty0 42--48, 2017.
\newblock ISSN 0370-2693.
\newblock \doi{https://doi.org/10.1016/j.physletb.2016.11.002}.
\newblock URL
  \url{https://www.sciencedirect.com/science/article/pii/S0370269316306633}.

\bibitem[Reisdorf()]{Reisdorf}
W.~Reisdorf.
\newblock Analysis of fissionability data at high excitation energies.
\newblock \emph{Zeitschrift für Physik A Atoms and Nuclei}, 300:\penalty0
  227--238.
\newblock \doi{10.1007/BF01412298}.
\newblock URL \url{https://doi.org/10.1007/BF01412298}.

\bibitem[Ignatyuk et~al.(1975)Ignatyuk, Smirenkin, and Tishin]{Ignatiuk}
A~V Ignatyuk, G~N Smirenkin, and A~S Tishin.
\newblock Phenomenological description of energy dependence of the level
  density parameter.
\newblock \emph{Yad. Fiz., v. 21, no. 3, pp. 485-490}, 3 1975.
\newblock URL \url{https://www.osti.gov/biblio/4175339}.

\bibitem[Rahmatinejad et~al.(2021)Rahmatinejad, Bezbakh, Shneidman, Adamian,
  Antonenko, Jachimowicz, and Kowal]{level1}
A.~Rahmatinejad, A.~N. Bezbakh, T.~M. Shneidman, G.~Adamian, N.~V. Antonenko,
  P.~Jachimowicz, and M.~Kowal.
\newblock Level-density parameters in superheavy nuclei.
\newblock \emph{Phys. Rev. C}, 103:\penalty0 034309, Mar 2021.
\newblock \doi{10.1103/PhysRevC.103.034309}.
\newblock URL \url{https://link.aps.org/doi/10.1103/PhysRevC.103.034309}.

\bibitem[Rahmatinejad et~al.(2022)Rahmatinejad, Shneidman, Adamian, Antonenko,
  Jachimowicz, and Kowal]{level2}
A.~Rahmatinejad, T.~M. Shneidman, G.~G. Adamian, N.~V. Antonenko,
  P.~Jachimowicz, and M.~Kowal.
\newblock Energy dependent ratios of level-density parameters in superheavy
  nuclei.
\newblock \emph{Phys. Rev. C}, 105:\penalty0 044328, Apr 2022.
\newblock \doi{10.1103/PhysRevC.105.044328}.
\newblock URL \url{https://link.aps.org/doi/10.1103/PhysRevC.105.044328}.

\bibitem[Cwiok et~al.(1987)Cwiok, Dudek, Nazarewicz, Skalski, and
  Werner]{dudek}
S.~Cwiok, J.~Dudek, W.~Nazarewicz, J.~Skalski, and T.~Werner.
\newblock Single-particle energies, wave functions, quadrupole moments and
  g-factors in an axially deformed woods-saxon potential with applications to
  the two-centre-type nuclear problems.
\newblock \emph{Computer Physics Communications}, 46\penalty0 (3):\penalty0
  379--399, 1987.
\newblock ISSN 0010-4655.
\newblock \doi{https://doi.org/10.1016/0010-4655(87)90093-2}.
\newblock URL
  \url{https://www.sciencedirect.com/science/article/pii/0010465587900932}.

\bibitem[Krappe et~al.(1979)Krappe, Nix, and Sierk]{krappe}
H.~J. Krappe, J.~R. Nix, and A.~J. Sierk.
\newblock Unified nuclear potential for heavy-ion elastic scattering, fusion,
  fission, and ground-state masses and deformations.
\newblock \emph{Phys. Rev. C}, 20:\penalty0 992--1013, Sep 1979.
\newblock \doi{10.1103/PhysRevC.20.992}.
\newblock URL \url{https://link.aps.org/doi/10.1103/PhysRevC.20.992}.

\bibitem[Cap et~al.(2022)Cap, Kowal, and Siwek-Wilczy\ifmmode~\acute{n}\else
  \'{n}\fi{}ska]{PRC-Pfus}
T.~Cap, M.~Kowal, and K.~Siwek-Wilczy\ifmmode~\acute{n}\else \'{n}\fi{}ska.
\newblock Diffusion as a possible mechanism controlling the production of
  superheavy nuclei in cold fusion reactions.
\newblock \emph{Phys. Rev. C}, 105:\penalty0 L051601, May 2022.
\newblock \doi{10.1103/PhysRevC.105.L051601}.
\newblock URL \url{https://link.aps.org/doi/10.1103/PhysRevC.105.L051601}.

\bibitem[Hagino et~al.(1999)Hagino, Rowley, and Kruppa]{Hagino}
K.~Hagino, N.~Rowley, and A.T. Kruppa.
\newblock A program for coupled-channel calculations with all order couplings
  for heavy-ion fusion reactions.
\newblock \emph{Computer Physics Communications}, 123\penalty0 (1):\penalty0
  143--152, 1999.
\newblock ISSN 0010-4655.
\newblock \doi{https://doi.org/10.1016/S0010-4655(99)00243-X}.
\newblock URL
  \url{https://www.sciencedirect.com/science/article/pii/S001046559900243X}.

\bibitem[Itkis et~al.(2011)Itkis, Kozulin, Itkis, Knyazheva, Bogachev,
  Chernysheva, Krupa, Oganessian, Zagrebaev, Rusanov, Goennenwein, Dorvaux,
  Stuttg\'e, Hanappe, Vardaci, and de~Go\'es~Brennand]{PhysRevC.83.064613}
I.~M. Itkis, E.~M. Kozulin, M.~G. Itkis, G.~N. Knyazheva, A.~A. Bogachev, E.~V.
  Chernysheva, L.~Krupa, Yu.~Ts. Oganessian, V.~I. Zagrebaev, A.~Ya. Rusanov,
  F.~Goennenwein, O.~Dorvaux, L.~Stuttg\'e, F.~Hanappe, E.~Vardaci, and
  E.~de~Go\'es~Brennand.
\newblock Fission and quasifission modes in heavy-ion-induced reactions leading
  to the formation of hs${}^{*}$.
\newblock \emph{Phys. Rev. C}, 83:\penalty0 064613, Jun 2011.
\newblock \doi{10.1103/PhysRevC.83.064613}.
\newblock URL \url{https://link.aps.org/doi/10.1103/PhysRevC.83.064613}.

\bibitem[Kozulin et~al.(2016)Kozulin, Knyazheva, Novikov, Itkis, Itkis,
  Dmitriev, Oganessian, Bogachev, Kozulina, Harca, Trzaska, and
  Ghosh]{PhysRevC.94.054613}
E.~M. Kozulin, G.~N. Knyazheva, K.~V. Novikov, I.~M. Itkis, M.~G. Itkis, S.~N.
  Dmitriev, Yu.~Ts. Oganessian, A.~A. Bogachev, N.~I. Kozulina, I.~Harca, W.~H.
  Trzaska, and T.~K. Ghosh.
\newblock Fission and quasifission of composite systems with
  $z=108\ensuremath{-}120$: Transition from heavy-ion reactions involving s and
  ca to ti and ni ions.
\newblock \emph{Phys. Rev. C}, 94:\penalty0 054613, Nov 2016.
\newblock \doi{10.1103/PhysRevC.94.054613}.
\newblock URL \url{https://link.aps.org/doi/10.1103/PhysRevC.94.054613}.

\bibitem[Hong et~al.(2021)Hong, Adamian, Antonenko, Jachimowicz, and
  Kowal]{Hong}
J.~Hong, G.~G. Adamian, N.~V. Antonenko, P.~Jachimowicz, and M.~Kowal.
\newblock Rate of decline of the production cross section of superheavy nuclei
  with $z=114--117$ at high excitation energies.
\newblock \emph{Phys. Rev. C}, 103:\penalty0 L041601, Apr 2021.
\newblock \doi{10.1103/PhysRevC.103.L041601}.
\newblock URL \url{https://link.aps.org/doi/10.1103/PhysRevC.103.L041601}.

\end{thebibliography}

\end{document}